\pdfoutput=1
\documentclass[twocolumn]{IEEEtran}

\def\BibTeX{{\rm B\kern-.05em{\sc i\kern-.025em b}\kern-.08em
    T\kern-.1667em\lower.7ex\hbox{E}\kern-.125emX}}

\usepackage{balance}
\usepackage{amsmath,amssymb,amsfonts,bm}
\usepackage{algorithmic}
\usepackage[linesnumbered,ruled]{algorithm2e}
\usepackage{url}
\usepackage{multirow}
\usepackage{graphicx}
\usepackage{cite}
\usepackage{color}
\usepackage{xspace}
\usepackage{csquotes}
\usepackage{lipsum}
\usepackage[labelfont=bf]{caption}
\usepackage{subcaption}

\usepackage{textcomp}

\usepackage{hhline}

\allowdisplaybreaks[4]

\newcommand{\blue}[1]{\textcolor{black}{#1}}

\newcommand{\etal}{\text{et~al.}\xspace}

\newcommand{\eat}[1]{}
\newcommand{\spara}[1]{\smallskip\noindent\textbf{#1}.}

\begin{document}


\title{Multi-modal Deep Analysis for Multimedia
}

\author{
Wenwu Zhu,~\IEEEmembership{Fellow,~IEEE}, Xin Wang,~\IEEEmembership{Member,~IEEE}, Hongzhi Li~\IEEEmembership{Member,~IEEE}

\IEEEcompsocitemizethanks{
	\IEEEcompsocthanksitem Manuscript received February 12, 2019; revised June 21, 2019; accepted August 28, 2019.
	This work is in part supported by the National Program on Key Basic Research
	Project under Grant 2015CB352300, in part by the China Postdoctoral Science Foundation under Grant BX201700136 and in part by the National Natural Science Foundation of China Major Project under Grant U1611461. ({\it Corresponding author: Xin Wang.})
	\IEEEcompsocthanksitem W. Zhu and X. Wang are with the
	Department of Computer Science and Technology, Tsinghua University, Beijing, China (e-mail: wwzhu@tsinghua.edu.cn; xin\_wang@tsinghua.edu.cn).
	\IEEEcompsocthanksitem
	H. Li is with Microsoft Research, Redmond, USA (e-mail: Hongzhi.Li@microsoft.com).
	\IEEEcompsocthanksitem
	Digital Object Identifier 10.1109/TCSVT.2019.2940647
}

}

\markboth{IEEE TRANSACTIONS ON CIRCUITS AND SYSTEMS FOR VIDEO TECHNOLOGY}%
{Shell \MakeLowercase{\textit{et al.}}: Bare Demo of IEEEtran.cls for IEEE Journals}

\maketitle

\begin{abstract}
With the rapid development of Internet and multimedia
services in the past decade, a huge amount of user-generated
and service provider-generated multimedia data become available.
These data are heterogeneous and multi-modal in nature,
imposing great challenges for processing and analyzing them.
Multi-modal data consist of a mixture of various
types of data from different modalities such as texts, images, videos, audios etc.
In this article, we present a deep and comprehensive overview for multi-modal
analysis in multimedia. We introduce two scientific research problems,
data-driven correlational representation and knowledge-guided fusion
for multimedia analysis. To address the two scientific problems,
we investigate them from the following aspects:
1) {\it multi-modal correlational representation}: multi-modal fusion of data across different modalities,
and 2){\it multi-modal data and knowledge fusion}: multi-modal fusion of data with domain knowledge.
More specifically, on data-driven correlational representation,
we highlight three important categories of
methods, such as multi-modal deep representation,
multi-modal transfer learning, and multi-modal hashing.
On knowledge-guided fusion,
we discuss the approaches for fusing knowledge with data and four exemplar
applications that require various kinds of domain knowledge,
including multi-modal visual question answering,
multi-modal video summarization, multi-modal visual pattern
mining and multi-modal recommendation.
Finally, we bring forward our insights and future research directions.
\end{abstract}

\begin{IEEEkeywords}
Multi-modal analysis, Data-driven correlational representation, Knowledge-guided data fusion
\end{IEEEkeywords}

\let\thefootnote\relax\footnotetext{Copyright \textcircled{c} 2019 IEEE. Personal use of this material is permitted. However, permission to use this material for any other purposes must be obtained from the IEEE by sending an email to pubs-permissions@ieee.org.}




\section{Introduction}
\label{sec:intro}

\IEEEPARstart{W}{e} ARE now living in the era of {\it Cyber}, {\it Physical} and {\it Human} (CPH) spaces.
The Moore Law illustrates that the CPU speed will double every 18 months, 
resulting in the ubiquity of computing;
the Bell Laws indicates that the chip size tends to reduce by half every 18 months, 
making devices including all types of sensors everywhere;
the Gilder‘s Law shows that the network bandwidth can double every 6 months, 
causing communications which connect human, computers and physical identities 
to be ubiquitous in our daily lives.
In short, data are everywhere in the era of {\it Cyber}, {\it Physical} and {\it Human} (CPH) spaces.
For example, various kinds of user-generated and service provider-generated data in social media
together with the growing popularity of other information sources such as cell phones and cameras
have produced a large amount of multi-modal multimedia data.
Multi-modal data consists of a mixture of various types of data such as texts, images,
audios, videos etc.
In the past decades, most researchers\eat{, in a very long period of time,} focus on analyzing data in a single modality, 
making uni-modal/single-medium analysis a well studied topic to date.
However, we need to study multi-modal data in real life,
which is particular important when we enter the ``Artificial Intelligence (AI) Epoch''. 
Discoveries in cognitive science~\cite{mcgurk1976hearing} have confirmed the fact that human are able to
perceive their surrounding environment through fusing the feedback from multiple sensory organs (eyes, nose, ears, etc.) together.
As such, the investigation of multi-modal analysis serves as a very promising direction in boosting the progress
of research in big data and AI.
Fortunately, the advent of multi-modal data brings us great opportunities for multi-modal analysis in
multimedia.
\begin{figure*}[h!]
	\centering
	\begin{tabular}{ccc}
		\includegraphics[width=0.29\linewidth, height = 4.8cm]{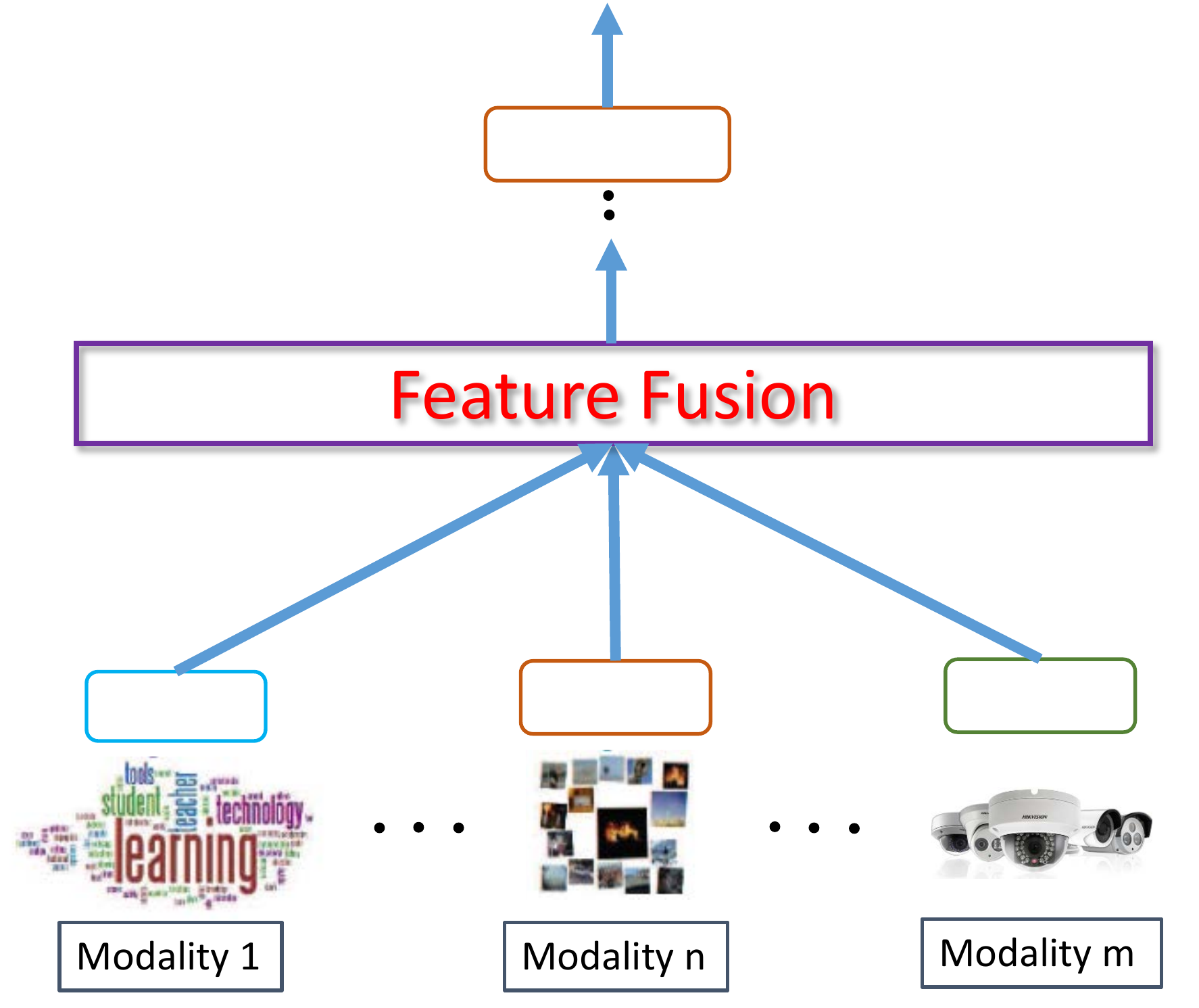} & 
		\includegraphics[width=0.29\linewidth, height = 4.7cm]{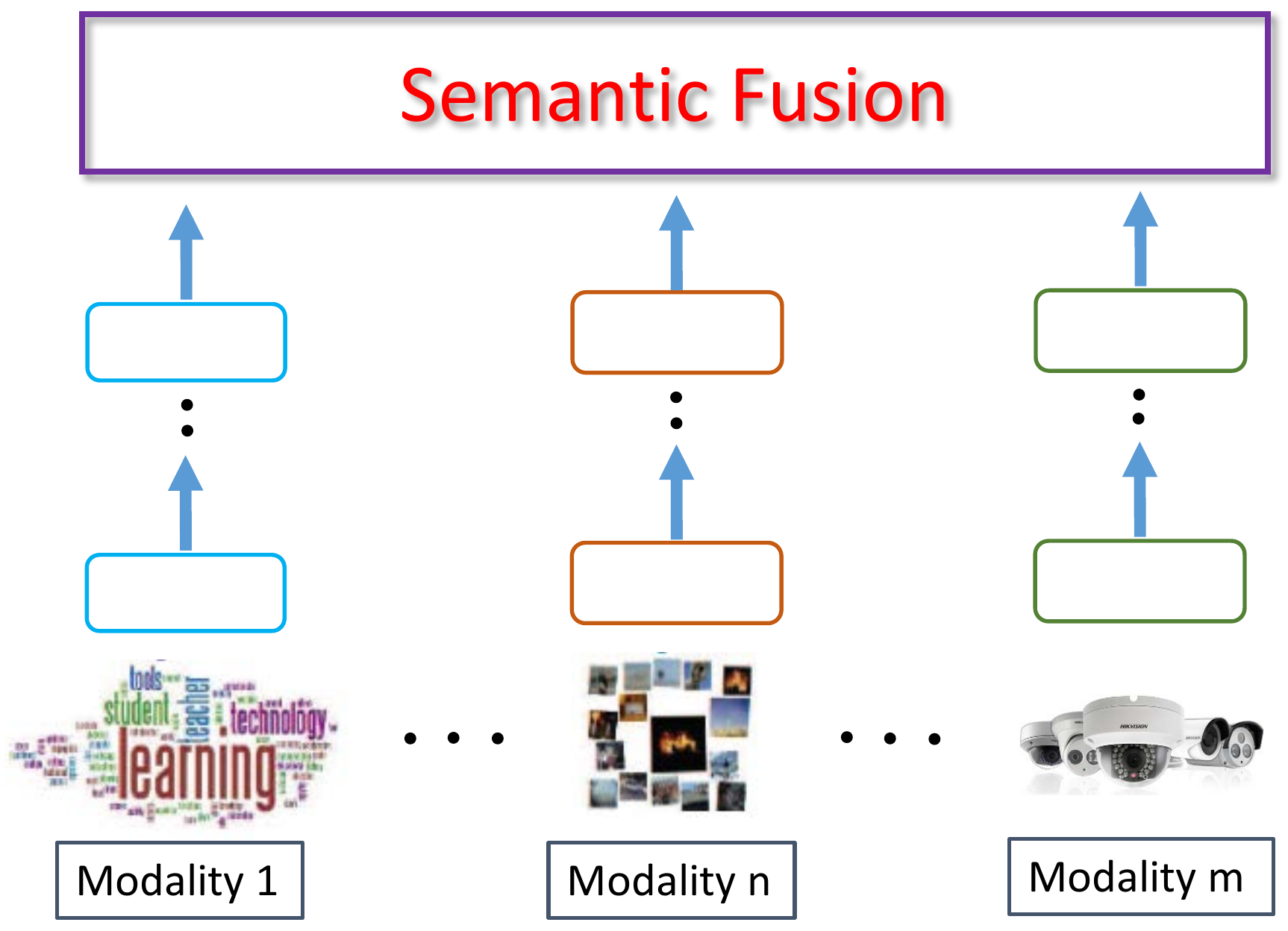} &
		\includegraphics[width=0.29\linewidth, height = 4.8cm]{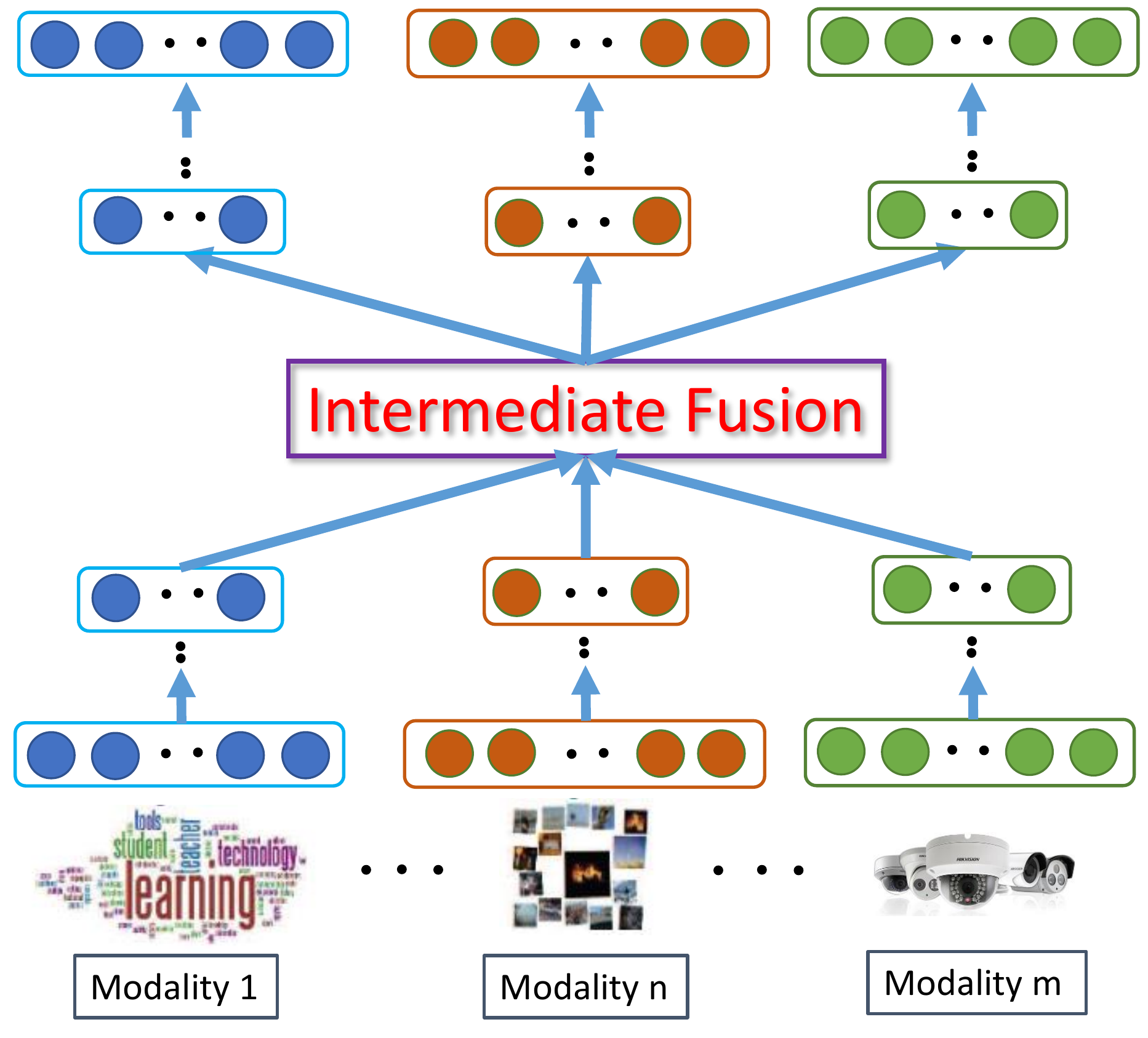} \\
		(a) Feature fusion & (b) Semantic fusion  & (c) Intermediate fusion 
	\end{tabular}
	\caption{Schematic diagram for semantic fusion methods and intermediate fusion methods}
	\label{fig:feature_semantic_intermediate_fusion}
\end{figure*}

Nevertheless, analyzing multi-modal multimedia data imposes great challenges. 
One scientific problem is how to jointly consider and fuse information from different modalities such that
multi-modal approaches are able to outperform uni-modal methods which utilize information from single 
modality separately. Traditional approaches for multi-modal analysis can be categorized into two groups: 
{\it feature fusion} and {\it semantic fusion}.
Feature fusion (also known as feature engineering) approaches simply conduct feature concatenations on 
raw features from different modalities, which is normally
achieved via manual operations and has very low efficiency, as is shown in Figure~\ref{fig:feature_semantic_intermediate_fusion}(a). 
Semantic fusion first analyzes information from single modality separately in the
beginning and conduct multi-modal fusion at semantic level, as is illustrated in Figure~\ref{fig:feature_semantic_intermediate_fusion}(b).
This type of methods can maintain the explainability in semantic fusion, but fails to make full use of the rich information hidden in
multi-modal.

Thanks to the success of deep neural network in computer science, a new type of approaches capable of
fusing information from different modalities in hidden space at intermediate level cuts a splendid figure
in multi-modal analysis, as is demonstrated in Figure~\ref{fig:feature_semantic_intermediate_fusion}(c). 
This type of methods can fully utilize the multi-modal
data through learning a correlational representation for different modality in a data-driven way.
\begin{figure}[h!]
	\centering
	\includegraphics[width=0.89\linewidth]{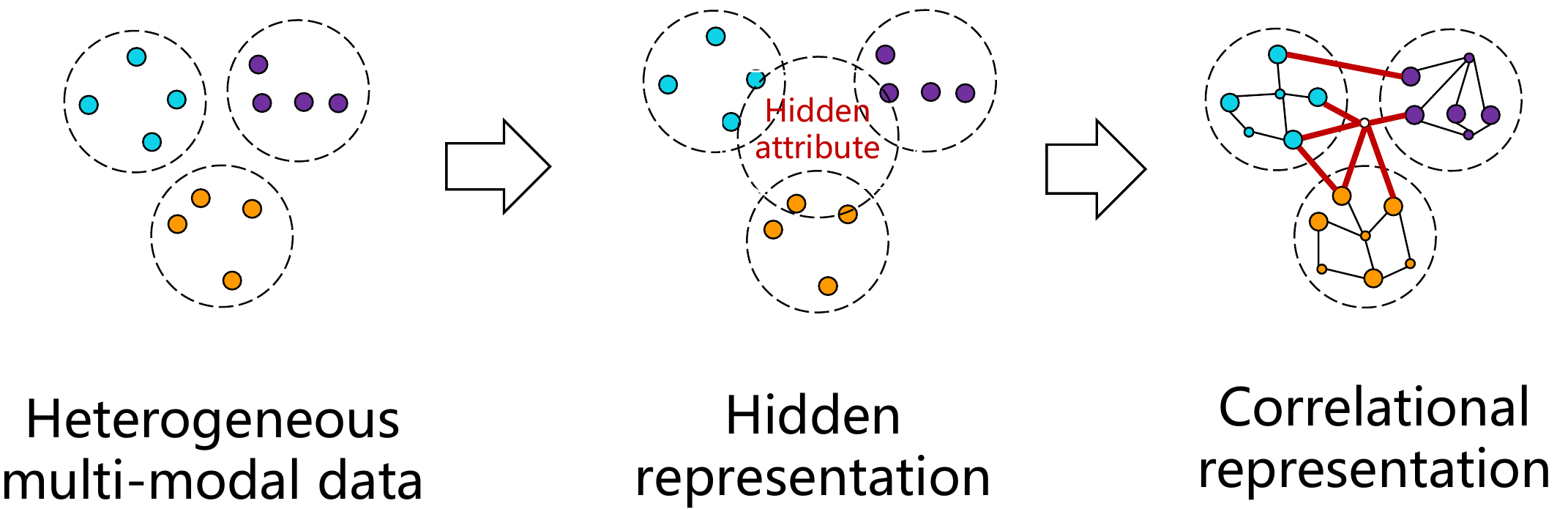}
	\caption{Correlational representation}
	\label{fig:cr2}
\end{figure}
Figure~\ref{fig:cr2} demonstrates a common way for multi-modal correlational 
representation, which is to map multi-modal data (leftmost) to a hidden representation (middle)
and/or correlational representation (rightmost). 
Quite a few methods including deep learning can be used to learn 
the hidden representation and further correlational mining techniques are necessary for the
correlational representation learning.

Though being capable of handling large-scale multi-modal data, 
the results obtained from data-driven approaches (e.g., deep neural networks)
can sometimes be unexplainable, which do not utilize
too much domain knowledge, leading to a huge exploration space and low accuracy.
Therefore it is very challenging to get explainable correlational representations from uncertain big data.
Human, on the other hand, is capable of utilizing domain knowledge to help making decisions, 
resulting in high explainability and accuracy.
As such, there exists a paradox between scalability and explainability, and it is desirable to 
figure out a balance which 
requires the best cooperative fusion between data and knowledge
between data-driven and knowledge-driven methods.

The goal of this article is two-fold. 
We first give a deep and comprehensive overview for multi-modal problems in multimedia  
from two aspects: 1) {\it data-driven correlational representation}: multi-modal fusion of data from different modalities and
2) {\it knowledge-guided data fusion}: multi-modal fusion of data with domain knowledge.
We then present our insights and thinkings on future directions for multi-modal research 
in the new era of artificial intelligence, and point out several promising research directions
including cross-modal reasoning, cross-modal cognition and cross-modal collective intelligence, for further investigation.

One natural scientific problem is how to find a hidden representation that can best correlate information
from different modalities. Several methodologies have the potential to tackle this challenge and we highlight three important categories
of data-driven approaches that focus on multi-modal correlational representation:
Multi-modal Deep Representation, Multi-modal Transfer Learning and Multi-modal Hashing.

Given an effective hidden correlation representation for different modality data, 
the next scientific research problem is how we can increase the explainability of 
data-driven approaches while maintaining their scalability
via the guidance of domain knowledge and take advantage of their superiority.
However, to the best of our knowledge, there have been no systematic or consolidated
methodologies for incorporating domain knowledge into the process of cross-modal learning.
We observe that there exist mainly three families of methods that may be suitable for knowledge-guided
cross-modal fusion, i.e., Bayesian Inference, Teacher-student Network and Reinforcement Learning.
We will elaborate our thoughts on why these three methodologies deserve further investigations for future research
later in the paper.

On the other hand, many existing approaches for multi-modal oriented problems
have unwittingly tried resorting to domain knowledge for the improvement of model performances.
Among these methods, some utilize domain knowledge in a naive or straightforward way while some
others may do it more sufficiently or elegantly. 
Although the existing literature is still in the preliminary stage,
we believe these trials deserve attentions from researchers in the community.   
For a clear elaboration on the 
existing ideas of knowledge-guided cross-modal data fusion, 
we pick up four exemplar multi-modal oriented applications 
that require various domain knowledge, and discuss their research directions
in terms of knowledge-guided multi-modal data fusion, i.e.,
Multi-modal Visual Question Answering, Multi-modal Video Summarization, Multi-modal Visual Pattern Mining and
Multi-modal Recommendation.

In a nutshell, we present our insights on the key problems for multi-modal analysis, 
review some representative state-of-the-art multi-modal approaches in multimedia 
and summarize their characteristics in essence.
Our discussions will center around the two mentioned scientific research problems in multi-modal analysis for multimedia. 
We discuss approaches focusing
on {\it data-driven multi-modal correlational representation} in Section~\ref{sec:correlation} and analyze several 
exemplar applications in {\it knowledge-guided multi-modal data fusion} in Section~\ref{sec:knowledge}.
We then highlight our insights on promising research directions that may lead the next breakthrough
in cross-modal intelligence, i.e., {\it cross-modal reasoning}, {\it cross-modal cognition} and {\it cross-modal collective intelligence}.
We share our opinions about why and how researchers should pay more attentions on these topics in the future in Section~\ref{sec:future}.
In the end, we conclude the whole paper in
Section~\ref{sec:conclusion}.

\section{Data-driven multi-modal Correlational Representation}
\label{sec:correlation}

In this section, we briefly introduce the concept and aim of multi-modal analysis, succinctly summarize 
multi-task and multi-view learning, two classic and well-documented techniques that target at learning
from multiple angles, followed by our comprehensive analysis on three important categories of approaches
for multi-modal correlational representation, i.e.,  multi-modal deep representation, multi-modal transfer learning 
and multi-modal hashing.

\subsection{Multi-task and Multi-view Learning}
Multi-modal/Cross-media correlational representation seeks a way to represent different modality data in a common space 
such that data from every modality becomes comparable with each other and as many properties in their
original spaces can be preserved in the common space as possible. As two classic methodologies, multi-task learning
and multi-view learning serve as two popular ways to consider the learning process from more than one angle. 

Multi-task learning aims to learn distinct tasks simultaneously by finding relationships among multiple tasks, 
which has been studied for roughly 20 years. 
One of the most important strategies in multi-task learning is to take both differences and connections among multiple
tasks into account simultaneously. This strategy has been widely used in multi-label classification , face recognition, and etc. 
Multi-task learning can be roughly divided into two categories:
\begin{enumerate}
	\item Methods forcing multiple tasks to share common parameters;
	\item Methods mining the common latent features among multiple tasks;
\end{enumerate}

Evgeniouand and Pontil~\cite{evgeniou2004regularized} propose {\it Regularized Multi-task Learning}, a representative model 
on common parameters which minimizes the regularization function during the learning process. 
Evgeniouand and Pontil combine the concept of multi-task learning with single-task SVM and illustrate the connections among
different single SVM tasks. They assume all tasks share a common central separation hyper plane which in turn determines
the final decision boundary for the current task through an offset parameter. As for methods mining the latent features,
Argyriou~\etal~\cite{argyriou2008convex} introduce a typical {\it Convex Multi-task Feature Learning} framework, laying
the foundation of many later multi-task learning algorithms. Jebara, in his overview paper~\cite{jebara2011multitask},
discusses four groups of multi-task learning algorithms in terms of feature selection, kernel selection, adaptive pooling
and graphical model structure. For more details, please refer to survey articles~\cite{zhang2017survey, ruder2017overview} .

Multi-view learning, as its name indicates, considers multiple views from the same input data through employing one function to model each view
and jointly optimizing all the functions so that the information of multiple views can be best exploited and the learning performance 
therefore can be dramatically improved.
Different from multi-task learning which input data may come from multiple tasks, multi-view learning takes distinct views of the
same task as input. For example, these different views can be face ID and fingerprint in recognition task, or
color and words in image representation task.
Multi-view learning can be categorized into three types: 
\begin{enumerate}
	\item Co-training: train models to achieve the maximization of the mutual consistency between two different views of the unlabeled data;
	\item Multi-kernel learning: combine different kernels corresponding to distinct views together to achieve a performance boost;
	\item Subspace learning: assume that there exists a common latent subspace shared by all views such that different view data
	can be generated from this shared latent subspace;
\end{enumerate}
Besides, there are two principles widely adopted to make sure that information from multiple views can be sufficiently utilized.
\begin{enumerate}
	\item Consensus principle (used by co-training): maximize the mutual consistency between two views 
	by requiring the two hypotheses to be as consistent as possible, i.e.,
	\begin{align}
	P(f^1 \neq f^2) \geq \textsf{max} \{ P_{err}(f^1), P_{err}(f^2) \} ,
	\label{eq:multi_view}
	\end{align}
	where $P(f^1 \neq f^2)$ is the disagreement rate between two hypotheses from the corresponding two views and 
	\blue{$P_{err}(f^1)$, $P_{err}(f^2)$ are error rates of single hypothesis $f^1$, $f^2$.}
	Thus the error rate of each single hypothesis is indirectly minimized through minimizing $P(f^1 \neq f^2)$. 
	
	\item Complementary principle: every distinct view has some unique information which is not possessed by others. Thus we may improve the 
	learning performance by making full use of complementary knowledge from different views can result in an improvement for the learning performance.
\end{enumerate}
Readers with interests may refer to overview papers~\cite{xu2013survey,sun2013survey} on multi-view learning for more detailed information.

\blue{Besides the ``pure'' multi-view learning, others have investigated metric fusion~\cite{wang2015unsupervised} or similarity learning~\cite{gao2018topic} based on the muli-view data as well.}
There are also some works combining ``multi-task'' and ``multi-view'' together whose 
details can be found in ~\cite{he2011graphbased,yan2013no,hong2013tracking,liu2016urban}.
We note that both multi-task and multi-view learning are not customized for multi-modal correlational representation.
This being the case, we highlight three promising groups of multi-modal methods designed specifically for multi-modal data and
discuss them in the rest of this section.

\subsection{Multi-modal Deep Representation}
\label{subsec_MMDR}
Before deep learning is widely used in computer vision and multimedia research works, muti-modal methods can be mainly divided into two groups:
\begin{itemize}
	\item Feature-fusion approaches~\cite{Hinton2006Reducing,Mart2014Deep}: aggregate features extracted from each modality and feed the aggregated features to the model (similar to the process of feature engineering);
	
	\item Semantic-fusion approaches~\cite{Kahou2013Combining,Simonyan2014Two,Wu2016Deep} :
	feed features from each modality into the model separately and combine the results from all the models to get the final results (similar to the methodology of ensemble learning);
\end{itemize}
A general comment on these two strategies is that feature-fusion is suitable for problems whose modalities share many correlated features
while the semantic-fusion approach fits for those who have significantly uncorrelated modalities.

The prevalent success of deep learning brings us a new option for multi-modal fusion --- intermediate-fusion.
Thanks to deep neural network that provides variable number of layers for latent representations, 
it becomes flexible to choose when and which layer(s) can be used to fuse data from different 
modalities~\cite{Yi2014Shared,Karpathy2014Large,Ding2015Robust,Tzeng2015Simultaneous,Neverova2016ModDrop}.

On the other side, it is also possible to categorize multi-modal methods based on whether they 
are discriminative, generative or both (hybrid).
Discriminative models~\cite{Karpathy2014Deep,Donahue2015Long,Ren2015Exploring,Ord2015Deep,Wu2016Deep,Kim2016Multimodal} 
usually learn conditional distributions of labels given features. 
Generative models~\cite{ngiam2011multimodal,srivastava2012learning,srivastava2012multimodal,Yu2014Medical,Goodfellow2014Generative,Kingma2014Auto,Huang2015Unconstrained,Reed2016Generative,Suzuki2016Joint,Pandey2017Variational}    
tend to learn their joint distributions.
And hybrid models~\cite{Amer2014Multimodal,Liu2015Multimodal,Amer2018Deep}
learn both conditional distributions and joint distributions by combining discriminative and generative parts.
We refer readers to a survey paper on multi-modal deep learning~\cite{Ramachandram2017Deep} for further information.
\blue{Different from the survey paper~\cite{Ramachandram2017Deep}, in this work we focus on multi-modal scenarios in areas related to multimedia,
	which includes but is not limited to deep neural network based architectures.}

\begin{figure}[h!]
	\centering
	\includegraphics[width=0.99\linewidth]{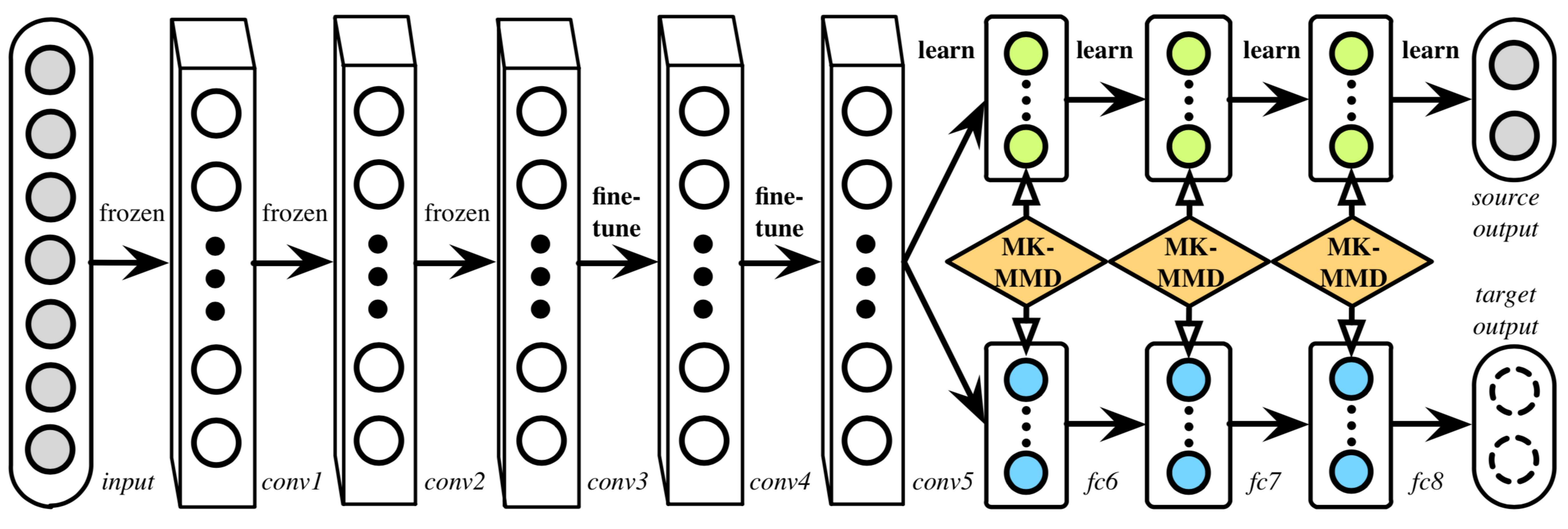}
	\caption{Deep Adaptation Network, figure from~\cite{Long2015Learning}}
	\label{fig:DAN}
\end{figure}

Next, we discuss two works~\cite{Long2015Learning,Long2017Deep} utilizing the idea of domain adaption to bridge the 
deep representations of different modalities, which is not covered by the referred survey paper. \eat{{\it Reproducing Kernel Hilbert Space} (RKHS)} 
We start from the classic work~\cite{Yosinski2014How} by Yosinski~\etal on the the transferability in deep neural networks.
The conclusion is that for a given deep neural network, deeper representation layers are more dependent on the specific task to be solved.
While the shallow layers are responsible for capturing more general features. This principle inspires us to adapt those deeper representation layers in deep learning for multi-modal tasks. As such, Tzeng~\etal propose a representative method (called DCC)~\cite{Tzeng2014Deep} to adapt deeper
layers by employing {\it Maximum Mean Discrepancy} (MMD)~\cite{sejdinovic2013equivalence} 
to reduce the disagreement between two modalities on the seventh layer (before softmax) of a eight-layer
AlexNet. DCC has two drawbacks: 1) it only adapts one single layer (i.e., the seventh layer) in the deep neural network (AlexNet). It may not be
enough as Yosinski~\etal~\cite{Yosinski2014How} point out that more than one layer is transferable and 2) it adopts a single-kernel MMD (SK-MMD)
which may not serve as the optimal kernel.
To tackle these weakness, Long~\etal propose a {\it Deep Adaptation Network} (DAN)~\cite{Long2015Learning} that 
adapts three deep layers simultaneously through a multi-kernel MMD (MK-MMD) which is capable of constructing the final kernel 
by combining multiple kernels together in {\it Reproducing Kernel Hilbert Space} (RKHS).
Figure~\ref{fig:DAN} illustrates the architecture of {\it Deep Adaptation Network} (DAN).

The objective of DAN then consists of two components:
\begin{enumerate}
	\item Deep adaptation which matches distributions of representation layers in multiple modalities,
	\item Optimal matching which maximizes two-sample test power by MK-MMD in RKHS.
\end{enumerate}

\begin{figure}[h!]
	\centering
	\includegraphics[width=0.99\linewidth]{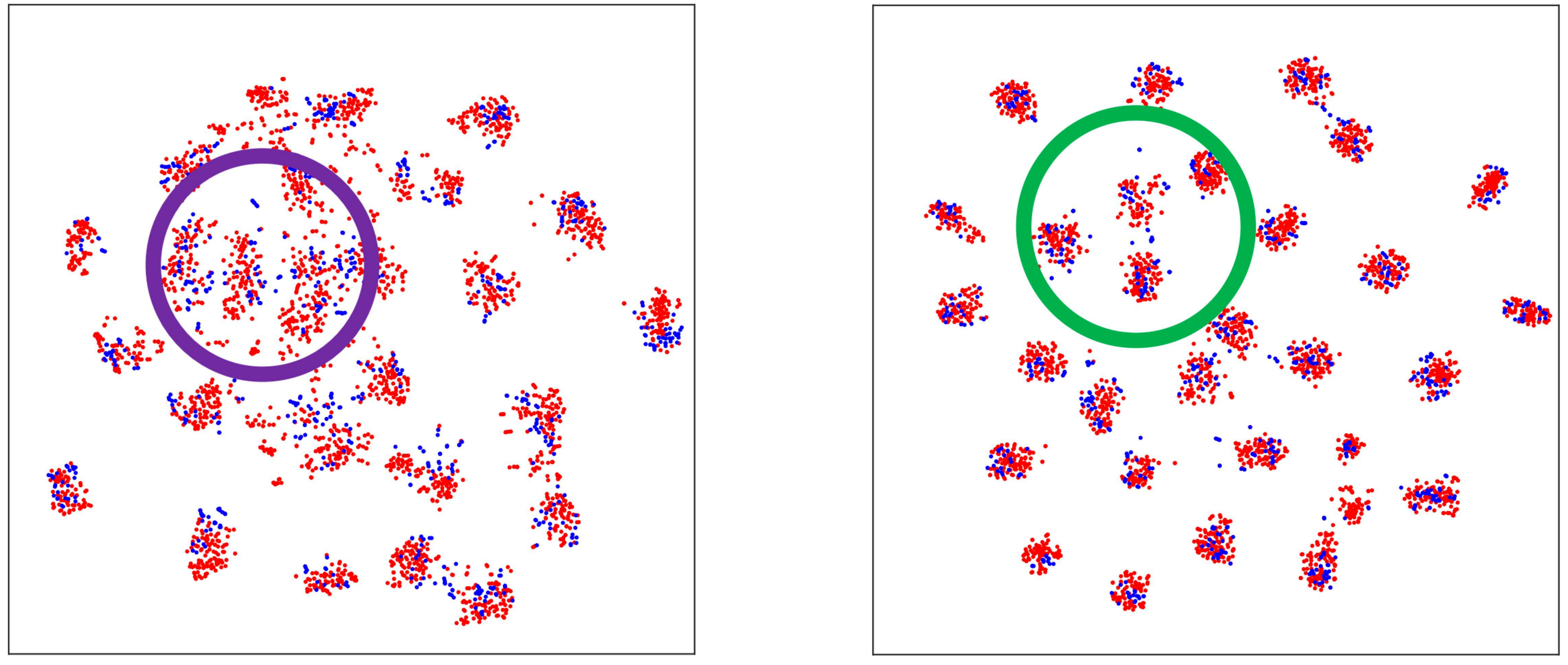} \\
	\hspace{-2.5mm} (a) Match Marginal Distributions      (b) Match Joint Distributions
	\caption{Matching Marginal Distributions v.s. Matching Joint Distributions, figure from~\cite{Long2017Deep}.}
	\label{fig:JAN_Margin_Joint}
\end{figure}

\begin{figure*}[t]
	\centering
	\begin{tabular}{cc}
		\hspace{-0mm}\includegraphics[width=.49\textwidth]{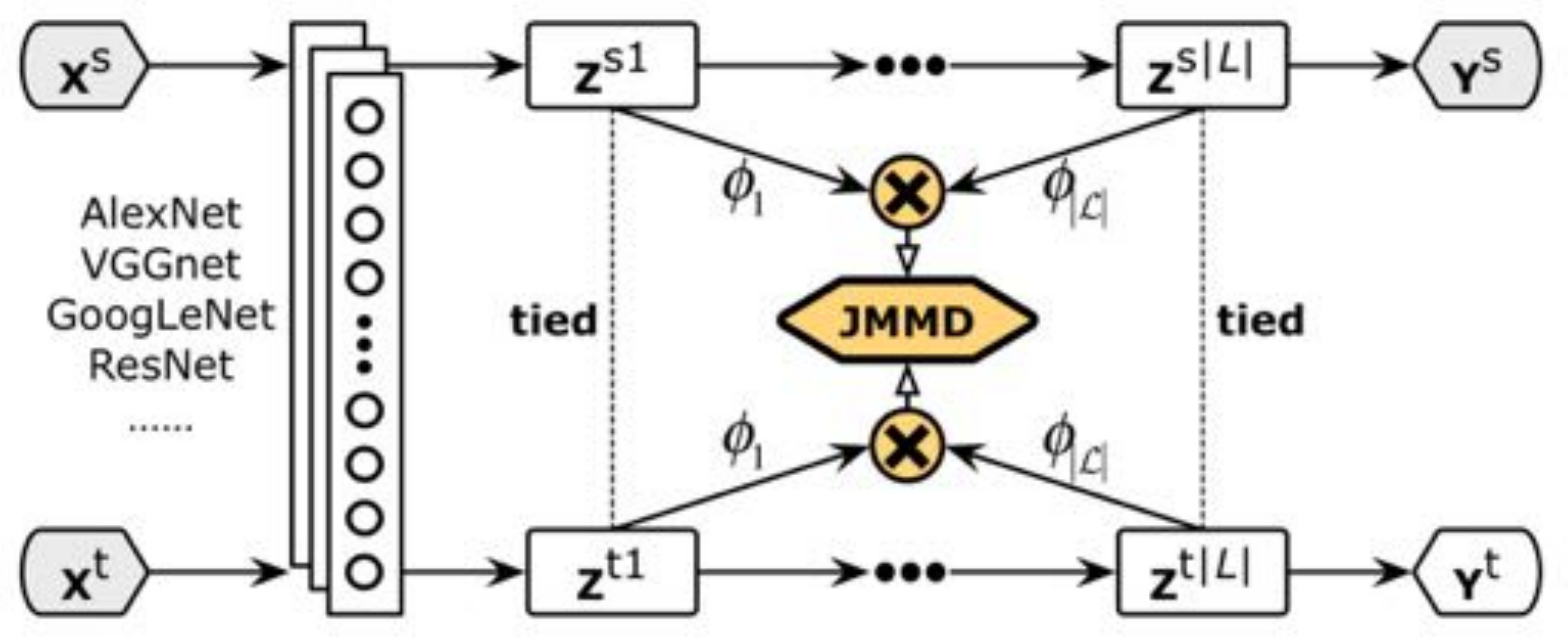} &
		\hspace{-5mm}\includegraphics[width=.49\textwidth]{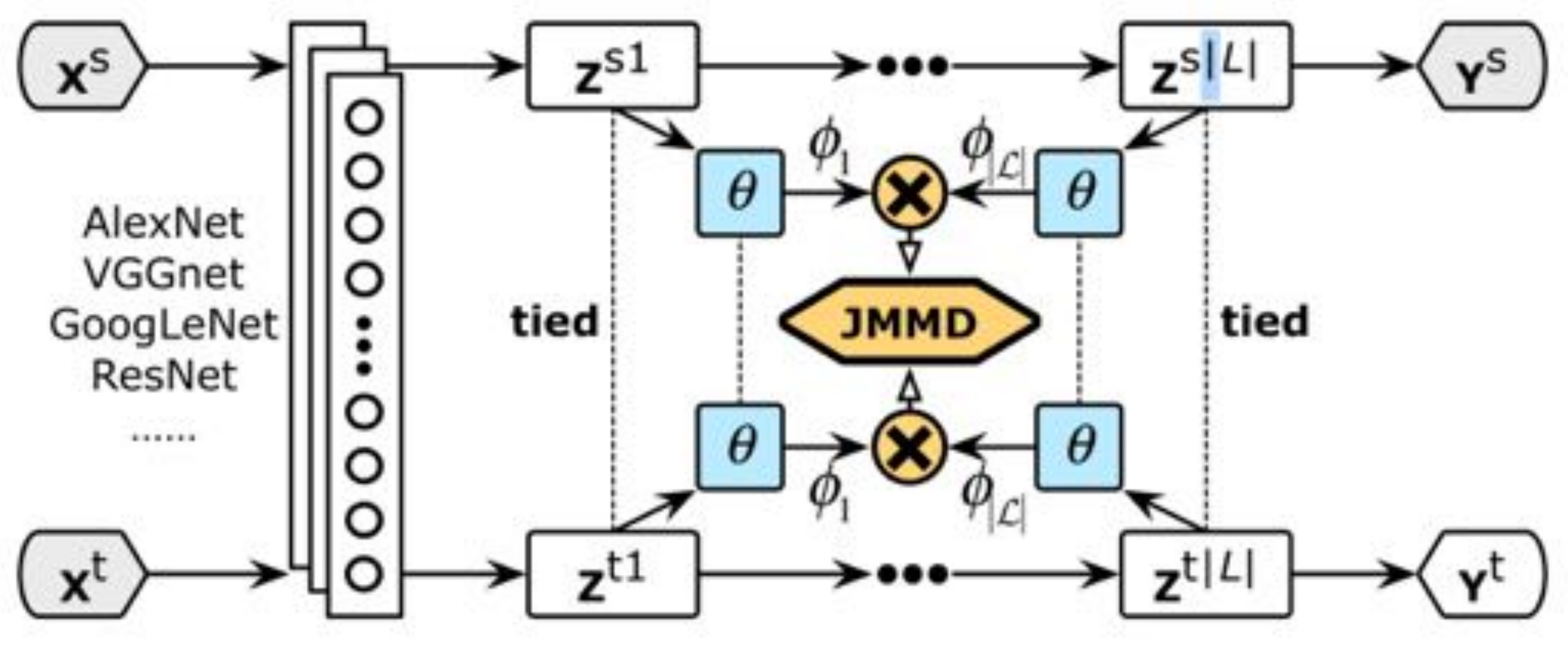} \\
		\hspace{-6mm}(a) Joint Adaptation Network (JAN) & \hspace{-2mm}(b) Adversarial Joint Adaptation Network (JAN-A) 
	\end{tabular}
	\caption{Joint Adaptation Network and its adversarial version, figure from~\cite{Long2017Deep}.}
	\label{fig:JAN}
\end{figure*}

DAN, on the other side, is also not perfect because it matches the marginal distributions $P({\bf x})$ and $Q({\bf x})$ rather
than the joint distributions $P({\bf x},y)$ and $Q({\bf x},y)$. 
As is shown in Figure~\ref{fig:JAN_Margin_Joint}, matching the joint distributions can achieve a better performance than matching
the marginal distributions. Thus a model based on {\it Joint Adaptation Network} (JAN)~\cite{Long2017Deep} is proposed by Long~\etal
to match the joint distributions between deep representations from different modailities. 
Figure~\ref{fig:JAN} illustrates the structures of Joint Adaptation Network (JAN) and its adversarial version (JAN-A).
\blue{As the RKHS is normally high-dimensional or even infinite-dimensional, Gaussian kernel mapping samples to infinite spaces is usually adopted
	as the kernel function, and the final bandwidth parameter is selected according to empirical experiences.}

We note that although these two works study the transferring representations between different (two) modalities, their learning
processes are bidirectional and the proposed models can be tested on any two modalities without a fixed requirement
of ``source'' or ``target'' domain in the experiments.
Therefore, we group these two works in the category of multi-modal deep representation rather than multi-modal transfer learning.
\blue{Besides domain adaptation, there are also works aiming at feature learning by means of deep neural networks, such as a recent
	work~\cite{liu2019cross} by Liu~\etal}

\subsection{Multi-modal Transfer Learning}
\label{subsec_MMTL}

In the past decade, researchers have developed plenty of good models that can achieve fairly good performances
on large amounts of labeled data including images, sentences etc. which is for the same
task and in the same domain (e.g., predicting image class labels given images and their corresponding labels as input for training). 
However, these models still suffer in situations 
containing new scenarios that the models have never taken into account in their training phases. 
For instance, a model trained on detecting pedestrians
during day-time may experience a deterioration in performance when being applied to detect bicyclists during night-time.
Transfer learning aims to enhance the ability of a model to generalize and transfer learned knowledge (from source domain) to 
new scenarios (target domains),
which has been an active research topic for quite a long time before the proliferation of deep neural network and we refer readers
to an excellent survey~\cite{pan2010survey} published in 2010 for details about early models.
Multi-modal transfer learning particularly focuses on transferring knowledge from one modality (source) to a 
different one (target). This method enables us to handle labeled data in a new modality through leveraging the existing
labeled data in the original modality.
\blue{Different from Pan's work~\cite{pan2010survey}, we focus on multi-modal transfer learning in multimedia through adding more recent and advanced technologies 
	including deep learning methods and so forth in this section.}

One benefit brought by deep neural network is the famous deep convolutional neural network (CNN) features trained on 
ImageNet~\cite{Krizhevsky2012ImageNet} which can be used as pre-trained features for new task(s) in the target domain. 
The credits should be given to CNN's capability of learning the basic components such as edges and shapes which serve as
general elements in images. Thus a straightforward way to handle a new task can be simply applying some pre-trained
CNN features on ImageNet to this new task, with parameters either fixed or slightly tuned under a very small learning rate~\cite{Razavian2014CNN}. 

However, CNN is designed specifically for image data (pixels) and what if we have data from other domains such as text or signal data?
The idea of domain adaptation which tries to preserve general knowledge that does not change in different domains can serve as an appropriate
candidate. 
Several works on natural language processing (NLP)~\cite{Glorot2011Domain,Chen2012Marginalized} and computer vision (CV)~\cite{Zhuang2015Supervised} 
have gained success through employing stacked denoising autoencoders to learn the domain-invariant deep representations. 

Besides, forcing the learned representations of source domain and target domain to be similar to each other may also be an option because this
procedure is able to remove domain-specific features while keeping common features shared across the two domains.
It is possible to achieve this goal through either applying the strategy to the initial representations before training~\cite{Iii2007Frustratingly,Sun2016Return} 
or ensuring the representations of source and target domains to be similar during the training process~\cite{Tzeng2014Deep,Bousmalis2016Domain}.

\begin{figure}[h!]
	\centering
	\includegraphics[width=0.99\linewidth]{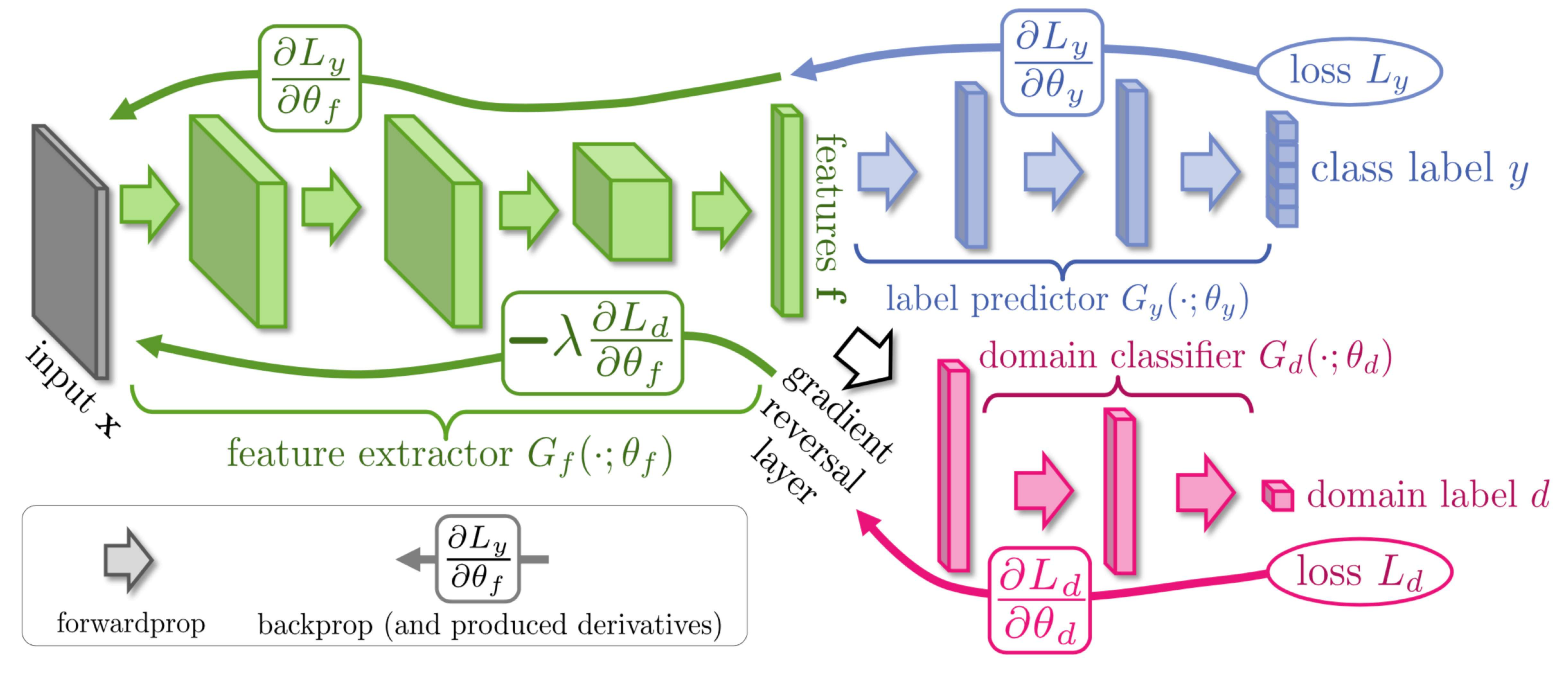}
	\caption{Indistinguishable domains with a gradient reversal layer (in pink), figure from~\cite{Ganin2015Unsupervised}.}
	\label{fig:UnDoAdaBack}
\end{figure}

The works~\cite{Ganin2015Unsupervised,Ganin2015Domain} by Ganin~\etal propose a novel setting which makes the deep feature
extraction part of the model produce features incapable of distinguishing between source and target domain. 
As is shown by the pink part in Figure~\ref{fig:UnDoAdaBack}, 
this can be done through adding a gradient reversal layer that multiplies the gradient by a certain negative constant during back propagation. 
In other words, the designed model in Figure~\ref{fig:UnDoAdaBack} is able to
minimize the label classification error in the source domain and fails to distinguish between different domains simultaneously, forcing the
feature extractor to generate features beneficial for knowledge transfer.

Aside from the traditional images and texts data that are widely used in transfer learning, other recent works
study multi-modal transfer learning based on various data including audio and video~\cite{Moon2016Multimodal}, 
head movement and co-speech~\cite{Navarretta2014Transfer}, Alzheimer’s disease (AD)
and mild cognitive impairment (MCI)~\cite{Cheng2015Multimodal,Cheng2016Multimodal} etc.

\begin{figure}[h!]
	\centering
	\includegraphics[width=0.99\linewidth]{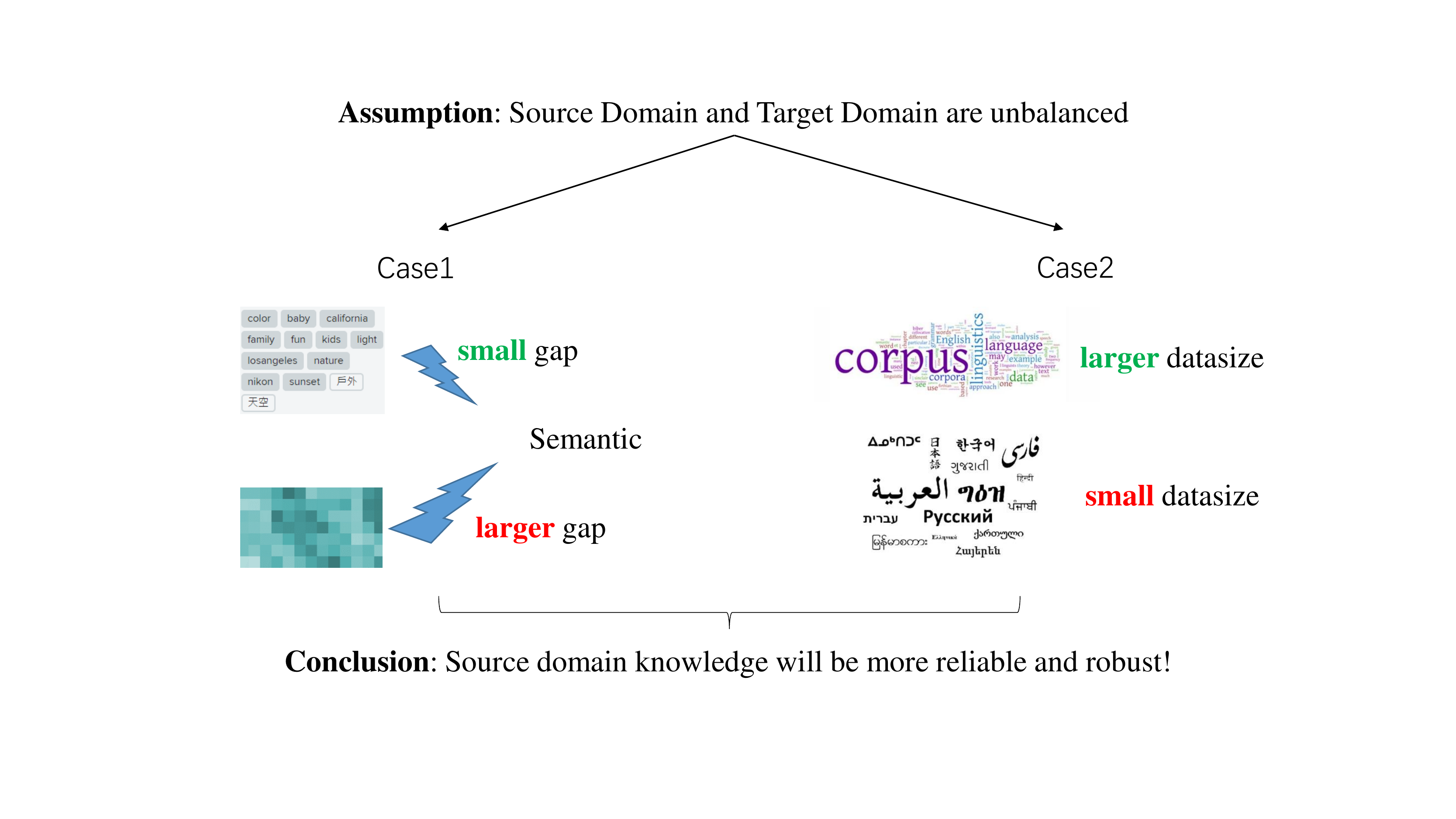}
	\caption{The unbalanced problem between the resource-rich source domain and resource-poor target domain.}
	\label{fig:DATN_unbalanced}
\end{figure}

\begin{figure}[h!t!]
	\centering
	\includegraphics[width=.88\linewidth]{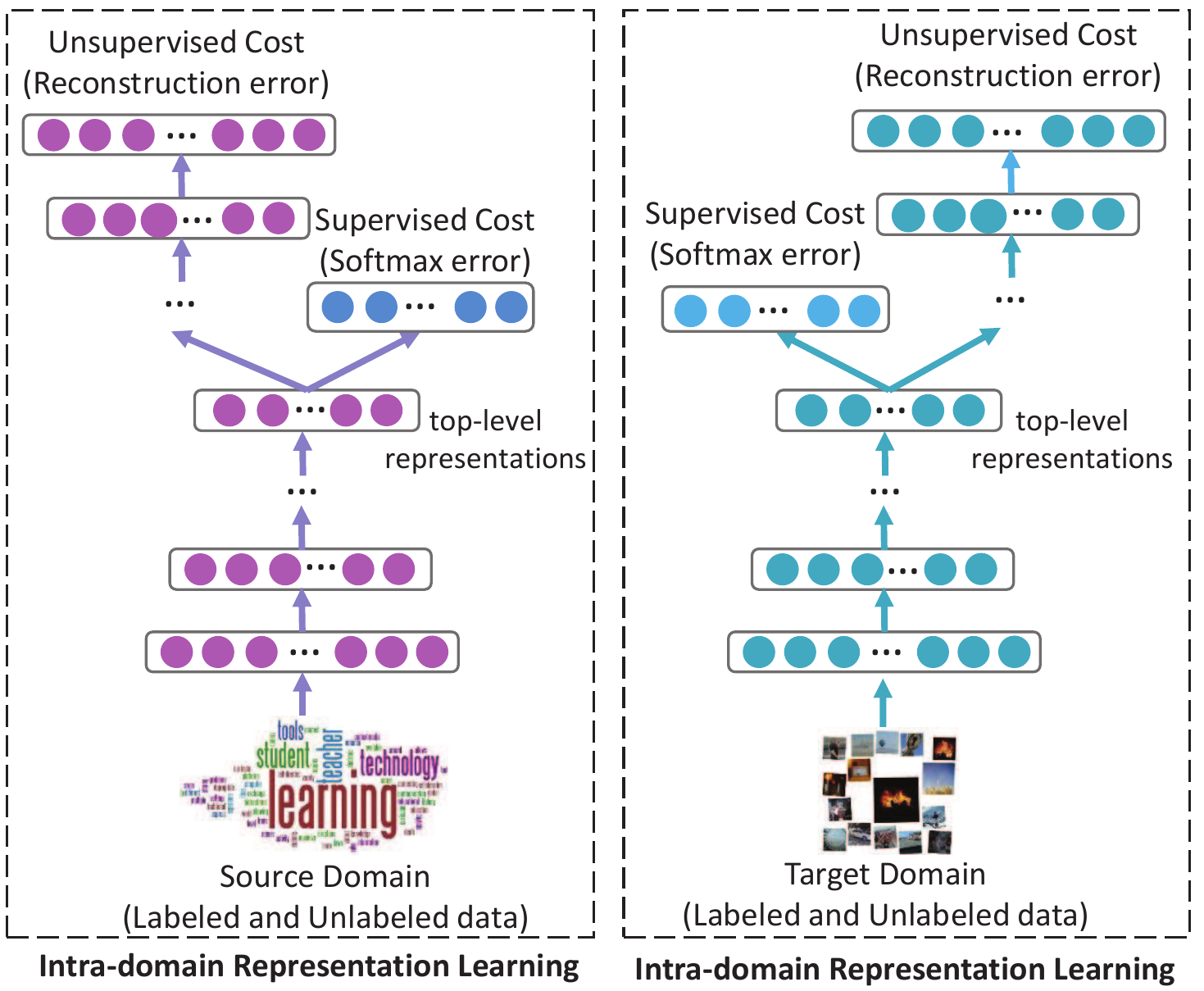} 
	\caption{Model initialization of DATN model, figure from~\cite{Wang2018Deep}.}
	\label{fig:DATN_model_intial}
\end{figure}

\begin{figure}[h!t!]
	\centering
	\includegraphics[width=.88\linewidth]{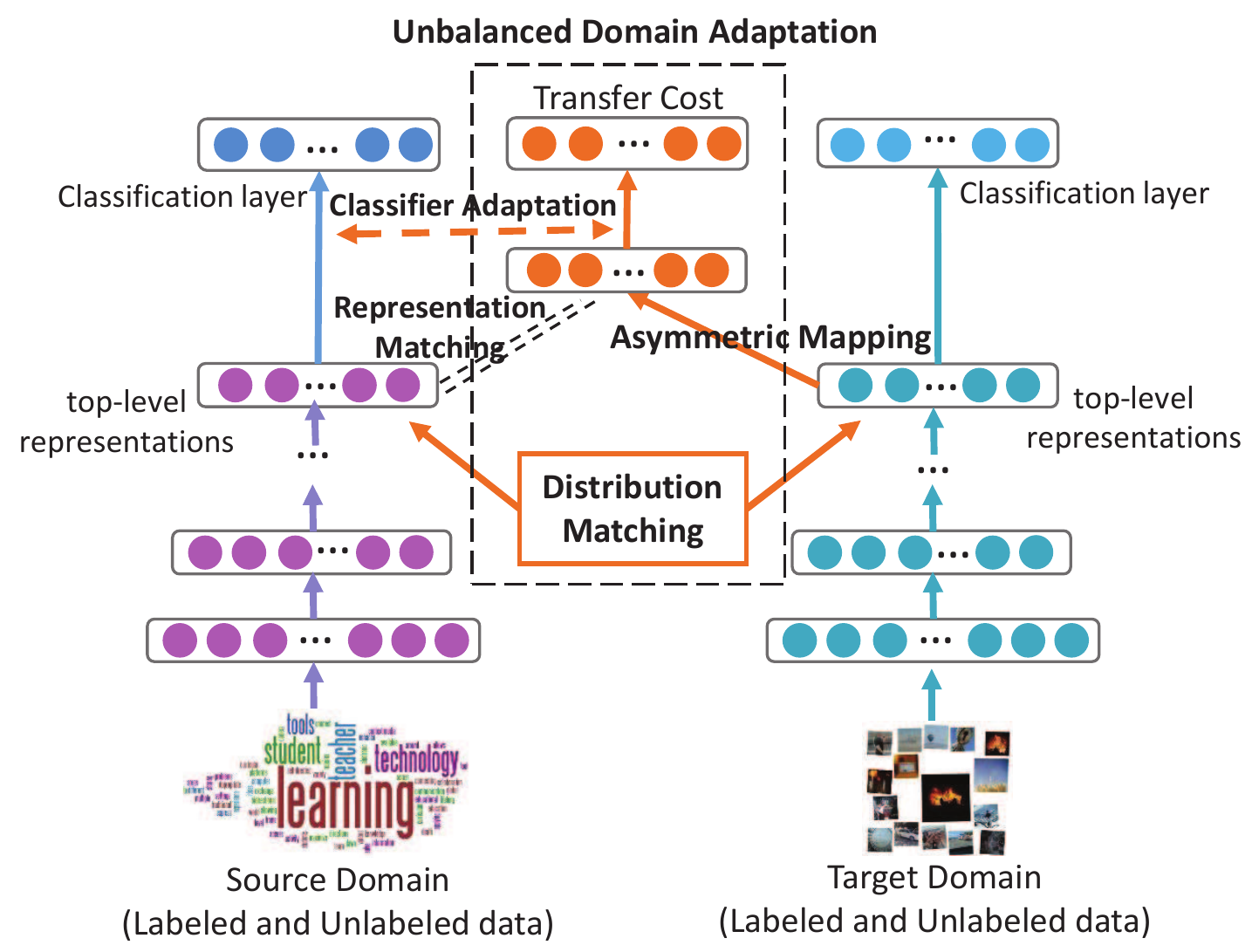} 
	\caption{Asymmetric transfer of DATN model, figure from~\cite{Wang2018Deep}.}
	\label{fig:DATN_model_transfer}
\end{figure}

In particular, a normal characteristic of multi-modal data in transfer learning is that the labeled data in the source task $\mathcal{T}_S$ 
is far more than that in target task $\mathcal{T}_T$ and thus the source domain is more resource-rich (reliable) than the target domain, 
resulting a very severe unbalanced problem as demonstrated by Figure~\ref{fig:DATN_unbalanced}.
To tackle this challenge, Wang~\cite{Wang2018Deep} in their recent work 
develop a deep asymmetric transfer network (DATN) that can adapt the classifier of source task to target task through
learning a transfer function which maps the deep representation in the target domain to that in the source domain.
The main framework of DATN is illustrated in Figure~\ref{fig:DATN_model_intial} and Figure~\ref{fig:DATN_model_transfer} 
where the initialization of deep representations 
in each modality is conducted separately through an autoencoder (shown in Figure~\ref{fig:DATN_model_intial}) 
and the asymmetric transfer together with adapting the source task classifier 
is achieved through a transfer function (shown in Figure~\ref{fig:DATN_model_transfer}).
The asymmetric transfer process consists of three parts:
\begin{itemize}
	\item Asymmetric mapping through transfer function $G$:
	\begin{eqnarray}
	\label{eq:transfuncloss}
	\begin{aligned}
	\mathcal{L}_{pair} & = \|Z_S^c-Z_T^c\cdot G\|_F^2+\lambda'\|G\|_F^2 
	\end{aligned}
	\end{eqnarray}
	
	\item Source classifier adaptation:
	\begin{align}
	\mathcal{L}_{trans} 
	=& -\frac{1}{n_T^L}{\sum_{i=1}^{n_T^L}\sum_{j=1}^k{1\{y_{T_i}=j\}\log{\frac{e^{\mathbf{z}_{T_i}^L \cdot G \cdot \mathbf{\vartheta}_{S_j}}}{\sum_{l=1}^k{e^{\mathbf{z}_{T_i}^L \cdot G \cdot \mathbf{\vartheta}_{S_l}}}}}}} 
	\label{eq:transloss}
	\end{align}
	
	\item Top-level distribution matching:
	\begin{eqnarray}
	\label{eq:unsuploss}
	\begin{aligned}
	\mathcal{L}_{unsup} &= MMD(Z_S,Z_T) \\
	&=\|\frac{1}{n_S}\sum_{i=1}^{n_S}{\mathbf{z}_{S_i}}-\frac{1}{n_T}\sum_{i=1}^{n_T}{\mathbf{z}_{T_i}}\|_2^2, 
	\end{aligned}
	\end{eqnarray}
	where $Z_S, Z_T$ denote the top-level deep representations for the source task, target task respectively
	and MMD refers to {\it Maximum Mean Discrepancy}~\cite{sejdinovic2013equivalence}.
\end{itemize}

By putting~\eqref{eq:transfuncloss},~\eqref{eq:transloss} and~\eqref{eq:unsuploss} together, the overall objectives can be expressed as follows:
\begin{eqnarray}
\label{eq:hetroloss}
\begin{aligned}
\mathcal{J}^{cross} &= \mathcal{L}_{pair}+\alpha \mathcal{L}_{trans}+ \beta \mathcal{L}_{unsup}+ \mathcal{L}_{reg},
\end{aligned}
\end{eqnarray}
where $\mathcal{L}_{reg}$ is the regularization term.

\subsection{Multi-modal Hashing}
\label{subsec_MMDH}


As the early works on multi-source hashing, multiple feature hashing~\cite{Song2011Multiple,song2013effective} and composite hashing~\cite{zhang2011composite} 
examine efficient hashing with multiple features or information sources taken into account. These works focus on the problem of returning the same
types of items as the queries, which though have a close relation to multi-modal hashing, are not specifically designed for retrieving different 
sorts of items from a given query.

In the setting of multi-modal hashing, we aim at retrieving a heterogeneous type of items (e.g., images) given a corresponding input 
query (e.g., texts describing the images).
Normally, multi-modal hashing maps data from different modalities
into some common space (e.g., Hamming space) in which the hash codes obtained from multi-modality data can be directly compared.

\begin{figure}[h!]
	\centering
	\includegraphics[width=0.99\linewidth]{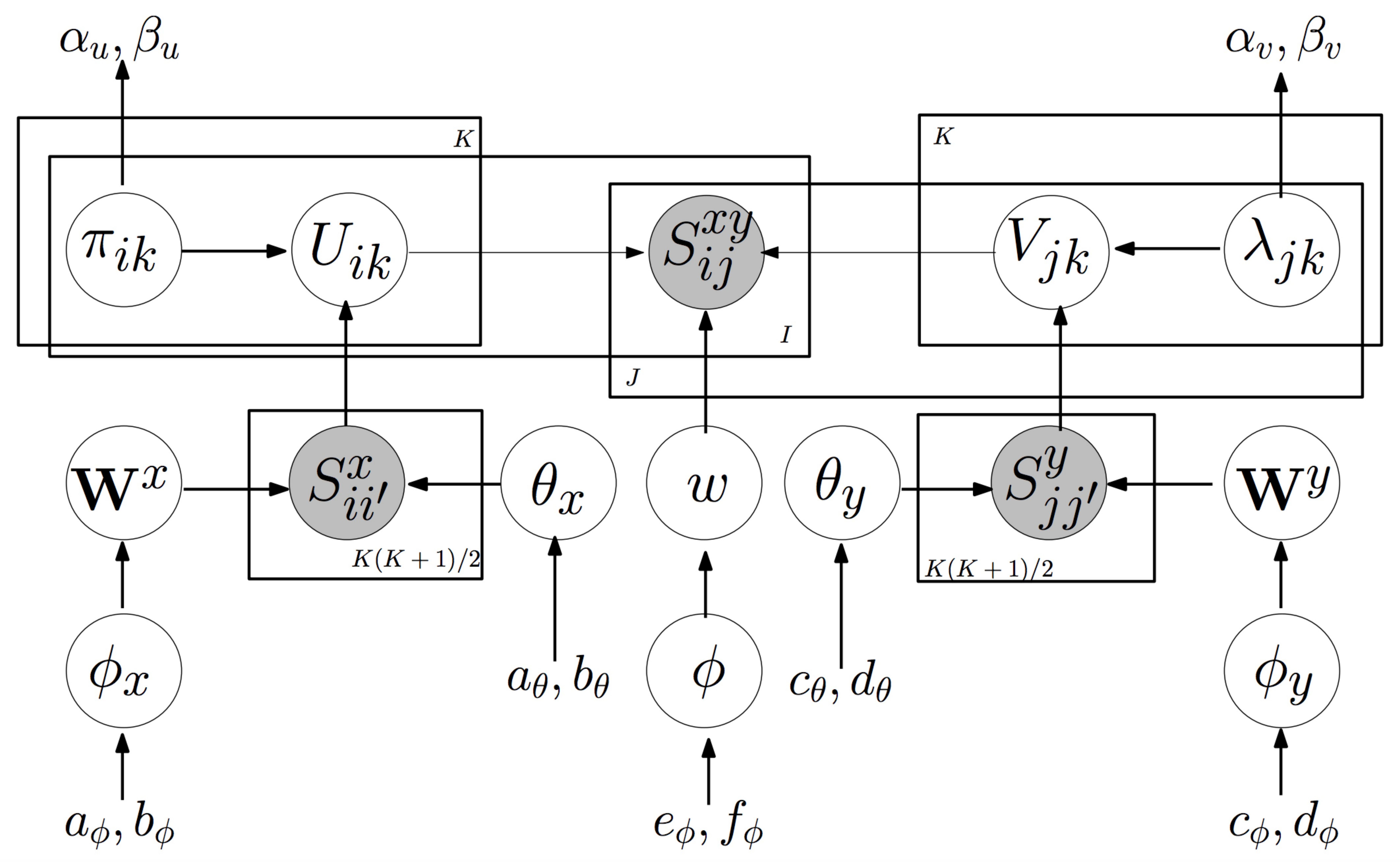}
	\caption{Graphic model of multi-modal latent binary embedding, figure from~\cite{zhen2012probabilistic}.}
	\label{fig:MLBE_graphic_model}
\end{figure}

\begin{figure}[h!]
	\centering
	\includegraphics[width=0.99\linewidth]{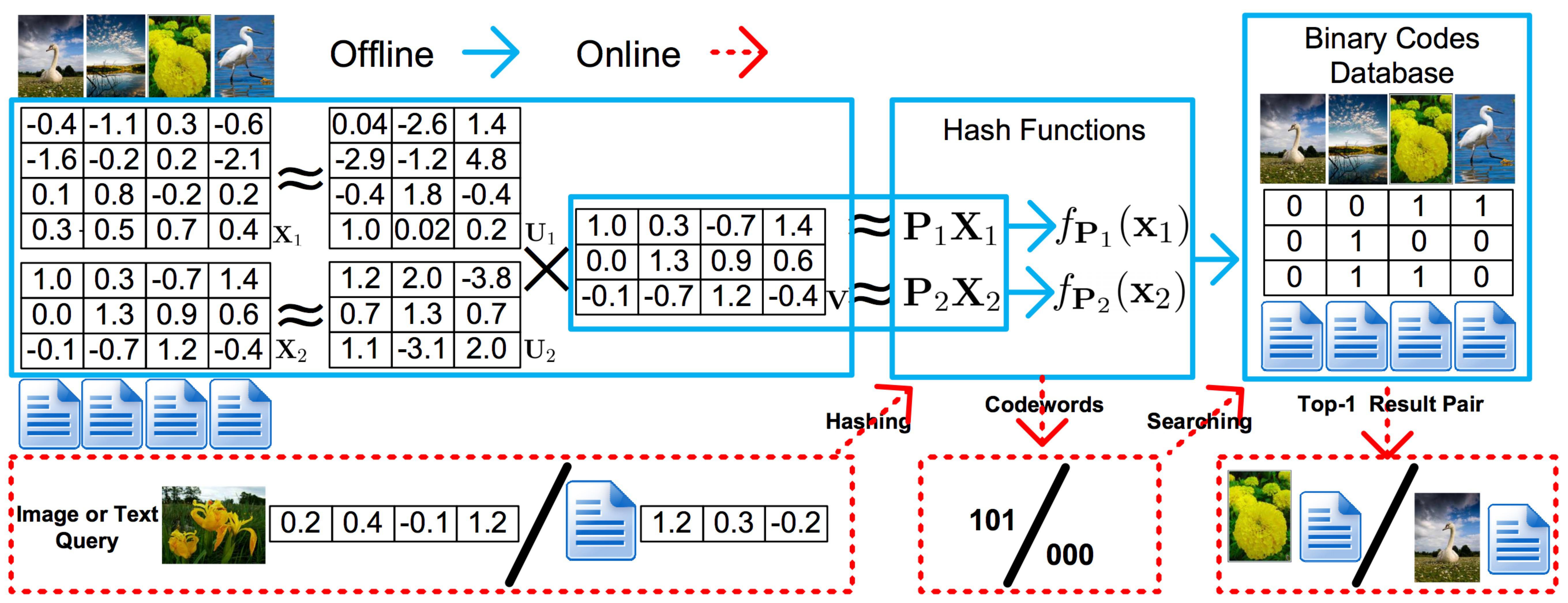}
	\caption{Framework of collective matrix factorization hashing, figure from~\cite{ding2014collective}.}
	\label{fig:CMFH}
\end{figure}

\begin{figure}[h!]
	\centering
	\includegraphics[width=0.99\linewidth]{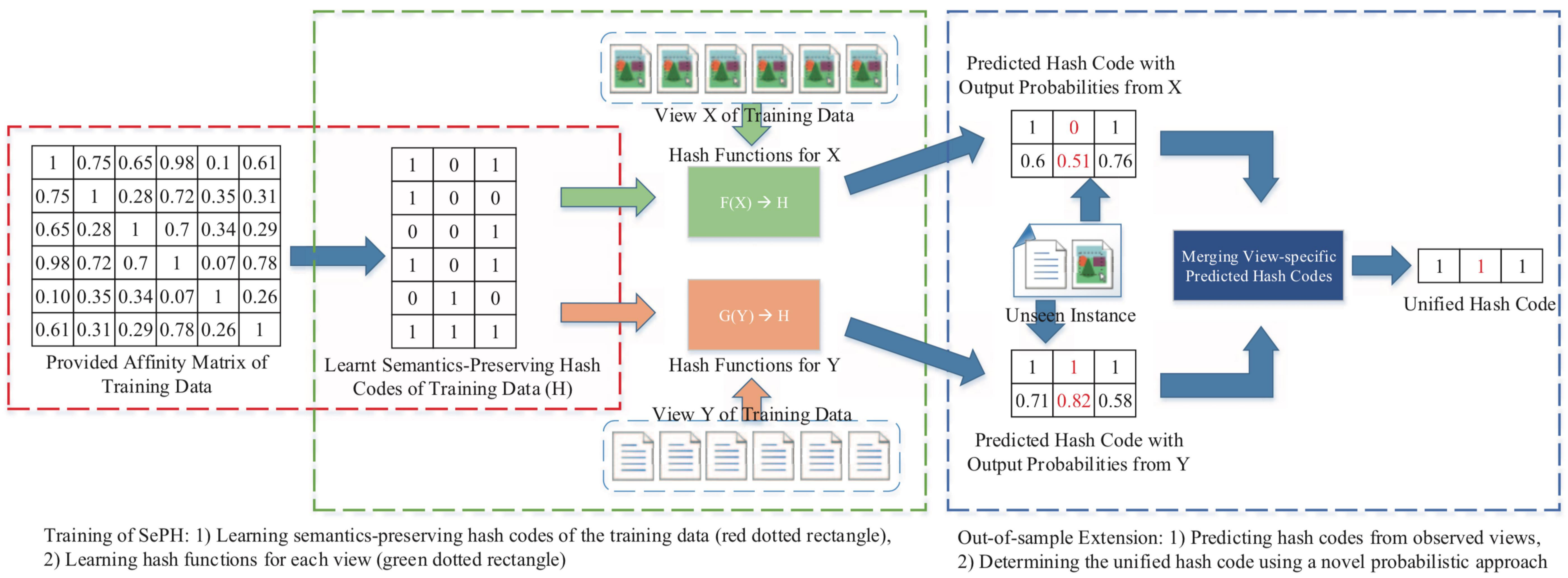}
	\caption{Framework of semantics preserving hashing with two views, figure from~\cite{lin2015semantics}.}
	\label{fig:SePH}
\end{figure}

Data from different modalities may share one unified hash code or possess separate hash codes in the new space.
Good multi-modal hashing models should be capable of designing good hash functions as well as efficiently bridging the gaps between 
different domains for fast and accurate similarity search across multiple modalities~\cite{bronstein2010data,kumar2011learning,kim2012sequential,zhen2012probabilistic,
	zhen2012co,zhu2013linear,song2013inter,ou2013comparing,ding2014collective,zhou2014latent,masci2014multimodal,zhang2014large,
	wu2014sparse,yu2014discriminative,hu2014iterative,lin2015semantics,wu2015quantized,moran2015regularised,Wang2015Learning,
	cao2016deep,cao2016correlation,jiang2017deep}.
In particular, cross view hashing~\cite{kumar2011learning} extends composite hashing to handle multi-view settings through summing over Hamming distance
for each view:
\begin{align}
d_{ij} = \sum^{K}_{k = 1} d (y_{i}^{(k)}, y_{j}^{(k)}) + \sum_{k=1}^{K}\sum_{k^{'}> k}^{K} d(y_{i}^{(k)}, y_{j}^{(k^{'})}).
\label{eq:distance_cvh}
\end{align}
Multi-modal latent binary embedding~\cite{zhen2012probabilistic} utilizes probability theory to learn hash function in the multi-modal setting
whose graphic model is shown in Figure~\ref{fig:MLBE_graphic_model}.
Co-regularized hashing, taking the regularization as an entry point, learns a multi-modal hash function with the help of a boosted co-regularization strategy,
whose objective function is as follows:
\begin{align}
\mathcal{O} &= \frac{1}{I}\sum_{i=1}^{I} l_{i}^{x} + \frac{1}{J}\sum_{j=1}^{J} l_{j}^{y} \nonumber \\
&+ \gamma \sum_{n=1}^{N}\omega_n l_{n}^{*} + \frac{\lambda_x}{2} \| \omega_x \|^2 + \frac{\lambda_y}{2} \| \omega_y \|^2 ,
\label{eq:CRH_objective}
\end{align}
where $l_{i}^{x}$ and $l_{j}^{y}$ are intra-modality losses and $l_{n}^{*}$ is inter-modality loss.
Motivated by the need for scalability and training hash functions on large scale multi-modal dataset, 
semantic correlation maximization hashing~\cite{zhang2014large} avoids explicit computation of pairwise similarity matrix 
through proposing a sequential hashing learning method with closed-form solution to each bit.
Collective matrix factorization hashing~\cite{ding2014collective} borrows the idea of collective matrix factorization to 
learn cross-modality hash functions by decomposing feature matrices from two different modailities (e.g., $X^{(1)}, X^{(2)}$) jointly
with the constraint $V_1 = V_2 = V$:
\begin{align}
\lambda \| X^{(1)} - U_1 V \|^2 + (1 - \lambda) \| X^{(2)} - U_2 V \|^2.
\label{eq:CMFH}
\end{align}
The whole framework of collective matrix factorization hashing is presented in Figure~\ref{fig:CMFH}.
Quantized correlation hashing~\cite{wu2015quantized} is the first to integrate the process of hash function learning with
quantization for multi-modal hashing by transforming multi-modal objective function into a single-modal formulation.
Semantics preserving hashing~\cite{lin2015semantics} maps the given affinity matrix $A$ into a probability distribution $\mathcal{P}$
and matches it with another probability distribution $\mathcal{Q}$ transformed from pairwise Hamming distances
between hash codes in the Hamming space through minimizing the KL-divergence between $\mathcal{P}$ and $\mathcal{Q}$.
The objective function is:
\begin{align}
\Phi = \text{min}_{\hat{H} \in \mathbb{R}^{n \times d_c}} \sum_{i \neq j} p_{ij} \text{log} \frac{p_{ij}}{q_{ij}} + \frac{\alpha}{C} \| |\hat{H}| - \mathbf{I} \|^2,
\label{eq:SePH_obj}
\end{align} 
where $\hat{H}$ is the relaxed hash code matrix and 
\begin{align}
q_{ij} = \frac{\big(1 + \frac{1}{4} \|\hat{H}_{i \cdot} - \hat{H}_{j \cdot}\|^2\big)^{-1}}{\sum_{k \neq m} \big(1 + \frac{1}{4}\|\hat{H}_{k \cdot} - \hat{H}_{m \cdot} \|^2 \big)^{-1} }, \nonumber
\end{align}
which utilizes the Student t-distribution of degree one to transform Hamming distances into probabilities.
Besides, $p_{ij} = \frac{A_{ij}}{\sum_{i \neq j} A_{ij}}$ where $A_{ij}$ is the element of $i$-th row and $j$-th column
in affinity matrix $A$, representing the given affinity between $i$ and $j$.
The overall structure of semantics preserving hashing is illustrated in Figure~\ref{fig:SePH}.

In general, inter-media hashing~\cite{song2013inter}, cross view hashing~\cite{kumar2011learning}, 
sequential spectral learning to hash~\cite{kim2012sequential}
are unsupervised hashing models extending spectral hashing~\cite{weiss2009spectral} to cross-modal scenario by defining the 
distance between documents in Hamming space and
aligning the hash codes from all modalities with the given inter-document similarity.
On the other hand, data fusion hashing~\cite{bronstein2010data}, semantic correlation maximization hashing~\cite{zhang2014large},
collective matrix factorization hashing~\cite{ding2014collective}, similarity-preserving hashing~\cite{masci2014multimodal},
sparse multi-modal hashing~\cite{wu2014sparse}, multi-modal latent binary embedding~\cite{zhen2012probabilistic},
semantics-preserving cross-view hashing~\cite{lin2015semantics} and co-regularized hashing~\cite{zhen2012co} 
all belong to supervised hashing approaches which take the pairwise similarity information
between two objects from different domains (modalities) as input and require the hash codes of these paired objects in Hamming space across different domains 
to be similar through the maximizing similarity-agreement criterion~\cite{bronstein2010data}, minimizing similarity-difference criterion~\cite{zhang2014large},
collective matrix factorization~\cite{ding2014collective} or inverted squared function~\cite{zhen2012co}.

\begin{figure}[h!]
	\centering
	\includegraphics[width=0.99\linewidth]{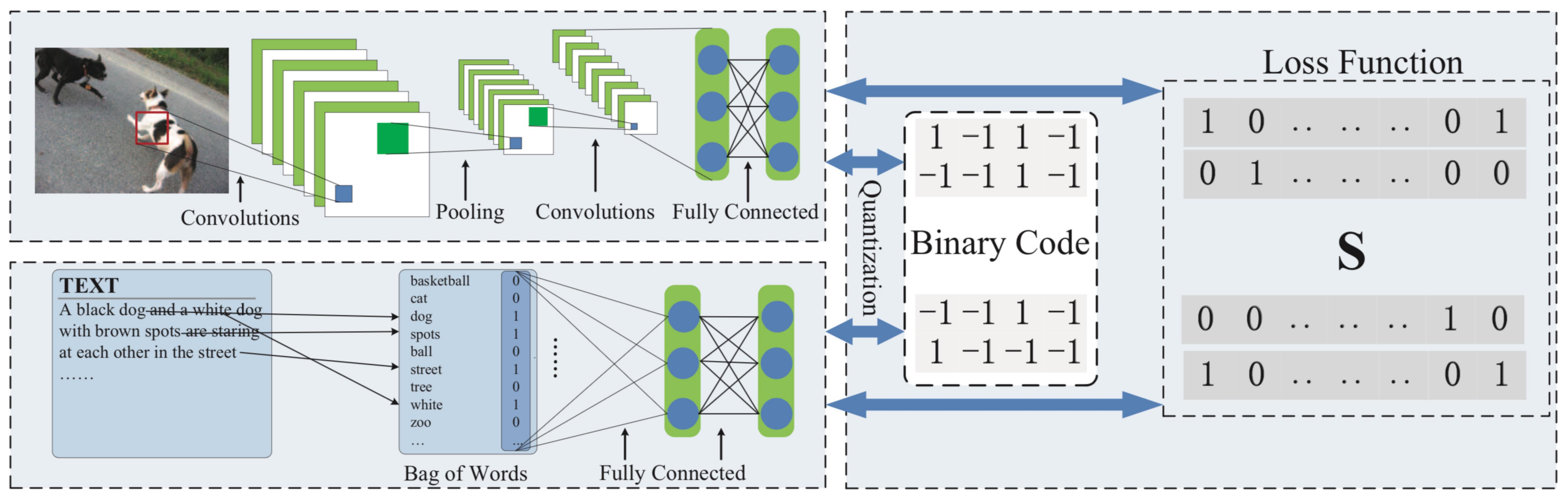}
	\caption{The learning architecture of deep cross-modal hashing, figure from~\cite{jiang2017deep}.}
	\label{fig:DCMH}
\end{figure}


Given the recent success of deep neural networks, there also have been several 
works~\cite{Wang2015Learning,cao2016deep,cao2016correlation,jiang2017deep,Liong2017Cross,Li2017Deep,Li2018Self,Zhang2018SCH,Wu2018Cycle} 
on combining hashing with deep structures for cross-modal similarity search.

Deep cross-modal hashing~\cite{jiang2017deep},whose framework is presented in Figure~\ref{fig:DCMH}, employs a convolutional 
neural network (CNN) that takes image data as input and a fully connected deep neural network that takes text data as input
to optimize the binary codes and parameters from two neural networks iteratively.
\begin{figure}[h!]
	\centering
	\includegraphics[width=0.99\linewidth]{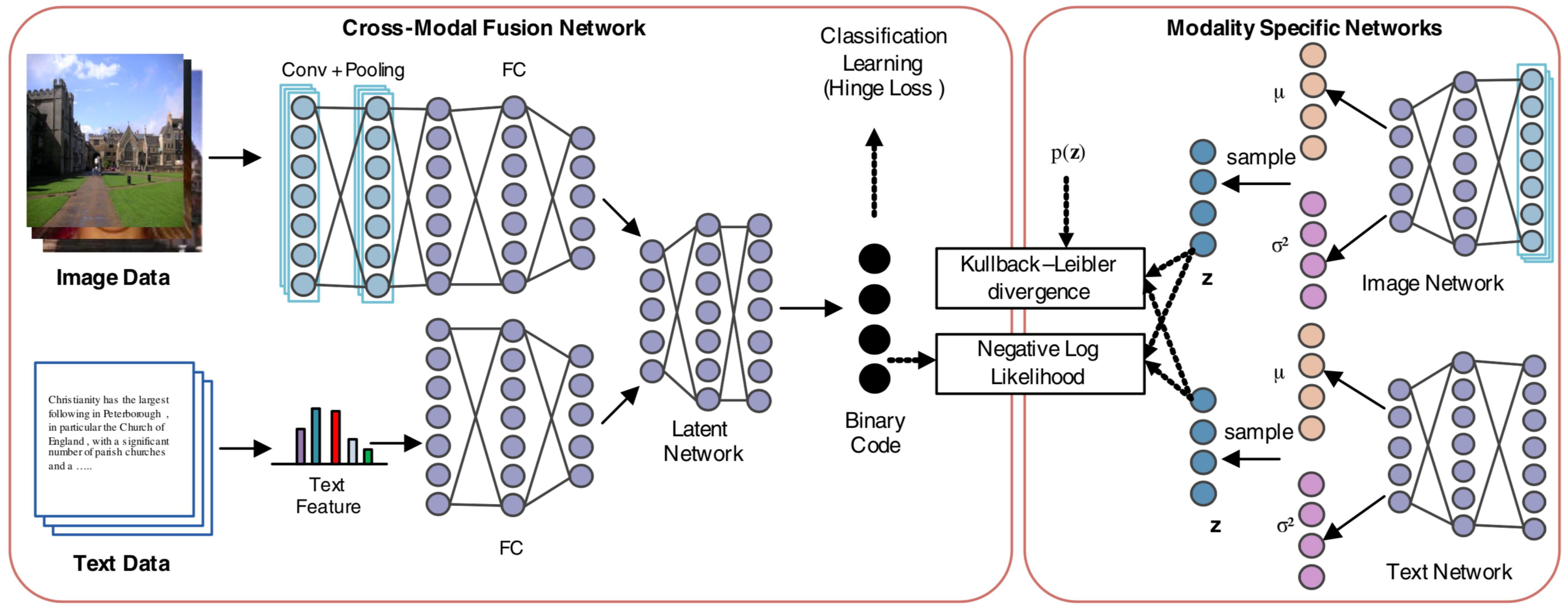}
	\caption{The end-to-end learning framework of cross-modal deep variational hashing, figure from~\cite{Liong2017Cross}.}
	\label{fig:CMDVH}
\end{figure}

\begin{figure}[htbp]
	\centering
	\includegraphics[width=0.99\linewidth]{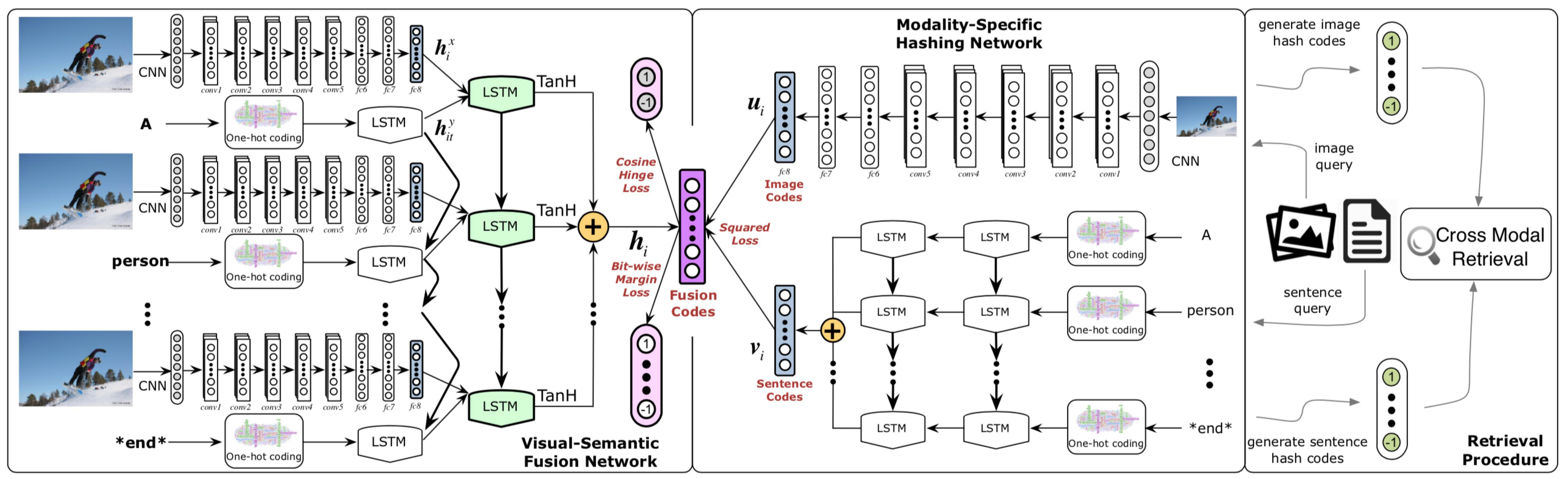}
	\caption{The architecture of deep visual-semantic hashing, figure from~\cite{cao2016deep}.}
	\label{fig:DVSH}
\end{figure}

A variational version of deep cross-modal hashing, cross-modal deep variational hashing~\cite{Liong2017Cross}, adopts a two-step learning procedure:
\begin{enumerate}
	\item Learn a fusion network and a joint binary code matrix shared by two modalities simultaneously through an alternative
	optimization procedure, which is similar to deep cross-modal hashing.
	\item Learn a modality specific neural network for each modality such that the top-level representation is
	as similar as possible to the binary codes obtained from the fusion network and also that the approximated posterior distribution
	can be as close as possible to the KL-divergence regularized prior distribution.
\end{enumerate}
For comparison with deep cross-modal hashing, we refer readers to Figure~\ref{fig:CMDVH} for the end-to-end learning framework of
cross-modal deep variational hashing.
As is shown in Figure~\ref{fig:DVSH}, deep visual-semantic hashing~\cite{cao2016deep} proposes an end-to-end 
image-sentence (each image is attached with at least one sentence) 
cross-modal hashing algorithm which utilizes convolutional neural network (CNN) to handle image data  and  long short term memory (LSTM)
to handle sentence data such that a joint embedding space for both modalities as well as a separate structure for each modality can be
learnt under the guidance of different losses including pairwise loss, cosine hinge loss, bit-wise margin loss, squared loss etc.

\begin{figure}[htbp]
	\centering
	\includegraphics[width=0.99\linewidth]{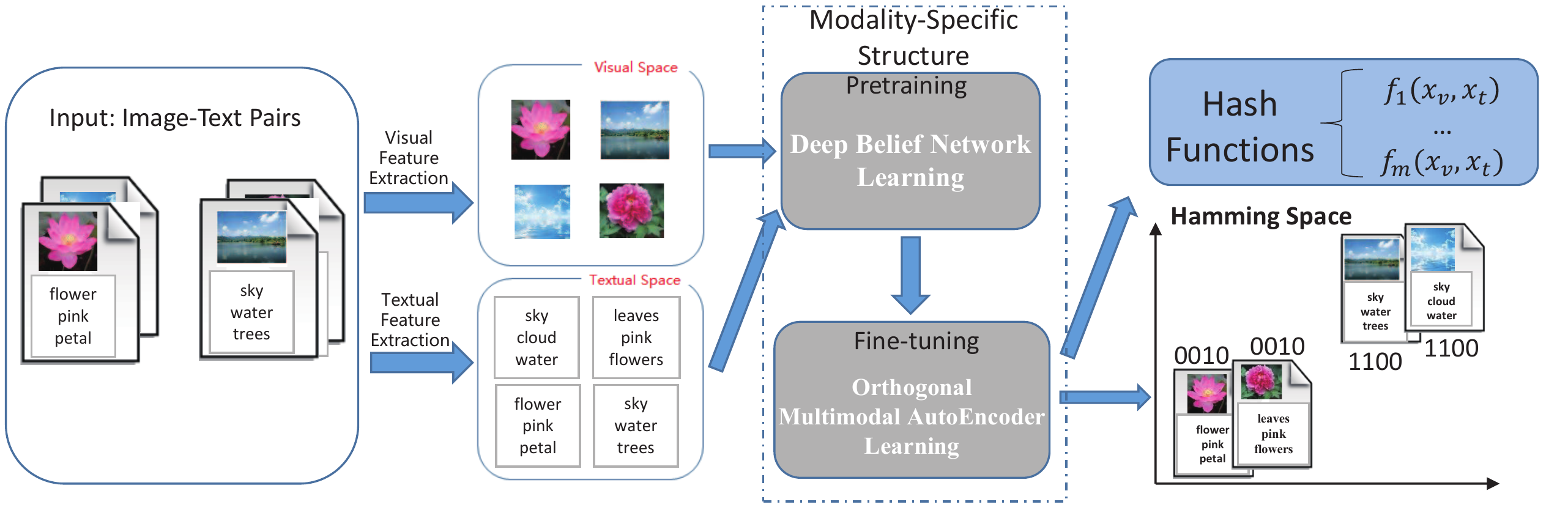}
	\caption{The flowchart of deep multi-modal hashing with orthogonal regularization, figure from~\cite{Wang2015Learning}.}
	\label{fig:DMHOR_flowchart}
\end{figure}

\begin{figure}[htbp]
	\centering
	\includegraphics[width=.8\linewidth]{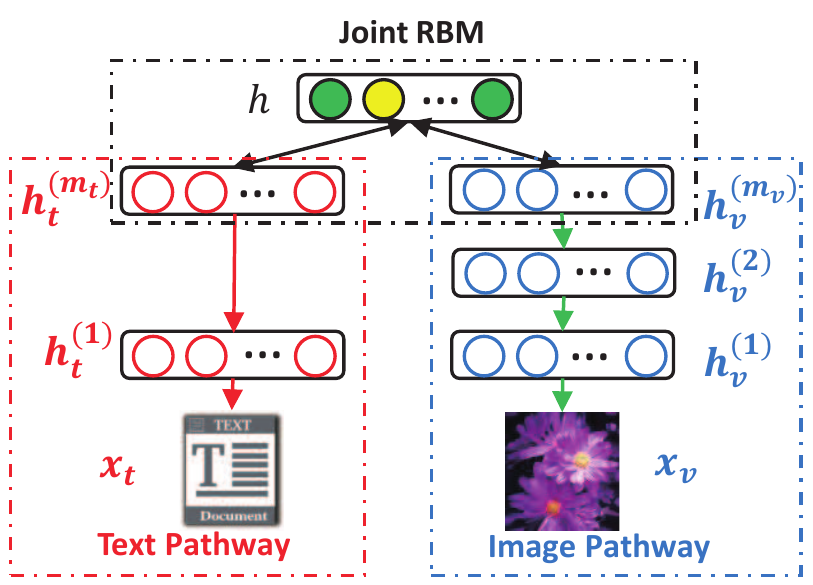}
	\caption{Model pretraining of DMHOR model, figure from~\cite{Wang2015Learning}.}
	\label{fig:DMHOR_model_pretrain}
\end{figure}

\begin{figure}[htbp]
	\centering
	\includegraphics[width=.8\linewidth]{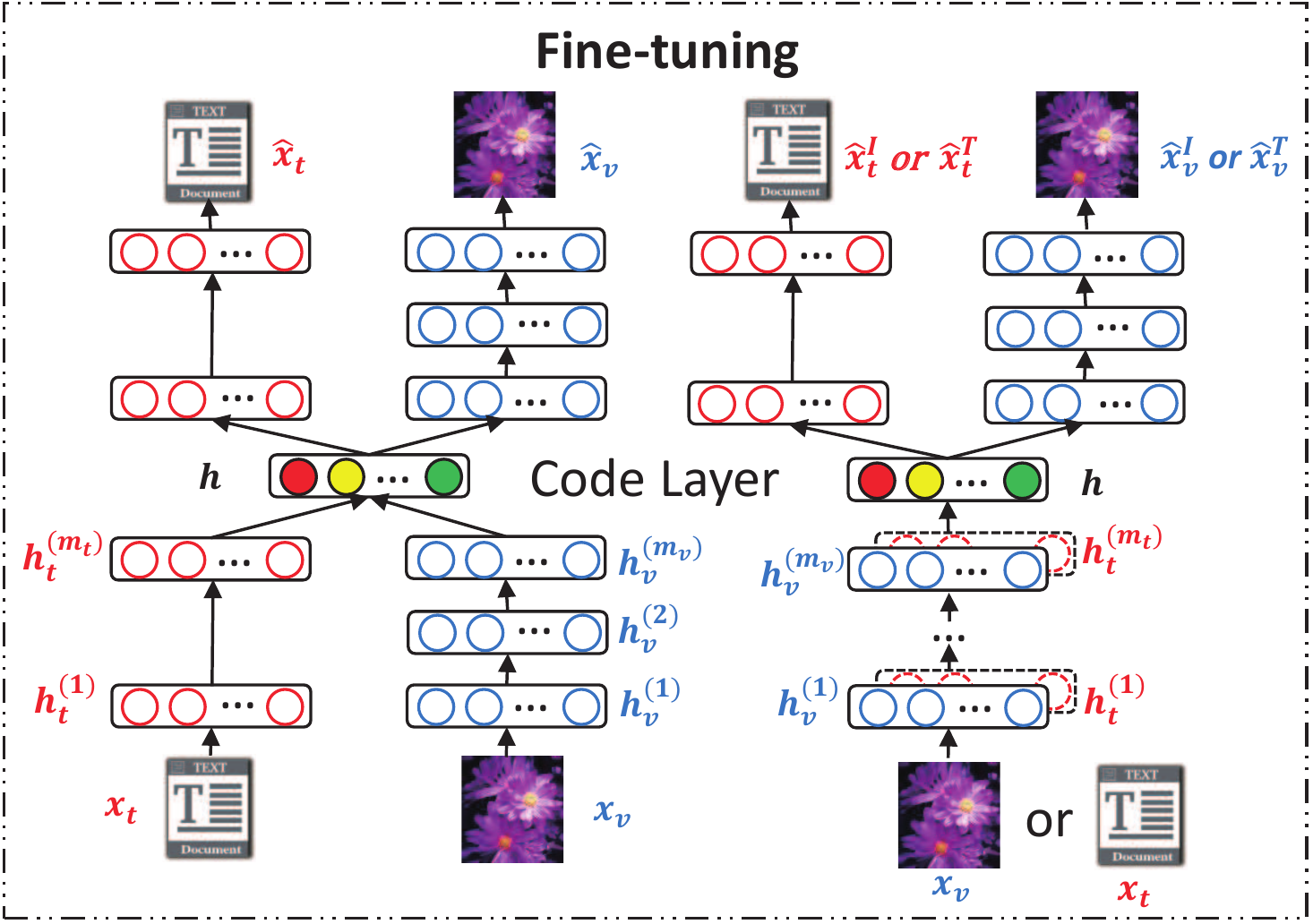} 
	\caption{Cross-modal fine tuning of DMHOR model, figure from~\cite{Wang2015Learning}.}
	\label{fig:DMHOR_model_fine_tune}
\end{figure}

Aside from CNN, some other works~\cite{Wang2015Learning,cao2016correlation,Li2017Deep} also feed data from different modalities
to autoencoder (AE) which serves as an adequate tool for both model initialization and unsupervised learning.
Particularly, deep multi-modal hashing with orthogonal regularization~\cite{Wang2015Learning} whose model flowchart is displayed
in Figure~\ref{fig:DMHOR_flowchart}, recognizes the phenomenon of redundant information in deep multi-modal representation and
proposes an orthogonal structure to reduce this redundancy with theoretical guarantee while keeping the learnt compact hash codes accurate.
As is illustrated in Figure~\ref{fig:DMHOR_model_pretrain}, a multi-modal deep belief network (DBN) consisting of several DBNs (one for each modality)
and a joint Restricted Boltzmann Machine (RBM) is first developed to correlate high-level representations of data from different modalities
for the purpose of pretraining.
Then, to learn an adequate multi-modal representation that preserves intra-modality and inter-modality simultaneously, a multi-modal autoencoder (MAE)
is developed to capture the joint correlations for different modalities and a cross-modal autoencoder (CAE) is explored to enable the 
reconstruction of representations in any modality from data in an arbitrary modality. The left part of Figure~\ref{fig:DMHOR_model_fine_tune} 
shows the structure of MAE whose loss function is shown as follows:
\begin{equation}
\label{eqn:DMAE}
L_{vt}(\mathbf{x}_v,\mathbf{x}_t;\theta) = {\frac{1}{2}(\|{\hat{\mathbf{x}}_v}-\mathbf{x}_v\|_2^2+\|\hat{\mathbf{x}}_v-\mathbf{x}_t\|_2^2)},
\end{equation}
where $\hat{\mathbf{x}}_v$ is the reconstruction of $\mathbf{x}_v$ and $\hat{\mathbf{x}}_t$ is the reconstruction of $\mathbf{x}_t$.
The right part of Figure~\ref{fig:DMHOR_model_fine_tune} demonstrates the structure of image-only CAE whose loss function can be expressed in the following:
\begin{equation}
\label{eqn:CAE}
L_{v\bar{t}}(x_v,x_t;\theta) = {\frac{1}{2}(\|\hat{\mathbf{x}}_v^I-\mathbf{x}_v\|_2^2+\|\hat{\mathbf{x}}_t^I-\mathbf{x}_t\|_2^2)},
\end{equation}
where the subscript $v\bar{t}$ denotes the input of the provided image pathway when the corresponding text pathway is absent. 
$\hat{\mathbf{x}}_v^I$ is the reconstruction of $x_v$ in the image pathway and $\hat{\mathbf{x}}_t^I$ is the reconstruction of $x_t$ in the text pathway.
The missing modality will be set to zero in the joining code layer for the calculation of $\hat{\mathbf{x}}_v^I$ and $\hat{\mathbf{x}}_t^I$. 
Thus the overall objective function with only two modalities (image and text) can be formulated as 
follows (loss function of text-only CAE $L_{\bar{v}t}$ can be formulated in a way similar to~\eqref{eqn:CAE}):
\begin{eqnarray}
\begin{aligned}
&\min_{\theta} \quad L_{MDAE}(X_v,X_t;\theta)  \\
=&\frac{1}{n}\sum\limits_{i=1}^n{(L_{vt}+L_{\bar{v}t}+L_{v\bar{t}})}+L_{reg} \\
&s.t. \quad\frac{1}{n} \tilde{H}^T\cdot \tilde{H} = I,
\end{aligned}
\end{eqnarray}
where $L_{reg}$ is a L2-norm regularizer term of weight matrix preventing overfitting and 
the constraint $\frac{1}{n} \tilde{H}^T\cdot \tilde{H} = I$ ensures the orthogonality of the hash codes
to reduce the redundant information.

Last but not least, the idea of adversarial training is also adopted in cross-modal hashing, 
such as semi-supervised cross-modal hashing~\cite{Li2018Self}, self-supervised adversarial hashing~\cite{Zhang2018SCH}, 
cycle-consistent deep generative hashing~\cite{Wu2018Cycle}.

\section{Knowledge-guided Multi-modal Analysis}
\label{sec:knowledge}
One intelligent aspect of human being is that we are able to make decisions
by resorting to domain knowledge from relevant fields or domains.
This motivates the advent of knowledge-guided multi-modal approaches which
adopt a more intelligent and promising multi-modal way through
utilizing complementary external domain knowledge to boost the model performance in multimedia.
In this section, we first present three types of methods adequate for the fusion of data and knowledge,
then discuss several exemplar applications that require knowledge-guided fusion.

\subsection{Approaches for Knowledge-guided Fusion }
There are three mainly families of methods that are suitable for knowledge-guided
cross-modal fusion, i.e., Bayesian Inference, Teacher-student Network and Reinforcement Learning,
which deserves further investigations for future research.
Bayesian theory~\cite{bernardo2009bayesian} has been a very popular tool in statistics. Bayesian
inference~\cite{dempster1968generalization,jensen1996introduction,box2011bayesian} aims to
simulate the inference ability of human through encoding some ``prior'' knowledge into the model.
Thus incorporating domain knowledge via Bayesian prior would be a good option for knowledge-guided
multi-modal fusion. Since the deep neural networks usually have quite complex structures, teacher-student network~\cite{hinton2015distilling} is
originally proposed to compress the deep model (student network)
via the guidance of a well-trained network (teacher network).
It has also been applied for information/knowledge transfer between image sets~\cite{luo2018graph},
RGB images and depth images~\cite{gupta2016cross}, as well as video sets~\cite{zhang2018better}.
Therefore, distilling useful domain knowledge through a teacher network and using it as guidance
in cross-modal data fusion could also be an appropriate direction.
Reinforcement learning~\cite{sutton1998introduction,kaelbling1996reinforcement} aims at
taking suitable actions to maximize rewards in certain situations.
It has been a well-established machine learning research topic
with wide applications, particularly in robotics~\cite{kober2012reinforcement} in the past decades.
As such, utilizing domain knowledge to guide the reward/feedback in reinforcement framework
seems to be another promising way to handle knowledge-guided multi-modal fusion.

\subsection{Exemplar Applications of Data and Knowledge Fusion}
Since different problems may require different domain knowledge, we discuss four exemplar research topics
covering visual question answering, video summarization, visual pattern mining and recommendation
from a knowledge-guided multi-modal perspective
for a better illustration.

\subsubsection{\bf Multi-modal Visual Question Answering}

Visual Question Answering (VQA) is a challenging task, 
which bridges Computer Vision (CV) and Natural Language Processing (NLP) via 
jointly understanding visual information and natural language.
Given an image and a related textual question, 
VQA systems are supposed to correctly answer the question based on the image, 
making VQA intrinsically cross-modal since it involves an image and a relevant question.
In order to achieve a joint deep understanding of visual and natural language, 
a VQA task is designed as a practical setting to evaluate the capability of an algorithm
for extracting high-level visual information and reasoning on the extracted information.
VQA is very challenging not only for its requirement of bridging visual and textual modalities
but also for the required versatile abilities ranging from object recognition and localization to 
high-level reasoning and common-sense knowledge learning. 
We will briefly describe the conventional cross-media architecture of VQA systems 
as well as several advanced techniques for connecting visual and textual modalities, 
followed by discussions on some issues in VQA systems and pioneering works that may 
lead the future research.

\begin{figure*}[!t]
	\begin{center}
		\includegraphics[width=0.8\linewidth]{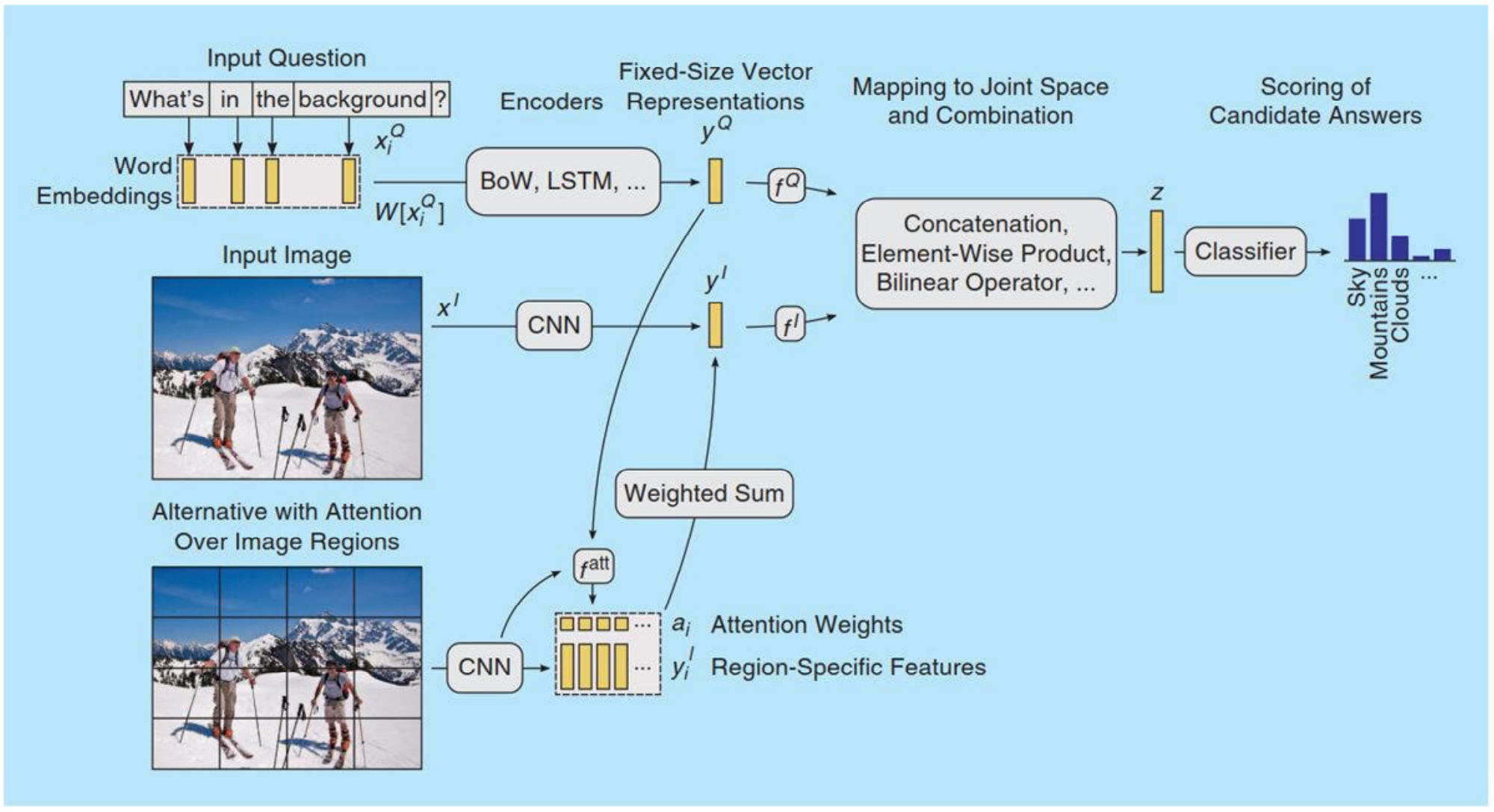} 
	\end{center}
	\caption{An illustration of the conventional VQA approach, figure from~\cite{teney2017visual}.}
	\label{fig:vqa-diagram}
\end{figure*}

Conventional approaches for VQA train a neural network using (image, question, answer) triplets as supervision
in an end-to-end way, establishing a mapping from the given image and question input to one of the candidate answers.
Here the core idea is to learn a unified embedding of image and question.
The input image will be passed through a convolutional neural network pretrained for image classification (e.g., ResNet) 
to obtain an image representation, i.e., a fixed-length vector. Meanwhile, each word in the textual questions will first 
be embedded into a continuous space by some well-established methods 
(e.g., one-hot encoding, or look up in a pretrained word-embedding matrix), and then the sequence of words will 
be encoded into a fixed-length vector through bag-of-words or recurrent neural network to 
capture the sequential relationships among words.
Upon obtaining the feature representations of image and question, each of them will be embedded into a common
space where the combination of image and question representation will then be conducted.
The embedding function is typically implemented as additional layers of neural networks, 
and straightforward options for combining the embedded features include concatenation and 
Hadamard (element-wise) multiplication in the common space.
This family of works can be regarded as the simplest cross-modal fusion methods. 
An illustration diagram is provided in Figure~\ref{fig:vqa-diagram} which is taken from~\cite{teney2017visual}. 

\begin{figure}
	\begin{minipage}{.5\linewidth}
		\begin{center}
			\includegraphics[width=.95\linewidth]{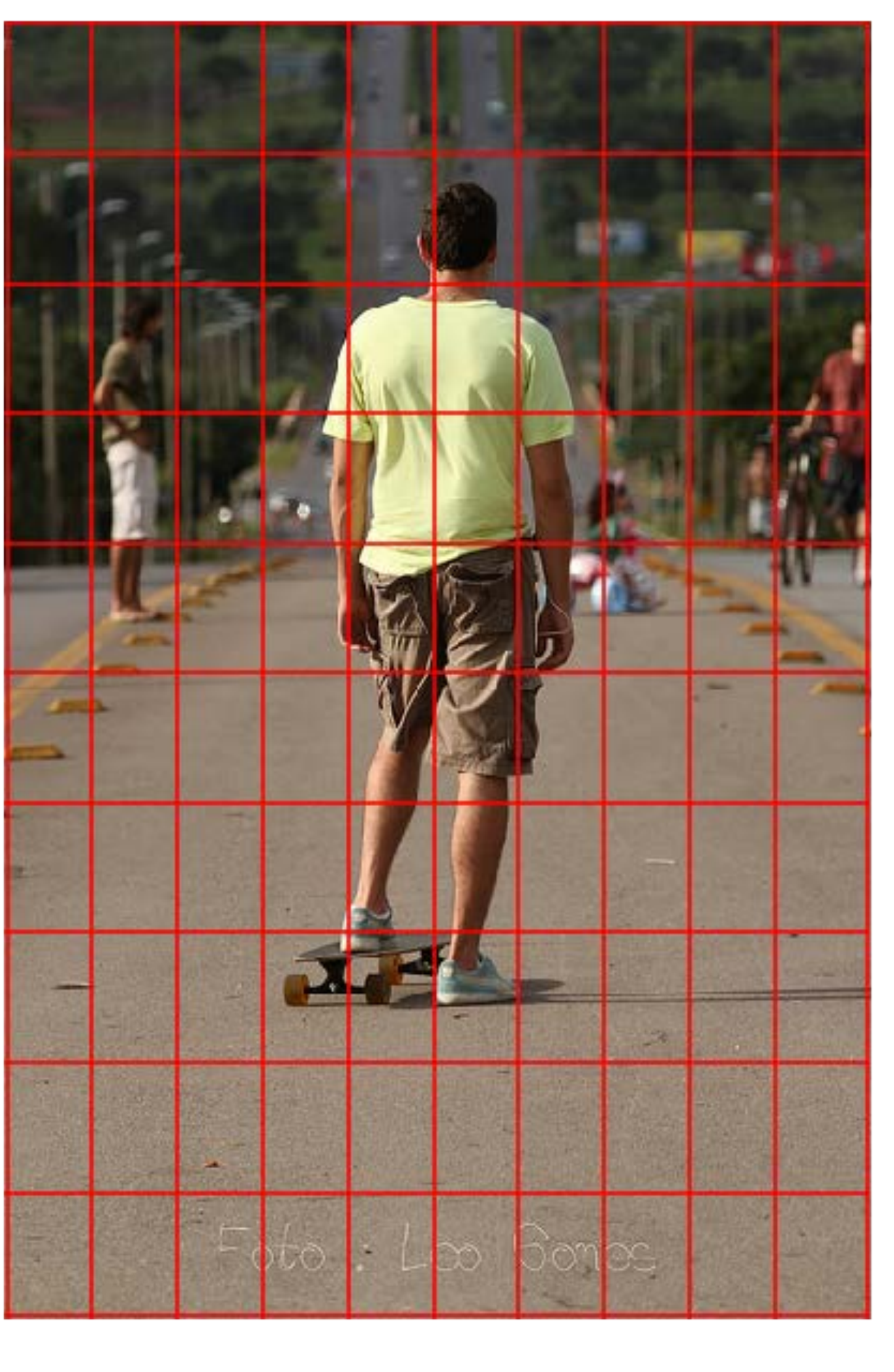}
		\end{center}
	\end{minipage}%
	\begin{minipage}{.5\linewidth}
		\begin{center}
			\includegraphics[width=.95\linewidth]{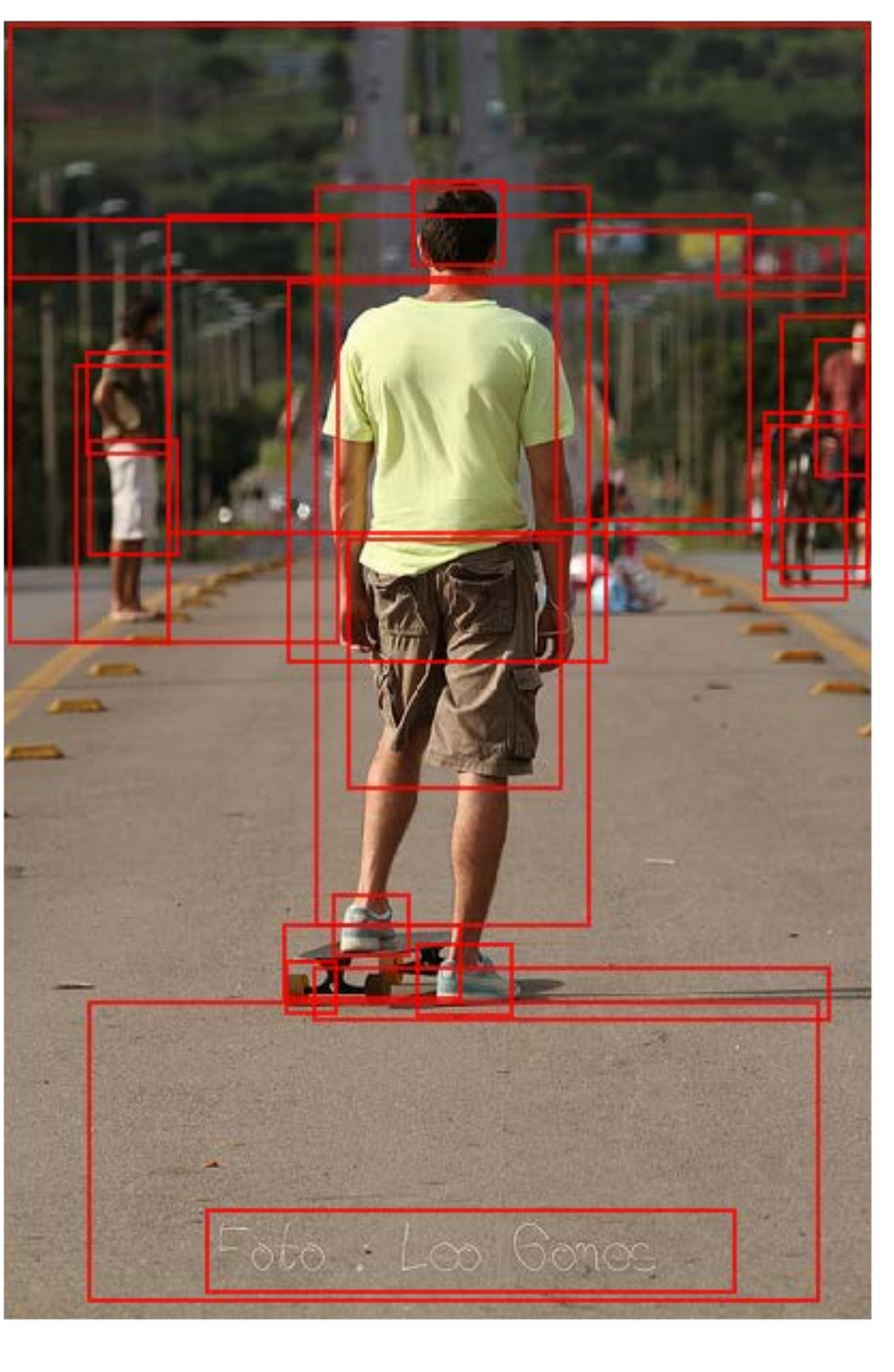}
		\end{center}
	\end{minipage}%
	\caption{\blue{Spatial-based versus object-based visual features, figure from~\cite{anderson2018bottom}.}}
	\label{fig:anderson2018bottom}
\end{figure}

Let us now turn to some advanced techniques used for modeling cross-modal interactions.
Upon understanding the visual world, humans have the ability to focus on specific region(s) instead of the entire scene.
Inspired by this  human-possessed ability, 
\textit{attention mechanism}~\cite{xu2015show} has been widely used in order to address the 
``where to look'' problem, resulting in one of the most effective improvement for various tasks including object
recognition, reading comprehension, image captioning and visual question answering etc. 
The core idea of attention is allowing the neural network to learn what regions to focus on, 
by means of modeling interactions between the content and side information in relevant regions. 
To adapt visual attention in VQA models, region-specific local image features are first extracted from 
an intermediate layer (before the last pooling operation) of a pretrained CNN.
Then a scalar attention weight for each region is calculated using both textual question and local visual features, 
which indicates the relevance of the given region and question. 
Finally, the image features can be represented as a weighted sum of the local visual region features.
As an essential component for many VQA models, 
quite a few variations of attention mechanism have been proposed in the literature for modeling the interactions between 
textual and visual modalities~\cite{yang2016stacked, lu2016hierarchical, yu2017multi, anderson2018bottom}. 
Yang~\etal~\cite{yang2016stacked} present a stacked attention network (SAN) which uses the semantic feature of the textual question 
as a query to search for those relevant visual regions through a multi-layer architecture. 
Lu~\etal~\cite{lu2016hierarchical} propose a hierarchical co-attention (HieCoAtt) model that 
combines ``visual attention'' and ``question attention''
via conducting a question-guided attention on image and a image-guided attention on question,
as is shown in Eq~\eqref{eq:cross-model_vqa_1}:
\begin{align}
&  \bm{H}^v = \tanh(\bm{W}_v \bm{V} + (\bm{W}_{q} \bm{Q})\bm{C}) , \nonumber\\
& \bm{H}^q = \tanh(\bm{W}_q \bm{Q} + (\bm{W}_{v} \bm{V}) \bm{C}^T), \nonumber \\ 
& \bm{a}^v = \textrm{softmax}(\bm{w}_{hv}^T \bm{H}^v) , \nonumber \\
& \bm{a}^q = \textrm{softmax}(\bm{w}_{hq}^T \bm{H}^q) , \nonumber \\
&\hat{\bm{v}} = \sum_{n = 1}^{N} a^v_n \bm{v}_n,  \quad  \hat{\bm{q}} = \sum_{t=1}^{T} a^q_t \bm{q}_t ,
\label{eq:cross-model_vqa_1}
\end{align}
where \blue{${H}^v$ and ${H}^q$ are latent deep representations of visual image features and textual question features respectively.}
$\bm{C} \in \mathcal{R}^{T \times N}$ is an affinity matrix whose entries represent the similarities between the question 
features $\bm{{Q}} \in \mathcal{R}^{d \times T}$ and image features $\bm{{V}} \in \mathcal{R}^{d \times N}$.
Actually the affinity matrix $\bm{C} \in \mathcal{R}^{T \times N}$ can also be regarded as a connection between the question attention 
space to the image attention space.
The attention weights for each image region $\bm{v}_n$ and word $\bm{q}_t$ are denoted
as $\bm{a}^v \in \mathcal{R}^N$ and $\bm{a}^q \in \mathcal{R}^T$, respectively.
\blue{
	Instead of performing attentions on spatial feature maps (e.g., $7 \times 7$ ResNet101~\cite{he2016deep} \textit{res5c} feature maps) as previous works, Anderson~\etal~\cite{anderson2018bottom} introduce a bottom-up visual attention mechanism
	that enables object-level attention based on image regions obtained through Faster R-CNN~\cite{ren2015faster}, as is shown in Figure~\ref{fig:anderson2018bottom}.
}

Rather than adopting the naive element-wise production or concatenation, another group of works resort to
the bilinear pooling model a well as its variations~\cite{fukui2016multimodal, kim2016hadamard, ben2017mutan, yu2017mlb} 
to achieve a great success by computing the outer product of two vectors to enable interactions among elements in both vectors. 
Denoting $\mathbf{v} \in \mathcal{R}^{d_v}$ and $\mathbf{q} \in \mathcal{R}^{d_q}$ as image(visual) and question feature vectors, 
the classification vector $\mathbf{y} \in \mathcal{R}^{|\mathcal{A}|}$ can be calculated by Eq~\eqref{eq:cross-model_vqa_2}:
\begin{align}
\mathbf{y} = \left( \bm{\mathcal{T}} \times_1  \mathbf{q} \right) \times_2  \mathbf{v},
\label{eq:cross-model_vqa_2}
\end{align}
where $\bm{\mathcal{T}} \in \mathcal{R}^{d_q \times d_v \times |\mathcal{A}|}$ is the parameter tensor, 
the operator $\times_i$ denotes the \emph{i-mode} product between a tensor and a matrix,
which suffers from high dimensionality $\left(d_q \times d_v \times |\mathcal{A}|\right)$. 
Fukui~\etal~\cite{fukui2016multimodal} propose a Multi-modal Compact Bilinear pooling (MCB) algorithm which 
adopts a sampling-based computation and projection method to reduce dimensionality while preserving the performance of full bilinear pooling. 
Kim~\etal~\cite{kim2016hadamard} present a Multi-modal Low-rank Bilinear pooling (MLB) model that forces 
the rank of the weight tensor to be low, as is shown in Eq~\eqref{eq:cross-model_vqa_3}:
\begin{align}
\mathbf{y} = \bm{P}^{\top} \left( \bm{W}_{q}^{\top}\mathbf{q} \circ \bm{W}_{v}^{\top}\mathbf{v}  \right) + \mathbf{b},
\label{eq:cross-model_vqa_3}
\end{align}
where $\bm{W}$, $\bm{P}$, $\bm{b}$ are model parameters and $\circ$ denotes the Hadamard product operator.
Yu~\etal~\cite{yu2017mlb} propose the Multi-modal Factorized Bilinear (MFB) pooling by utilizing some tricks in matrix factorization 
to improve the convergence rate and reduce the number of parameters. 
By combining low-rank matrix constraint with Tucker decomposition, 
i.e., $\bm{\mathcal{T}} = \left( \left( \bm{\mathcal{T}}_{c}\times_1 \bm{W}_q \right) \times_2 \bm{W}_v \right) \times_3 \bm{W}_o$, 
Ben~\etal~\cite{ben2017mutan} introduce MUTAN, and the combination is expressed in Eq~\eqref{eq:cross-model_vqa_4}:
\begin{align}
\mathbf{y} = \left( \left(\bm{\mathcal{T}}_{c} \times_1 \left(\mathbf{q}^\top \bm{W}_q \right)\right) \times_2 \left(\mathbf{v}^\top \bm{W}_v \right)\right) \times_3 \bm{W}_{o},
\label{eq:cross-model_vqa_4}
\end{align}
where $\bm{W}_q \in \mathcal{R}^{d_q \times t_q}$, $\bm{W}_v \in \mathcal{R}^{d_v \times t_v}$, 
$\bm{W}_o \in \mathcal{R}^{|\mathcal{A}| \times t_o}$, and $\bm{\mathcal{T}}_c \in \mathcal{R}^{t_q \times t_v \times t_o}$.


Recent studies have pointed out that current VQA models heavily rely on biases in different datasets and
many existing methods overly exploit these biases to ``correctly'' answer questions without considering the real visual information. 
For example, a model may answer ``2'' to any question starting with ``How many'' without really counting the numbers 
because the model learns (from biases) that answering ``2'' is the best guess for this dataset. 
As a consequence, even ``blind'' model can achieve satisfying results without well understanding the questions and images. 
Many efforts, such as building more balanced datasets~\cite{balanced_vqa_v2, agrawal2018don} and 
enforcing more transparent model designs, have been made to alleviate this issue.

\begin{figure*}[!t]
	\begin{center}
		\includegraphics[width=0.8\linewidth]{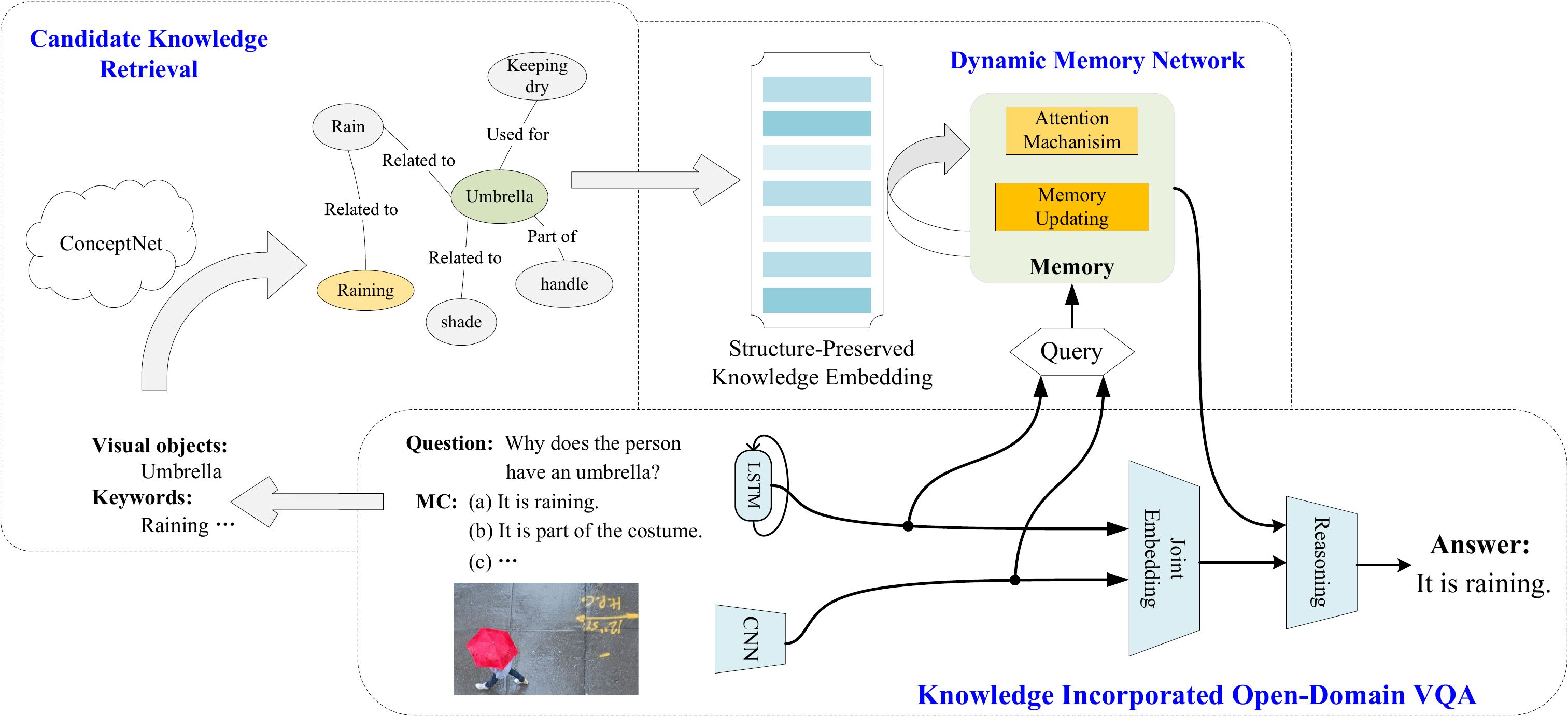} 
	\end{center}
	\caption{The architecture of a knowledge-incorporated VQA system, figure from~\cite{li2017incorporating}.}
	\label{fig:li2017incorporating}
\end{figure*}

\begin{figure}[htbp]
	\centering
	\includegraphics[width=0.499\textwidth]{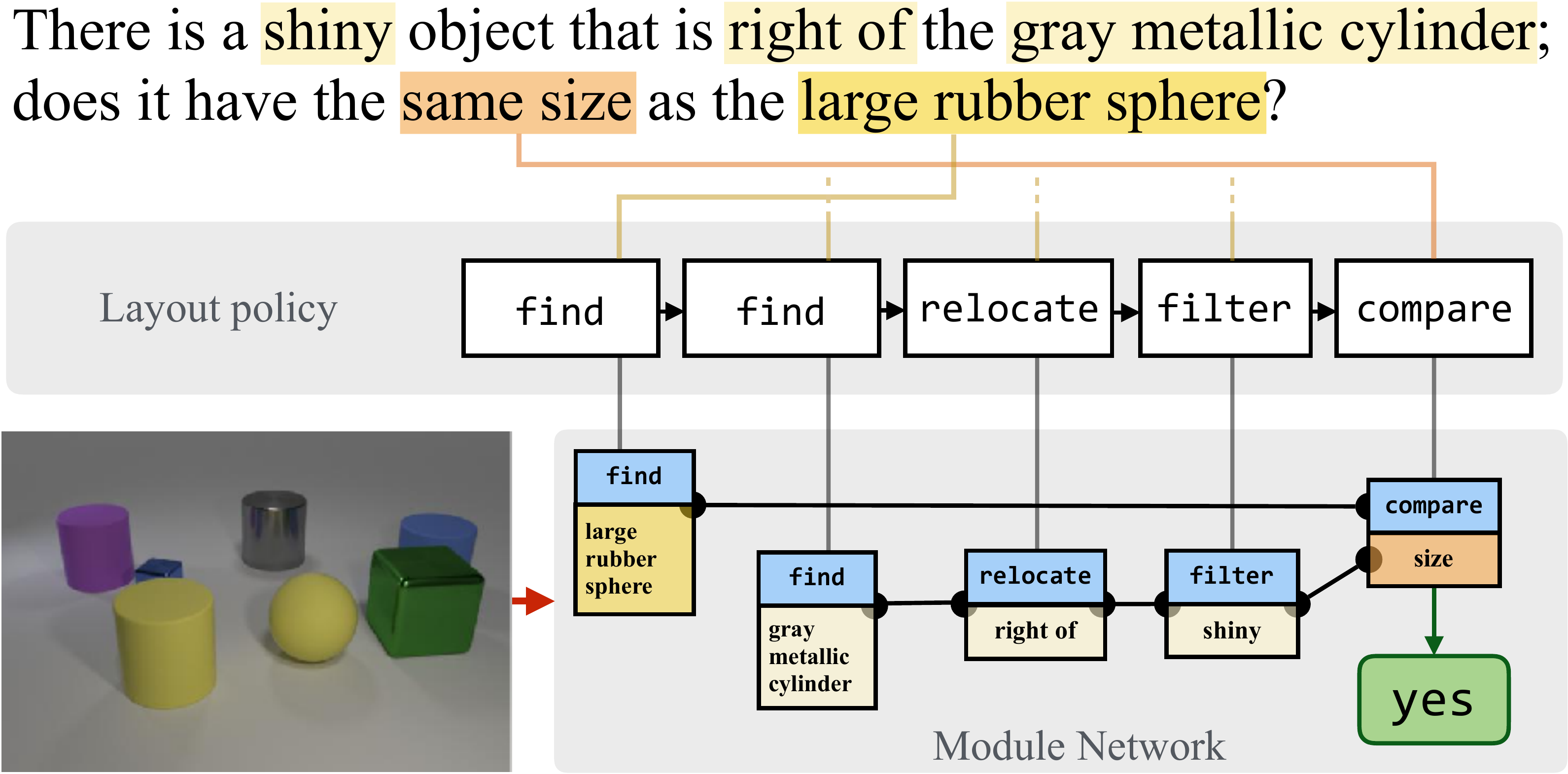} 
	\caption{\blue{An illustration of an end-to-end module network, figure from~\cite{hu2017learning}.}}
	\label{fig:hu2017learning}
\end{figure}

{\it Multi-modal fusion.}
Instead of building models merely based on visual and textual features via deep neural networks, 
several works seek for structural representations to handle the multi-modal nature in VQA. 
A series of works related to compositional models~\cite{andreas2016neural, hu2017learning, johnson2017inferring, mascharka2018transparency, cao2018visual} 
have shown exciting visual reasoning abilities on synthetic datasets. 
Their fundamental ideas are to compose instance-specific networks based on compositional structures of questions
via a collection of jointly trained neural modules. This can be regarded as a process of multi-modal information fusion 
where the question information is encoded inside the network architecture. 
\blue{
	An example of neural module networks is shown in Figure~\ref{fig:hu2017learning}.
}

\begin{figure}
	\centering
	\includegraphics[width=0.499\textwidth]{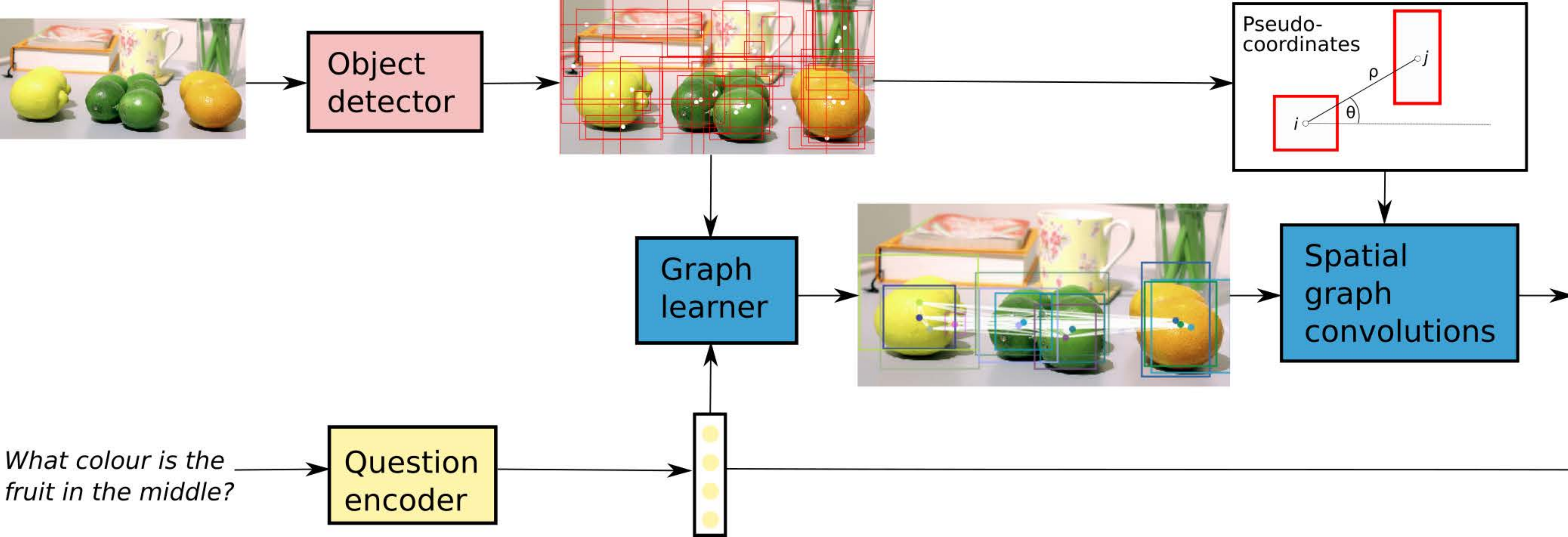} 
	\caption{\blue{Overview of a graph-based approach for VQA, figure from~\cite{norcliffe2018learning}.}}
	\label{fig:norcliffe2018learning}
\end{figure}

Another promising attempt is to exploit graph-structured 
representations in VQA~\cite{teney2016graph,norcliffe2018learning}, where object relations and language structures are represented as graphs whose
structure information can be further explored via techniques such as graph convolutional networks (GCN).
\blue{
	As is shown in Figure~\ref{fig:norcliffe2018learning}, Norcliffe-Brown~\etal~\cite{norcliffe2018learning} propose a graph-based approach for visual question answering.
	This work exploits a graph convolution-based method~\cite{monti2017geometric} to learn new visual representations from spatial graphs, where graph nodes are bounding boxes for object detections and graph edges are learned via an attention-based ``Graph Learner'' component.
	The graph convolution operator is defined at kernel $k$ for node $i$ as:
	\begin{align}
	\mathbf{f}_{k}(i) = \sum_{j \in \mathcal{N}(i)}{w_{k}(\mathbf{u}(i,j))}\mathbf{v}_j \alpha_{ij}, \qquad k=1,2,...,K 
	\end{align}
	where $\mathbf{u}(i,j)$ is a pseudo-coordinate function describing the relative spatial positions of vertex $i$ and $j$,
	$w_{k}(\mathbf{u})$ is the $k_{th}$ convolution kernel, $\mathcal{N}(i)$ denotes the neighbourhood of vertex $i$,
	$\mathbf{v}_j$ is the associated feature vector of vertices $j$, $\alpha_{ij}$ is the edge weight produced by the ``Graph Learner'' component.
	In the end, the convolutional feature of vertex $i$ is obtained through a concatenation over the $K$ kernels.
}

{\it Incorporating domain knowledge.}
In some situations, visual questions are not answerable by analyzing the questions and visual information themselves alone. 
Correctly answering visual questions may require extra information ranging from common-sense to expert domain knowledge, 
which is far beyond what the training dataset can provide. 
Thus it will be attractive to incorporate useful domain knowledge retrieved from other sources into VQA systems. 
Several pioneering works~\cite{wang2015explicit, wang2017fvqa} explore explicit reasoning on visual concepts and 
supporting facts in structural knowledge base, where raw visual signals are transformed into semantic symbols.
In contrast to above symbolic-based methods, Li~\etal~\cite{li2017incorporating} propose a Knowledge-incorporated Dynamic Memory Network (KDMN) framework
which incorporates massive domain knowledge into a semantic space to answer visual questions. 
Figure~\ref{fig:li2017incorporating} provides a general picture for KDMN framework which consists of three main modules,i.e., retrieval, fusion, inference. 
In retrieval module, an appropriate number of candidate knowledge triplets are retrieved from the external large-scale KB 
through analyzing the visual content and textual question. By treating the retrieved knowledge triplets as SVO phrases in fusion module, 
the authors utilize an LSTM to capture the semantic meanings and embed the knowledge into memory slots, as is shown in the following Eq~\eqref{eq:cross-model_vqa_5},
\begin{align}
C^{(t)}_{i} &= \text{LSTM}\left(\bm{L}[w^{t}_{i}], C^{(t-1)}_{i}\right), t=\{1,2,3\},  \\
\bm{M} &= \left[C^{(3)}_{i}\right],
\label{eq:cross-model_vqa_5}
\end{align}
where $w^{t}_{i}$ is the $t_{\text{th}}$ word of the $i_{\text{th}}$ SVO phrase, $\bm{L}$ is the word embedding matrix 
and $C_{i}$ is the internal state of LSTM cell when forwarding the $i_{\text{th}}$ SVO phrase. 
The memory bank $\bm{M}$ is designed to store a large amount of knowledge embedding. 
With the guidance of visual and textual features, those embeded knowledge triples are then fed into 
a Dynamic Memory Network~\cite{xiong2016dynamic} 
to obtain a distilled episodic memory vector in an iterative manner as follows:
\begin{align}
\mathbf{q}&= Query\left(\mathbf{f}^{(I)},\mathbf{f}^{(Q)},\mathbf{f}^{(A)}\right), \\
\mathbf{c}^{(t)}&=Attention\left(\bm{M}; \mathbf{m}^{(t-1)},\mathbf{q} \right), \\
\mathbf{m}^{(t)}&=Update\left(\mathbf{m}^{(t-1)},\mathbf{c}^{(t)},\mathbf{q}\right),
\label{eq:cross-model_vqa_6}
\end{align}
where $Query$ creates a context-aware query vector $\mathbf{q}$, 
$Attention$ condenses the knowledge into a context vector $\mathbf{c}^{(t)}$ in the $t_{\text{th}}$ iteration,
and $Update$ distills information into an episodic memory vector $\mathbf{m}^{(t)}$ iteratively.
The final episodic memory vector $\mathbf{m}^{(T)}$ can be jointly utilized with visual features to inference the answer. 

Compared with approaches based on simple explicit reasoning, methods incorporating external discrete knowledge 
not only maintain the superiority of deep models but also acquire the ability to exploit external knowledge for more complex reasoning.

\subsubsection{\bf Multi-modal Video Summarization}
Video summarization is an important and challenging research direction in computer vision (CV).
It aims to produce a short video summary which contains a small portion of the video segments,
so as to give users a synthetic and useful visual abstract of the video content.
A great number of uni-modal approaches have been proposed to solve the problem of video summarization, among which
unsupervised methods~\cite{cong2012towards,Lu2014A,lee2012discovering,yao2016highlight} normally pick frames or shots
from videos with some manually designed visual criteria and supervised methods~\cite{zhang2016video,sharghi2016query} tend to
directly leverage human-edited summary examples to learn video summarization models as well as dig the specific visual patterns for video summaries. 
Besides the visual features, it has also been observed that videos are often paired with abundant information from other modalities, such as audio signals, text descriptions and so on. All the modality information is aligned or complementary with each other, and capable of reflecting video contents in different aspects. Simultaneously considering different modality information of videos can provide video summarization model with a more comprehensive view.
Therefore, various multi-modal video summarization methods are proposed based on this idea and we remark that video summarization can also be treated as one application of multi-modal fusion.


\begin{figure}
	\centering
	\includegraphics[width=0.499\textwidth, height=7cm]{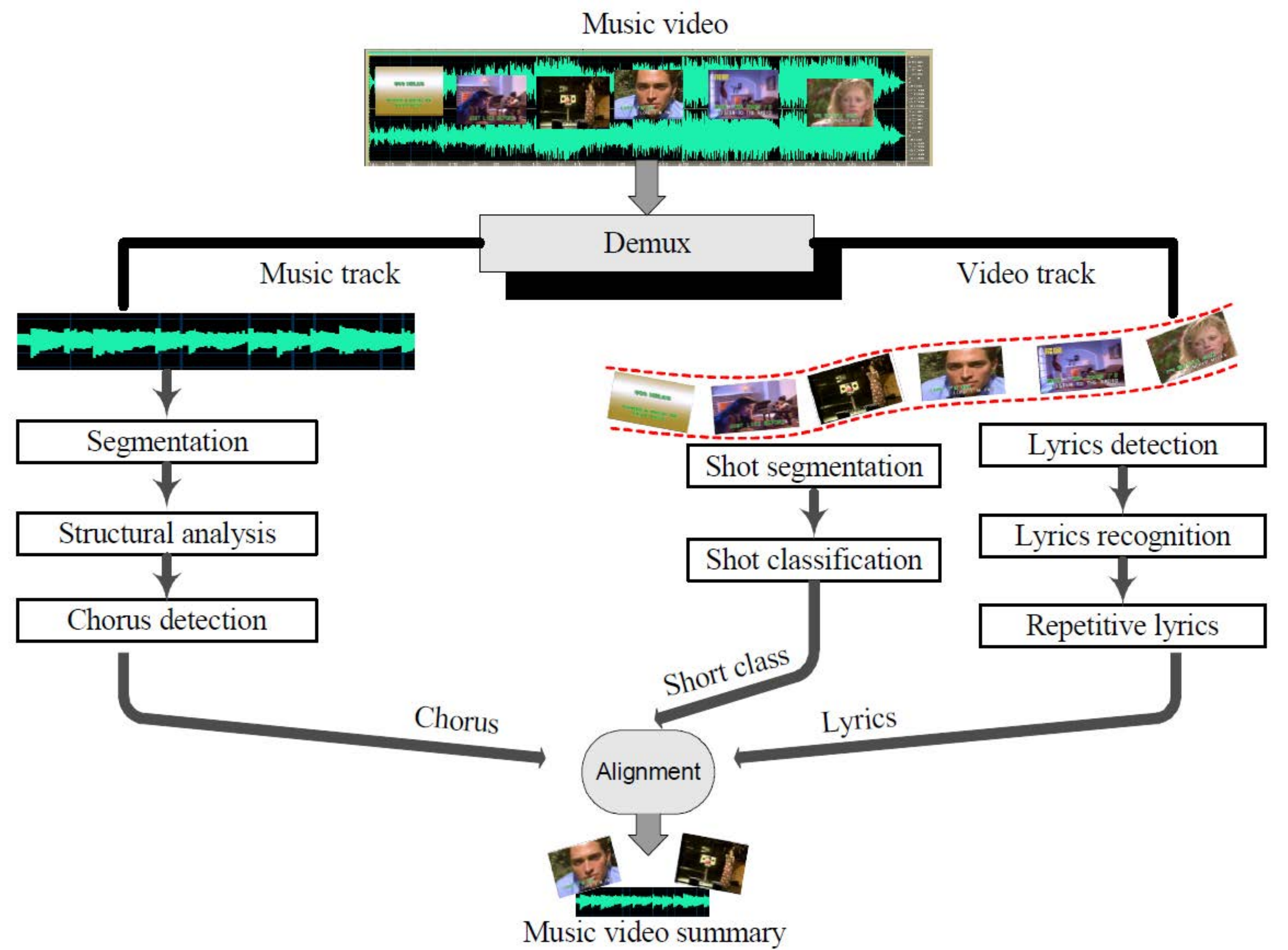} 
	\caption{Workflow of the music video summarization, figure from~\cite{Xu2005Automatic}.}
	\label{fig:music_video_summary}
\end{figure}

\begin{figure}
	\centering
	\includegraphics[width=0.45\textwidth]{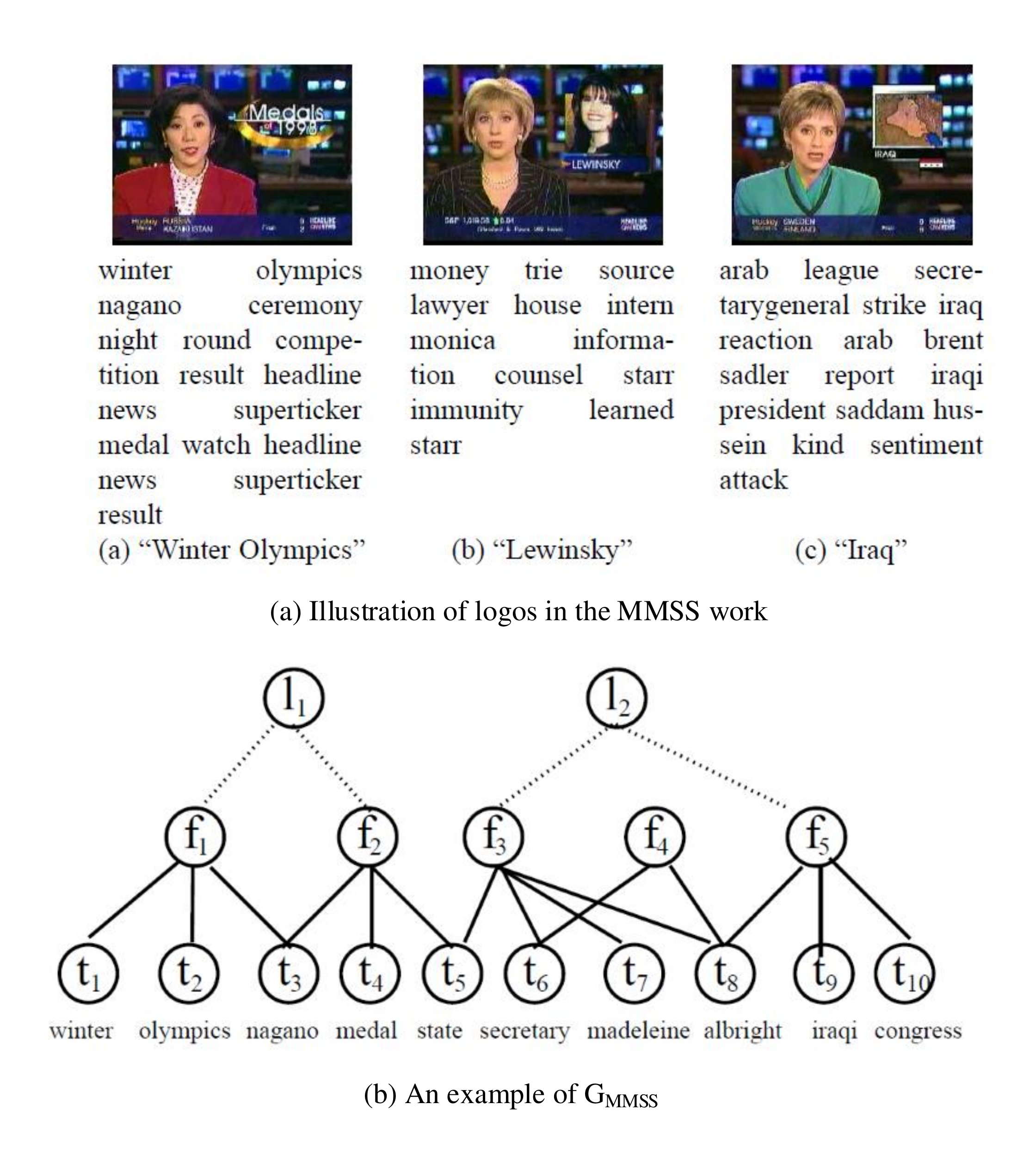} 
	\caption{\blue{Illustration of multi-modal story-oriented video summarization (MMSS), figure from~\cite{Pan2004MMSS}.}}
	\label{fig:MMSS_summary}
\end{figure}

{\it Conventional multi-modal video summarization.}
Conventional multi-modal video summarization methods mainly focus on summarizing movies or music videos.
These methods often detect and synthesize low-level visual/audio/textual cues from video itself to assess the saliency,
representativeness or quality of different video parts, and then extract those informative parts to create the final video summary.
Xu~\etal~\cite{Xu2005Automatic} propose a music video summarization method based on audio-visual-text analysis and alignment.
As is shown in Figure~\ref{fig:music_video_summary}, they first separate the music video into a music track and a video track. For the music track, the chorus is detected based on music structure analysis. For the video track, the (video) shots are segmented and classified into close-up face shots and non-face shots, followed by extraction of the lyrics and detection of the most repeated lyrics from these shots. The music video summary is generated based on the alignment of boundaries of the detected chorus, shot class and the most repeated lyrics from the music video.
Pan~\etal~\cite{Pan2004MMSS} introduce a multi-modal story-oriented video summarization (MMSS) model through
encoding both textual and scene information, as well as logos which link shots of a story as a graph. \blue{As is shown in Figure~\ref{fig:MMSS_summary}(a), broadcast news production commonly shows a small icon beside an anchorperson to represent the story. The same icon is usually reused later in the shots about the follow-up development of the story, as an aid for the viewers to link current coverage to past coverage. These icons are called ``newslogos''. The property of logos makes them a robust feature for linking separated footages of a story. Based on the above observations, Pan~\etal~\cite{Pan2004MMSS} build a $G_{MMSS}$ graph as shown in Figure~\ref{fig:MMSS_summary}(b), which is a three-layer graph with three types of nodes and two types of edges. The three types of nodes are logo-node, frame-node and term-node, corresponding to the logos, keyframes (each representing a shot), and terms, respectively. The two types of edges are the term-occurrence edge and the ``same-logo'' edge. In logo story summarization, frames and terms forming the summary are selected based on their ``relevance'' to the query object, the logo (node) of the story. The strategy of random walk with restarts (RWR) is used to obtain a story-specific relevance ranking among the terms and shot key frames in the graph $G_{MMSS}$, then the frames (i.e., nodes) and terms (nodes) with the highest RWR scores will be selected as the story summary.}
Evangelopoulos~\etal~\cite{Evangelopoulos2009Video} formulate
the detection of perceptually important video events on the basis of saliency models for the audio,
visual and textual information conveyed in a video stream.
Audio saliency $S_a$ is assessed by cues that quantify multi-frequency waveform modulations.
Visual saliency $S_v$ is measured through a spatio-temporal attention model driven by intensity,
color and motion. Text saliency $S_t$ is extracted by part-of-speech tagging on the subtitle information
from videos.
The various modality curves are integrated into a single attention curve by a weighted linear combination
of the audio, visual and text saliency,
\begin{align}
S_{avt} = w_a S_a + w_v S_v + w_t S_t,
\label{eq:cross-modal_video_summarization_1}
\end{align}
where the presence of an event may be identified in one or multiple domains.
This multi-modal saliency curve is the basis of bottom-up video summarization algorithms which
refine results from uni-modal or audiovisual-based skimming.

{\it Multi-modal video summarization for online videos with various side information.}
With the massive growth of video websites and social networks,
the problem of summarizing online web videos has attracted more and more attentions from researchers.
Different from traditional offline videos, online web videos are surrounded with various kinds of side information
such as tags, titles, descriptions and so on, which carries rich domain knowledge.
This domain knowledge often highlights crucial video contents
that people focus on and therefore is quite vital for improving the performances of video summarization algorithms.
Several multi-modal video summarization methods
link web videos with their domain knowledge to analyze video contents and
then generate the video summaries.

\begin{figure}
	\centering
	\includegraphics[width=0.98\linewidth]{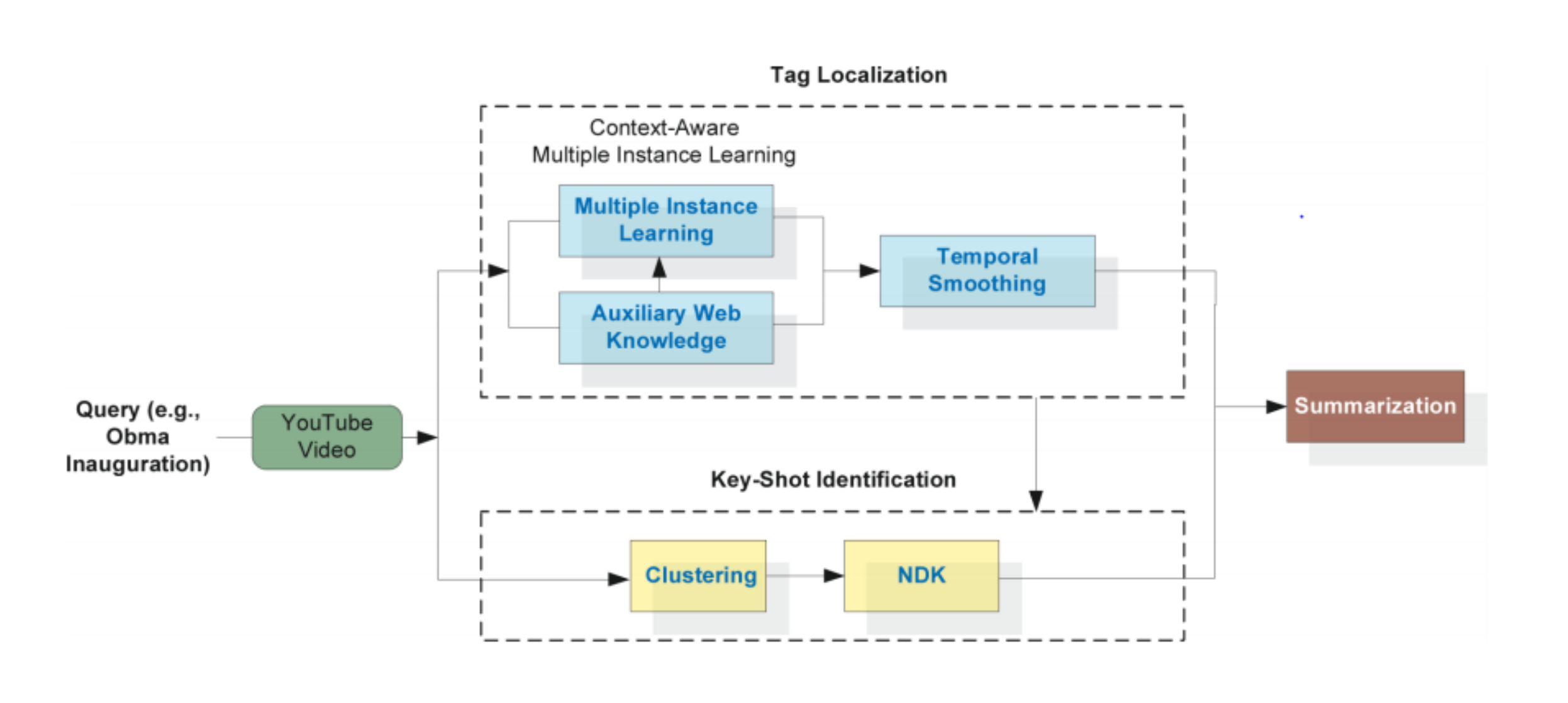}
	\caption{\blue{Schematic illustration of the event driven web video summarization approach, figure from~\cite{Wang2012Event}.}}
	\label{fig:eventdriven_summary}
\end{figure}

Wang~\etal~\cite{Wang2012Event} present an approach for event-driven video summarization
by tag localization and key-shot mining. 
\blue{As is illustrated in Figure~\ref{fig:eventdriven_summary},} they first localize the tags associated with each video into its shots, where the conditional probability that a shot contains a tag $t_k$ is defined as:
\begin{align}
v^k_{ij} = P_t(y_{ij}| f_{ij}) = \frac{1}{1+exp(-(w_kf_{ij}+b_k))},
\label{eq:cross-modal_video_summarization_tag1}
\end{align}
where $f_{ij}$ is the feature vector of the $j$th shot of the $i$th video. $w_k$ and $b_k$ are the parameters to be learned by the multiple instance learning. After obtaining the relevance scores of the shots with respect to all tags, the relevance score of each shot with respect to an event query can then be estimated. Denote $v^k$ as the relevance score of a shot with respect to the $k$th tag, then the relevance score of this shot with respect to an event query can be defined as follows:
\begin{align}
y = \frac{1}{K} \sum_k sim(q,t_k) v^k,
\label{eq:cross-modal_video_summarization_tag2}
\end{align}
where $q$ is the query and $sim(q,t_k)$ is the similarity between query $q$ and tag $t_k$. Finally, a set of key-shots having high relevance scores can be identified by exploring the repeated occurrence characteristics of key sub-events.

\begin{figure}
	\centering
	\includegraphics[width=0.98\linewidth]{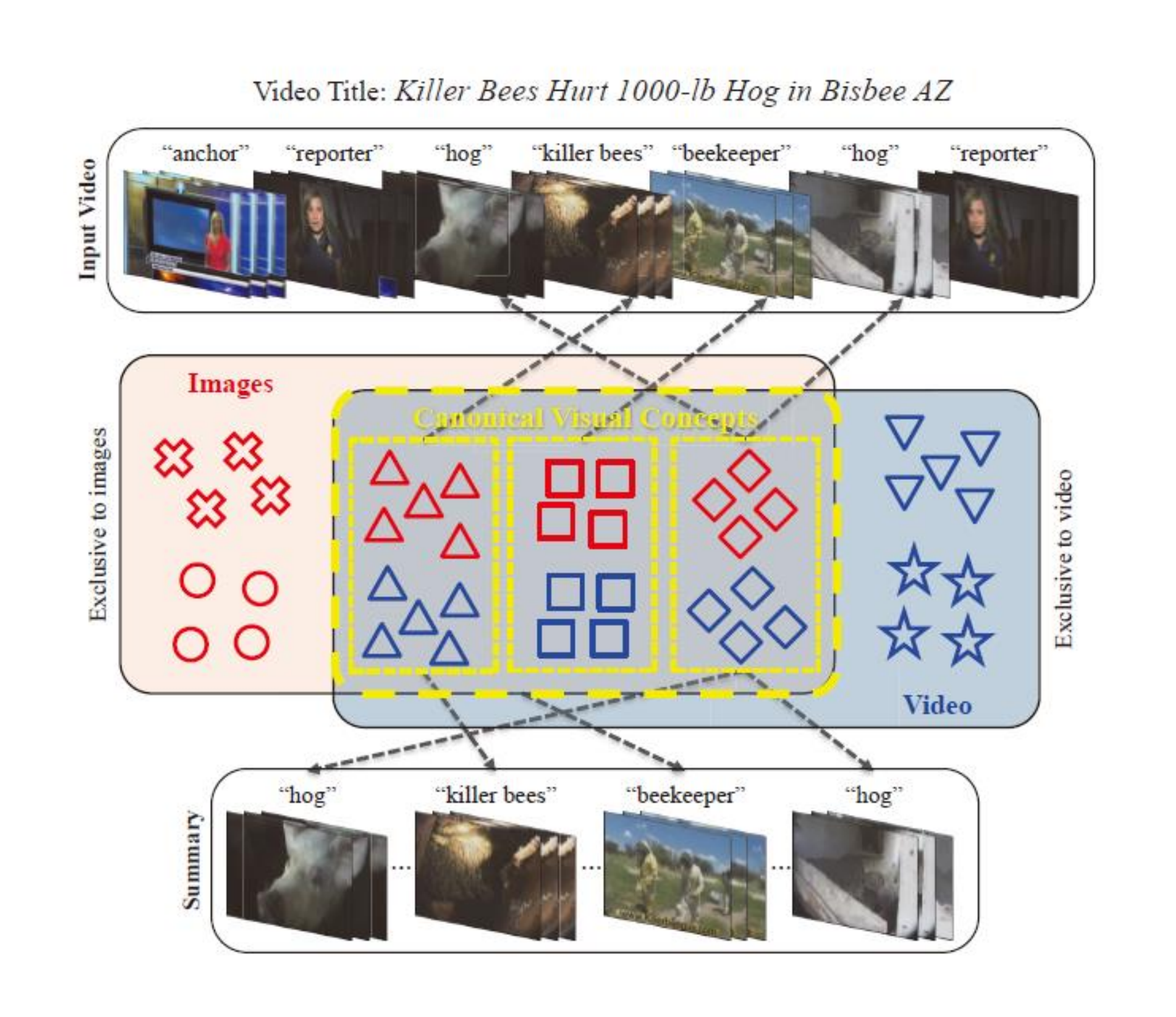}
	\caption{\blue{An illustration of title-based video summarization, figure from~\cite{Song2015TVSum}.}}
	\label{fig:tvsum_summary}
\end{figure}

Song~\etal~\cite{Song2015TVSum} observe that a video title is often carefully
chosen to be maximally descriptive of its main topic,
and thus images related to the title can serve as a proxy for important visual concepts of the main topic.
Therefore, \blue{as is depicted in Figure~\ref{fig:tvsum_summary},} 
they leverage video titles to retrieve web images through image search engines and develop a co-archetypal
analysis technique which learns canonical visual concepts shared between videos and web images.
Specifically, suppose $X = [x_1, \cdots ,x_n] \in R^{d \times n}$ is a matrix of $n$ video frames
with each column $x_i \in R^d$ representing a frame with a certain set of image feature descriptors.
$Y = [y_1, \cdots , y_m] \in R^{d \times m}$ is a matrix of $m$ retrieved images defined in a similar way.
The learning of canonical visual concepts $Z = [z_1, \cdots , z_p] \in R^{d \times p}$ between $X$ and $Y$
should satisfy the following two geometrical constraints:
\begin{enumerate}
	\item Each video frame $x_i$ and image $y_i$ should be well approximated by a convex combination of latent variables $Z$.
	\item Each latent variable $z_j$ should be well approximated jointly by a convex combination of video frames $X$ and by a convex combination of images $Y$.
\end{enumerate}
The co-archetypal analysis is thus formulated as an optimization problem that finds a
solution set $\Omega = \left\{A^X, B^X, A^Y, B^Y \right\}$ by the following objective:
\begin{align}
\min_{\Omega} ||X-ZA^X||_F^2 + ||Y-ZA^Y||_F^2 + \gamma ||XB^X-YB^Y||_F^2,
\label{eq:cross-modal_video_summarization_song_summary}
\end{align}
where $A^X=[\alpha_1^X, \cdots, \alpha_n^X] \in R^{p \times n}$, $B^X = [\beta_1^X, \cdots, \beta_p^X] \in R^{n \times p}$,
and similarly $A^Y \in R^{p \times m}$, $B^Y \in R^{m \times p}$.
The first geometrical constraint is reflected by the first two terms in Eq~\eqref{eq:cross-modal_video_summarization_song_summary},
and the second constraint is reflected by the last term, assuming $Z = XB^X = YB^Y$.
Upon learning the canonical visual concepts $Z$ as well as the corresponding coefficient matrix $A$ and $B$,
video matrix $X$ can be factorized into $XBA$, and the importance score of the $i$th video frame can then be derived as follows:
\begin{align}
score(x_i) = \sum_{j=1}^n B_i \alpha_j,
\label{eq:cross-modal_video_summarization_song_score}
\end{align}
which is the total contribution of the corresponding elements of $BA$ in reconstructing the original signal $X$.
With this frame importance measurement, video frames of higher important scores are concatenated in
chronological order to form the video summaries.

\begin{figure}
	\centering
	\includegraphics[width=0.98\linewidth]{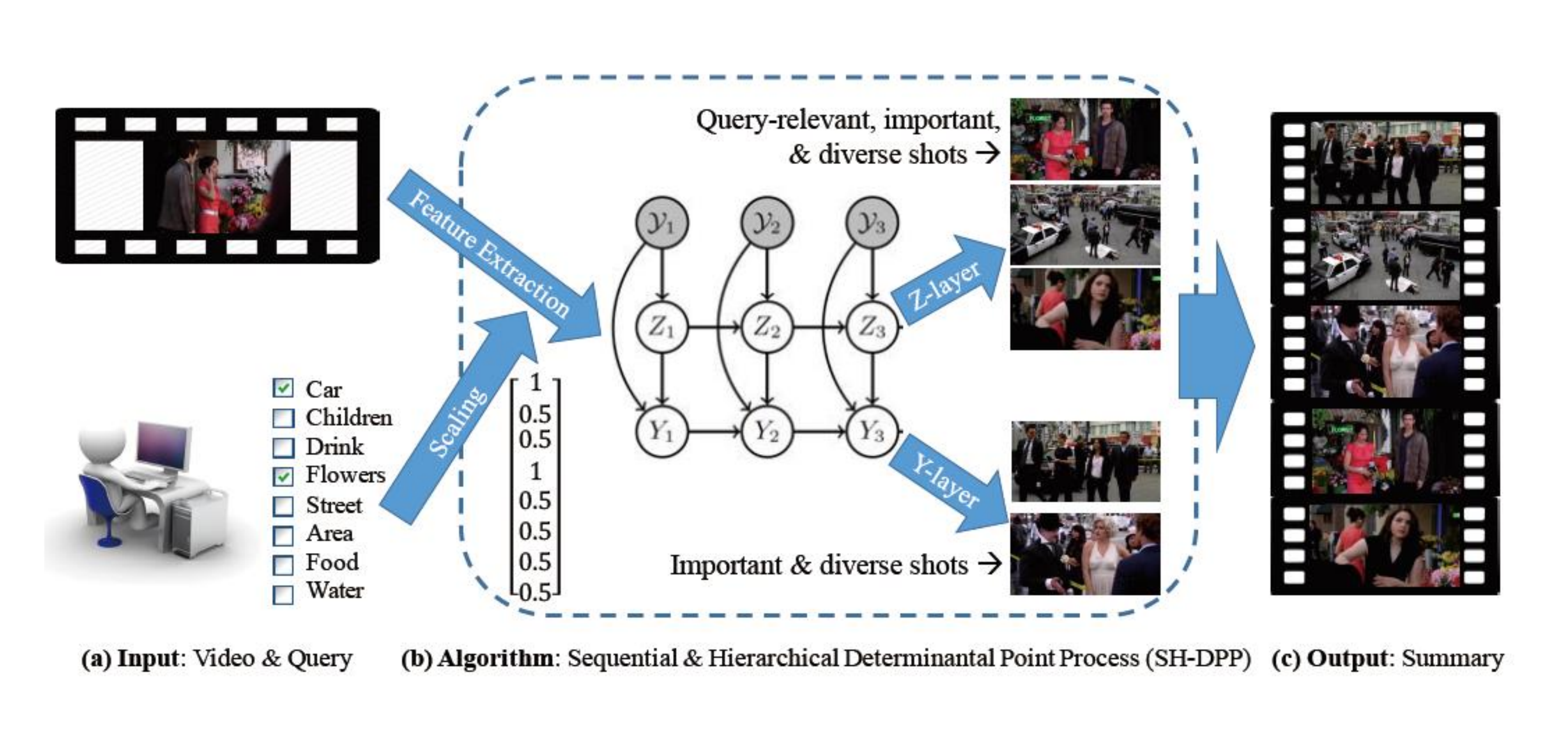}
	\caption{\blue{The workflow of query-focused extractive video summarization, figure from~\cite{sharghi2016query}.}}
	\label{fig:queryfocused_summary}
\end{figure}

\begin{figure}
	\centering
	\includegraphics[width=0.90\linewidth]{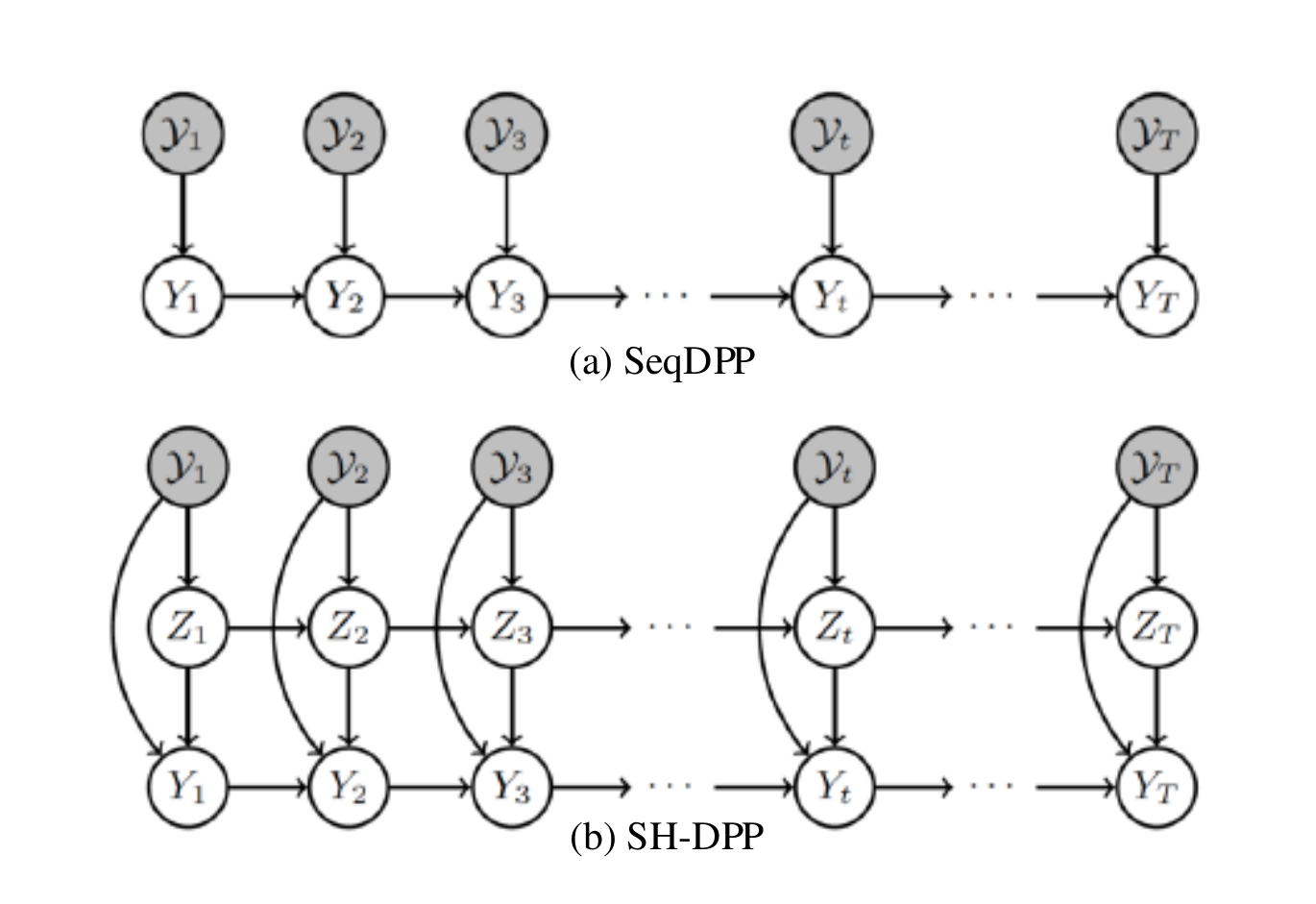}
	\caption{\blue{The graphical models of SeqDPP~\cite{gong2014diverse} (top) and SH-DPP (down), figure from~\cite{sharghi2016query}.}}
	\label{fig:SHDPP_model}
\end{figure}

\begin{figure*}
	\centering
	\includegraphics[width=0.98\textwidth]{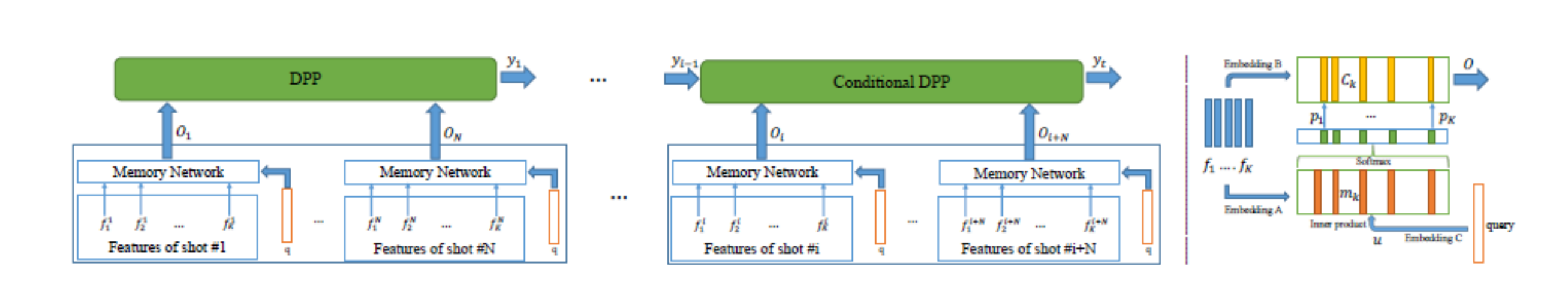}
	\caption{\blue{The overview for query-focused video summarizer with memory network, figure from~\cite{sharghi2017query}.}}
	\label{fig:memoryDPP_model}
\end{figure*}

\blue{
	Sharghi~\etal~\cite{sharghi2016query} propose a query-focused extractive video summarization problem, which aims to generate video summaries based on user provided textual queries. To solve the proposed problem, they develop a probabilistic model, i.e., Sequential and Hierarchical Determinantal Point Process (SH-DPP), where the decision to include one shot in the summary
	jointly depends on the shot's relevance to the user query and its importance in the context of video. The overall workflow for SH-DPP is shown in Figure~\ref{fig:queryfocused_summary}. Specifically, SH-DPP is established on a Sequential Determinantal Point Process (SeqDPP) method~\cite{gong2014diverse}, which firstly partitions a video into $T$ consecutive disjoint sets, $\cup_{t=1}^T \mathcal{Y}_t=\mathcal{Y}$, where $\mathcal{Y}_t$ represents a set consisting of only a few shots and stands as the ground set of time step $t$. The SeqDPP model is defined as follows (Figure~\ref{fig:SHDPP_model}(a) depicts its graphical model),
	\begin{align}
	P_{SEQ}(Y|\mathcal{Y})=P(Y_1|\mathcal{Y}_1)\prod_{t=2}^T P(Y_t|Y_{t-1},\mathcal{Y}_t), \ \ \mathcal{Y} = \cup_{t=1}^T \mathcal{Y}_t,
	\label{eq:seqDPP_1}
	\end{align}
	where $P(Y_t|Y_{t-1},\mathcal{Y}_t)$ is a conditional DPP to ensure the diversities between items selected at time step $t$ (by $Y_t$) and those selected in the previous time step (denoted as $Y_{t-1}$). In order to incorporate user queries into the video summarization procedure, the SH-DPP model (as is shown by the graphical model in Figure~\ref{fig:SHDPP_model}(b)) leverages the query information to guide the determinantal point process for video shot selection:
	\begin{equation}
	\begin{split}
	& P_{SH}(\{Y_1,Z_1\},\cdots, \{Y_T,Z_T\}|q,\mathcal{Y}) \\
	= & P(Z_1|q,\mathcal{Y}_1)P(Y_1|Z_1,\mathcal{Y}_1) \\
	& \prod_{t=2}^{T}P(Z_t|q,Z_{t-1},\mathcal{Y}_t)P(Y_t|Z_t,Y_{t-1},\mathcal{Y}_t).
	\end{split}
	\label{eq:seqDPP_2}
	\end{equation}
	The SH-DPP first utilizes the subset selection variables $Z_t$ to select the query-relevant video shots. Depending on the results from $Z_t$ and $Y_{t-1}$, the variable $Y_t$ in the last layer selects video shots to further summarize the remaining content in the video segment $\mathcal{Y}_t$. Since annotating ground-truth for selection variables $Z_t$ needs annotators to determine which query appears in which video shot, query-focused video summarization heavily relies on the user supervision for SH-DPP. 
}
\blue{
	Besides SH-DPP, Sharghi~\etal~\cite{sharghi2017query} further propose a query-focused video summarizer which employs memory network to parameterize the sequential determinantal point process. As is shown in Figure~\ref{fig:memoryDPP_model}, unlike the
	hierarchical model in~\cite{sharghi2016query}, the query-focused video summarizer does not require the costly user supervision on ``which queried concept appears in which video shot" or any pre-trained concept detectors.
}

\begin{figure}
	\centering
	\includegraphics[width=0.8\linewidth]{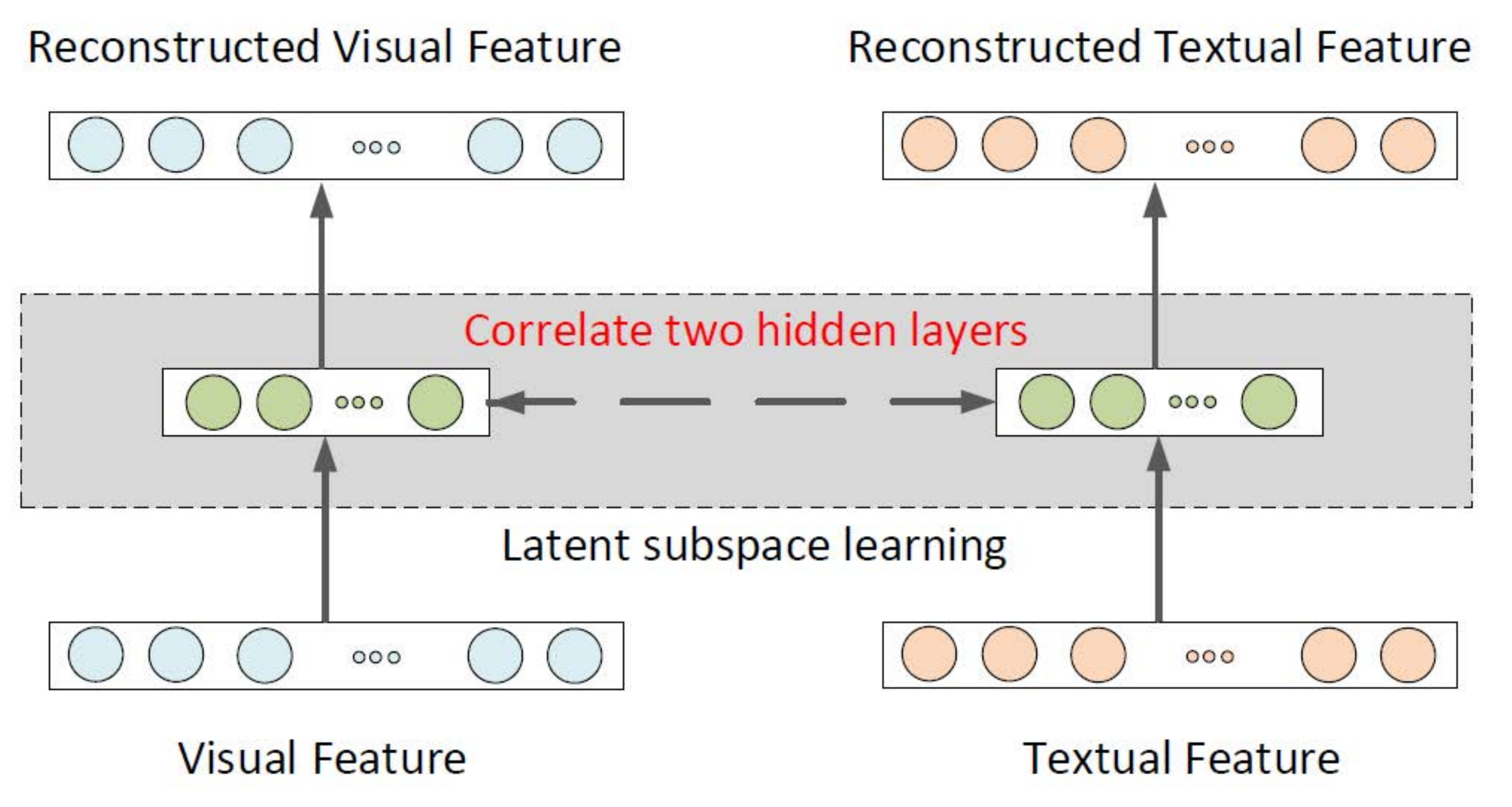}
	\caption{The architecture of multi-modal auto-encoders, figure from~\cite{Yuan2017Video}.}
	\label{fig:dsse_autoencoder}
\end{figure}

\begin{figure*}
	\centering
	\includegraphics[width=0.925\textwidth]{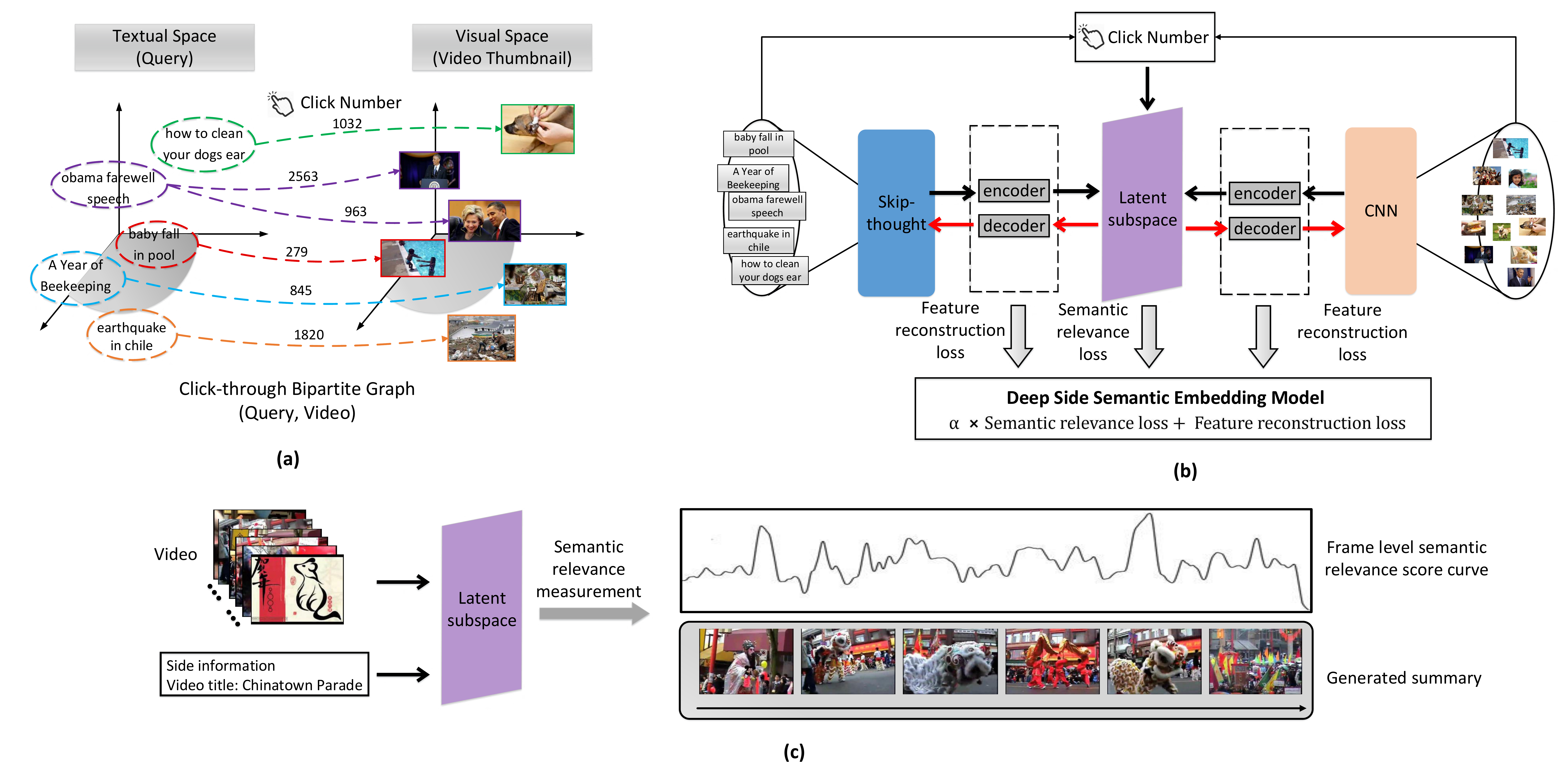}
	\caption{The overall framework of DSSE model, figure from~\cite{Yuan2017Video}.}
	\label{fig:dsse_whole}
\end{figure*}

Yuan~\etal~\cite{Yuan2017Video} present a Deep Side Semantic Embedding (DSSE) model to generate video summaries
by leveraging domain knowledge obtained from side information (e.g, captions, descriptions, queries) of online web videos.
The basic idea of DSSE is to construct a latent subspace with the ability of directly comparing domain knowledge and video frames.
In this latent subspace, the authors hope that the common information between videos and domain knowledge can be learned more completely
and the semantic relevance between them can be effectively measured.
As is shown in Figure~\ref{fig:dsse_autoencoder}, a latent subspace is constructed by correlating the hidden layers of two uni-modal auto-encoders
which embed the video frames and domain knowledge respectively.
Meanwhile, there are two components in the objective function of DSSE, i.e, $L_{rel}$ which
learns the semantic relevance and $L_{rec}$ which learns the feature reconstruction:
\begin{align}
L_{rel}(I_f,I_g;\Theta) = ||f(I_f;\Theta_f) - g(I_g;\Theta_g)||_2^2,
\label{eq:cross-modal_video_summarization_dsse_lrel}
\end{align}
\begin{align}
L_{rec}(I_f,I_g;\Theta) = ||\tilde{I_f} - I_f||_2^2 + ||\tilde{I_g} - I_g||_2^2,
\label{eq:cross-modal_video_summarization_dsse_lrec}
\end{align}
where $I_f$ represents the visual features of the video frames and $I_g$ represents the textual features of domain knowledge.
Accordingly, $f(I_f;\Theta_f)$ is the hidden representation of $I_f$ in the visual auto-encoder
and $f(I_g;\Theta_g)$ is the hidden representation of $I_g$ in the textual encoder.
$\tilde{I_f}$ and $\tilde{I_g}$ denote the reconstructed features.
$L_{rel}$ requires that the matched video frames and domain knowledge be close to each other in the latent subspace and
$L_{rec}$ preserves the useful original characteristics from different modalities/media in the common latent space.
By jointly minimizing $L_{rel}$ and $L_{rec}$ as follows:
\begin{align}
\min_{\Theta} \    \alpha L_{rel}(I_f,I_g;\Theta)+ L_{rec}(I_f,I_g; \Theta),
\label{eq:cross-modal_video_summarization_dsse}
\end{align}
the semantic relevance between video frames and domain knowledge can be measured in the hidden layers of the multi-modal auto-encoders
and semantically meaningful parts are selected from videos to generate video summaries by minimizing their distances to domain knowledge
in the constructed latent subspace. The whole picture of DSSE model is demonstrated in Figure~\ref{fig:dsse_whole}.

\subsubsection{\bf Multi-modal Visual Pattern Mining}
\begin{figure}
	\centering
	\includegraphics[width=0.98\linewidth]{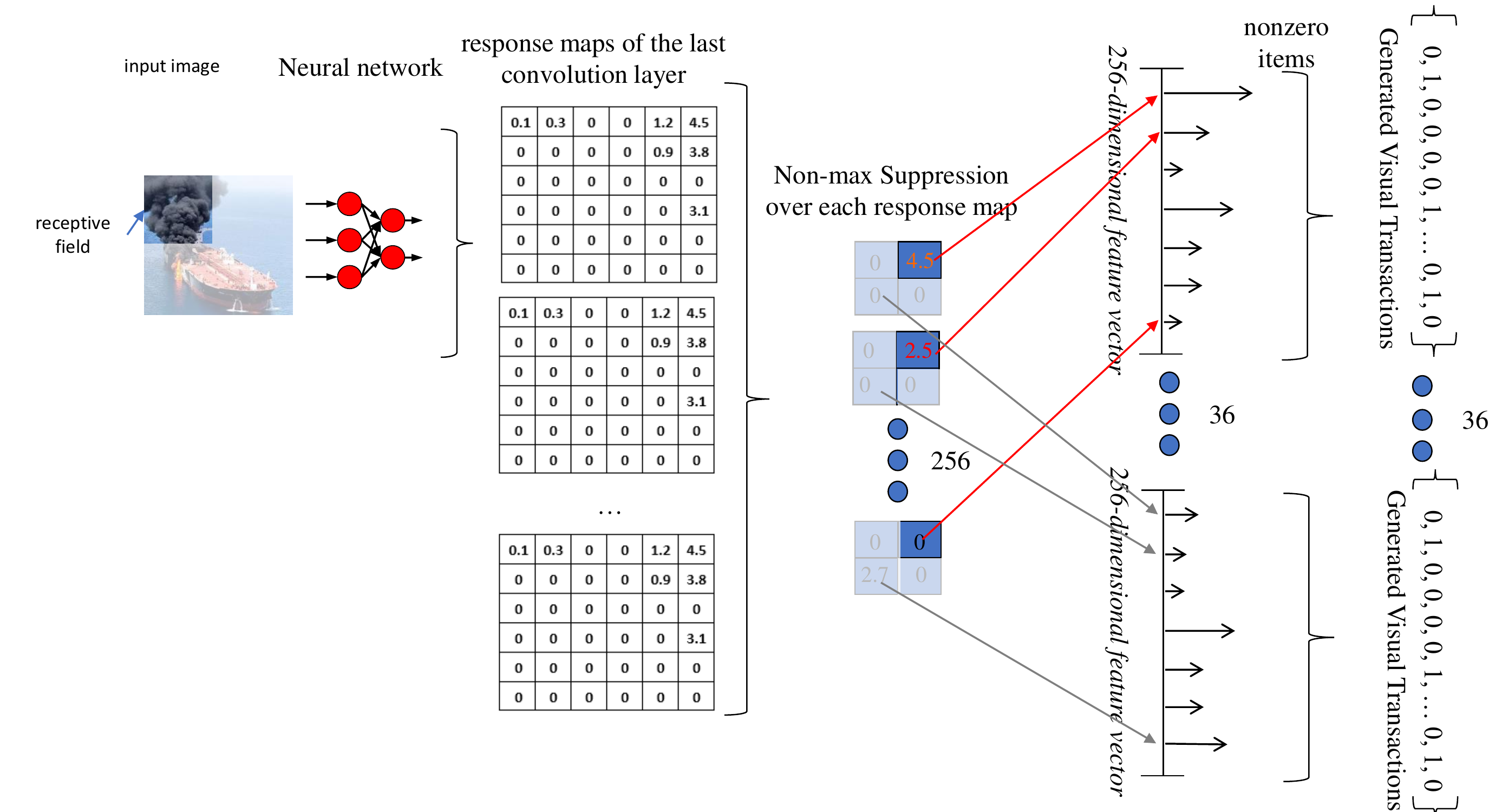} 
	\caption{\blue{The visual transaction generation pipeline utilizing the last convolutional layer of a convolutional neural network. This pipeline is used to obtain representations of each image that can localize the presence of a pattern within the image. figure from~\cite{li2016event}.}}
	\label{fig:visualtrans}
\end{figure}

Knowledge base is a collection of entities, attributes and the relations between them.
knowledge base schema is the structure of knowledge base and used to guide how the knowledge base is built.
It is often constructed manually using experts with specific domain knowledge for the field of interest.
Many tasks such as automatic content extraction highly depend on knowledge base.
However, the current approaches ignore visual information that could be used to build or populate these structured ontologies.
Preliminary work on visual knowledge base construction only explores limited basic objects and scene relations.
A few novel multi-modal pattern mining approaches are proposed
in~\cite{li2016event,li2018patternnet,zhang2015cross,lu2016cross,zhang2017improving}, towards constructing
a high-level ``event'' schema semi-automatically, which has the capability to extend text-only methods for schema construction.
A large unconstrained corpus of weakly-supervised image-caption pairs related to high-level events is utilized to both
discover visual aspects of an event, and name these visual components automatically.
Li~\etal~\cite{li2016event} leverage the activation signal of the convolution filters to encode the visual content,
and utilize the skip-gram language model to encode the textual information.
The association rule mining algorithm is introduced to jointly model the visual
and textual information from multi-modal data.
The encoded visual and textual contents are considered as transactions in
association rule mining algorithm. 
\blue{The visual transactions generation pipeline can be found in Figure~\ref{fig:visualtrans}.}

\begin{figure}
	\centering
	\includegraphics[width=0.98\linewidth]{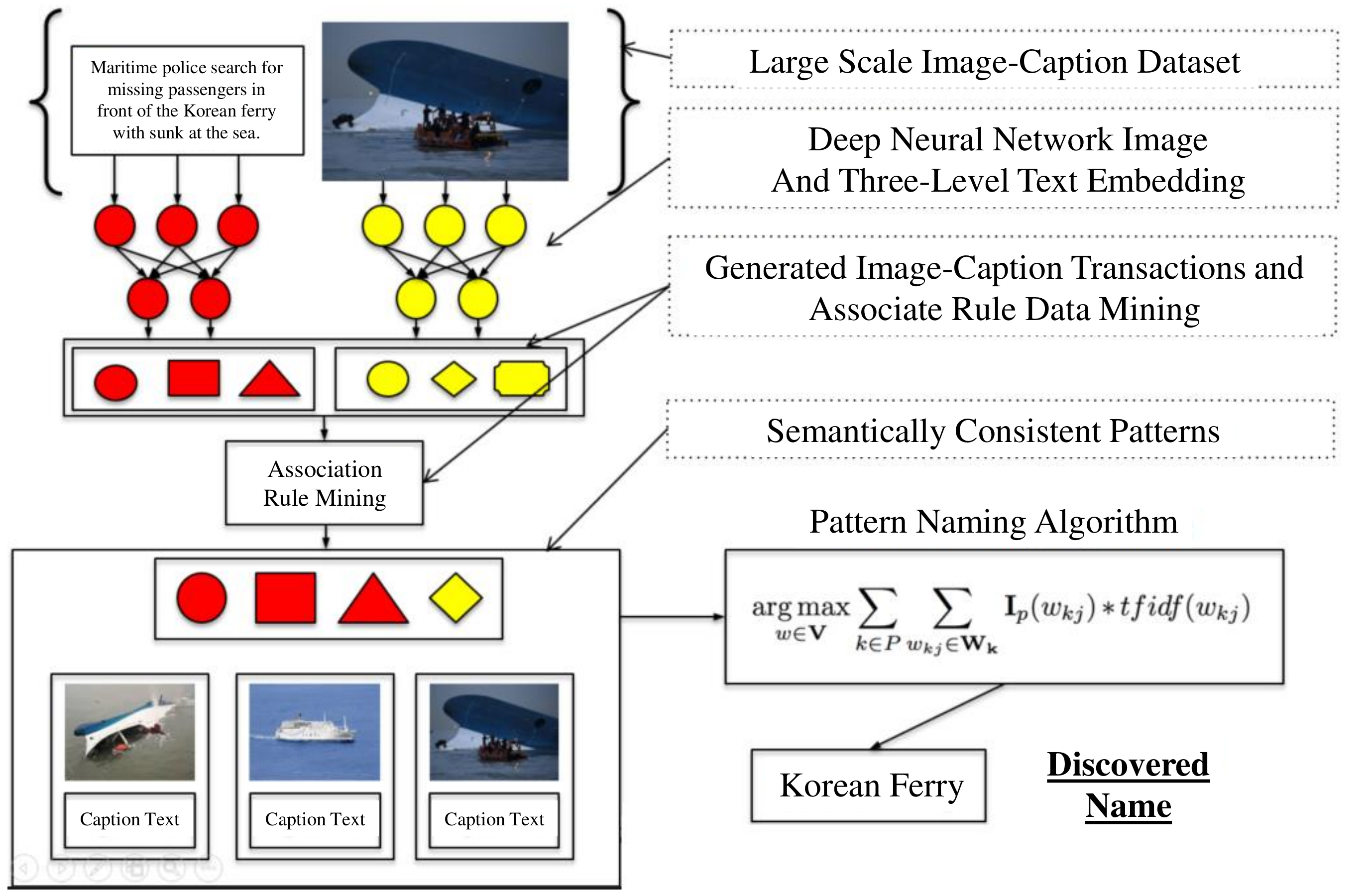} 
	\caption{Multimodal pattern discovery and naming pipeline, figure from~\cite{li2016event}.}
	\label{fig:patternmining}
\end{figure}

To discover the event related multi-modal patterns for knowledge base construction,
two criteria, \textit{representative} and \textit{discriminative}, are defined to find the high quality multi-modal visual patterns.
\textit{Discriminative} means the patterns discovered from a category should not be found in other categories.
\textit{Representative} means the discovered patterns should be commonly available in the category.
In association rule mining algorithm, representative property is defined by
support rate of a transaction, as is shown in~\eqref{eq:st}, and the discriminative property is defined by confidence rate,
as is shown in~\eqref{confidence}:

\begin{equation}
s(t^*) = \frac{|\{T_{a} |t^* \subseteq T_{a},  T_{a} \in S \}|}{m} ,
\label{eq:st}
\end{equation}

\begin{equation}
\label{confidence}
c(t^* \rightarrow y) = \frac{s(t^* \cup y)}{s(t^*)} ,
\end{equation}
where $T_a$ is a transaction, $t^*$ is a set of items and $y$ is the target category.
The discovered association rules can be converted to multi-modal visual patterns by the algorithm in~\cite{li2016event}.
Mathematically, the two pattern mining requirements can be defined as:

\blue{
	\begin{align}
	&c(t^* \rightarrow y) \geq c_{min} , \nonumber \\
	&s(t^*) \geq s_{min} , \nonumber \\
	&t^* \cap \mathbf{I} , \neq \emptyset , \nonumber \\
	&t^* \cap \mathbf{C} , \neq \emptyset ,
	\label{eq:contstraints}
	\end{align}
	where $y$ is the event category, $c_{min}$ is the threshold of minimum confidence rate, $s_{min}$ is the threshold of minimum support rate, $\mathbf{I}$ is the visual transactions, and $\mathbf{C}$ are the text transactions. Each multi-modal pattern $t^*$ has a set of visual items and a set of text patterns. The end-to-end multi-modal pattern discovery and naming framework can be found in Figure~\ref{fig:patternmining}.
}

Multi-modal pattern mining approach can be used as a bridge to fill the gap between text analysis and visual analysis.
Zhang~\etal~\cite{zhang2015cross,zhang2017improving} use the multi-modal visual pattern mining framework proposed
in~\cite{li2016event,zhang2014scalable} to improve the knowledge and event extraction problem in
Natural Language Processing community. Compare to the traditional text only event extraction approach,
multi-modal approach introduces the discovered domain knowledge from visual domain and achieve significantly better performance.

\subsubsection{\bf Multi-modal Recommendation}
With the explosive growth of various online social networks and multimedia sites, people are now getting used to
engaging on different medias simultaneously to satisfy their diverse information need~\cite{chen2012more}. 
It is reported that each user on average has $5.54$ social media accounts and is actively using $2.82$ social
platforms/media\footnote{\noindent GWI social report:
	http://www.globalwebindex.net/blog/internet-users-have-average-of-5-social-media-accounts}. 
The cross-modal information jointly reflects each individual's interest and preference. 
Therefore, organically transferring or associating cross-modal information is of significant importance 
in serving people intelligently~\cite{jiang2016little}.

Existing multi-modal recommendation works can be grouped from two angles, e.g., categorization 
according to association knowledge and categorization according to the entire model structure.

\begin{figure}[tb]
	\centering
	\includegraphics[width=0.8\linewidth]  {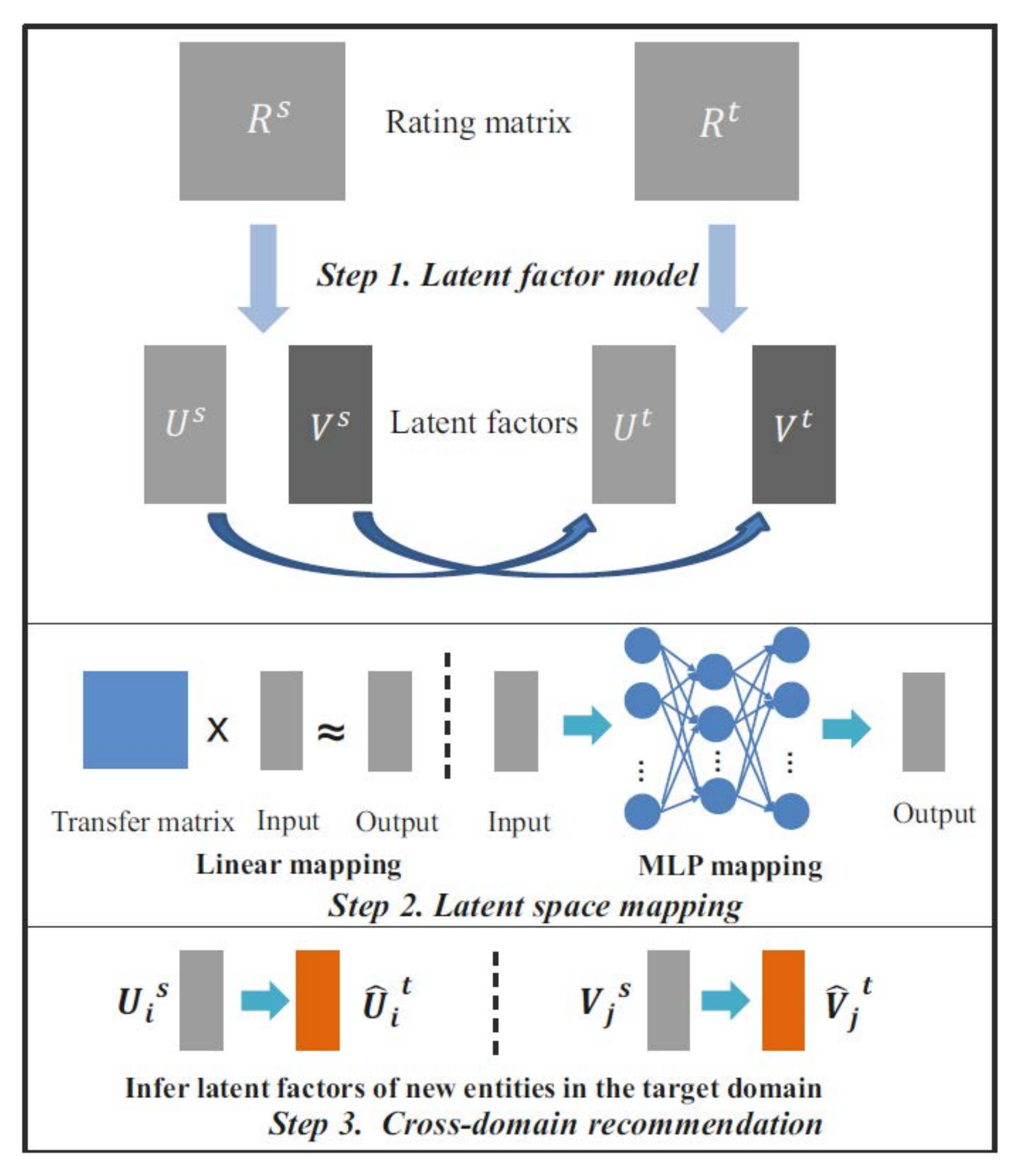}
	\caption{\blue{Illustrative diagram of the EMCDR framework in which linear transfer and MLP are adopted as mapping functions (MLP mapping is proved to perform better according to the experiment results), figure from~\cite{maaten2008visualizing}.}}
	\label{fig:emcdr}
\end{figure}

{\it Grouping by what knowledge to associate.}
When we look through existing multi-modal models in terms of the association knowledge, 
one group of methods follow a user-centric way, which focuses on cross-modal information of overlapped users.
A straight forward solution is to treat cross-modal association as a linear transfer problem, 
and pursue an explicit transfer matrix based on regression~\cite{yan2014mining, jiang2015social, Man2017Cross}. 
The objective function for this type of models can be expressed as follows:

\begin{equation}
\min_{\mathbf{W}} {\left\Vert \mathbf{W} \mathbf{U}^1 - \mathbf{U}^2 \right\Vert}_{\rm F}^2 + \lambda {\left\Vert \mathbf{W} \right\Vert}_2 ,
\label{eq:cross-model_recommendation_1}
\end{equation}
where $\mathbf{U}^i=[\bm{u}_1^i,\bm{u}_2^i, \cdots ,\bm{u}_{\vert U \vert}^i]$. 
The corresponding columns are the same user's representations on two platforms/media. 
$\lambda$ is the weighting parameter and the above ridge regression problem has an 
analytical solution. Instead of pursuing hard transfer, Yan~\etal~\cite{yan2014mining} propose a 
topic association framework based on latent attribute sparse coding. 
They also show that bridging information across different media 
in common latent space outperforms explicit matrix-oriented transfer. 
The objective function of the above association framework is shown in~\eqref{eq:cross-model_recommendation_2}:

\begin{equation}
\begin{aligned}
\min_{\mathbf{D}^1, \mathbf{D}^2, \mathbf{S}} & {\left\Vert \mathbf{U}^1 - \mathbf{D}^1 \mathbf{S} \right\Vert}_{\rm F}^2 + {\left\Vert \mathbf{U}^2 - \mathbf{D}^2 \mathbf{S} \right\Vert}_{\rm F}^2 + \lambda {\left\Vert \mathbf{S} \right\Vert}_1 \\ s.t. & {\left\Vert \bm{d}_i^Y \right\Vert}_2^2 \le 1, {\left\Vert \bm{d}_j^T \right\Vert}_2^2 \le 1, \forall i, j ,
\label{eq:cross-model_recommendation_2}
\end{aligned}
\end{equation}
where $\mathbf{D}^i$ includes user factors, and $\mathbf{S}$ includes user attribute representations. 
The constraint ${\left\Vert d \right\Vert}_2^2 \le 1$ aims to prevent $\mathbf{D}$ from being arbitrarily large. 
$\mathcal{L}1$-norm penalty is adopted to encourage a compact and sparse attribute distribution space for users. 
This problem can be efficiently solved by the sparse coding algorithm proposed in~\cite{lee2007efficient} 
after a few transformations.
\blue{
	As is shown in Figure~\ref{fig:emcdr}, Man~\etal~\cite{maaten2008visualizing} propose an embedding and mapping framework EMCDR 
	in which user representations on different platforms are first obtained through matrix factorization and then 
	mapped via linear mapping or multi-layer perceptron (MLP). 
}

The optimization problem can be formalized as:

\begin{equation}
\min_{\theta} \sum_{\bm{u} \in \mathbf{U}} {\left\Vert f_{mlp} (\bm{u}^1;\theta) - \bm{u}^2\right\Vert}_2^2 ,
\label{eq:cross-model_recommendation_3}
\end{equation}
where $f_{mlp} (\cdot;\theta)$ is the MLP mapping function, and $\theta$ is its parameter set.
Abel~\etal~\cite{abel2011analyzing} aggregate user profiles on Flickr, Twitter, Delicious, 
and propose a solution for the cold-start problem in recommendation. 
By utilizing the overlapped users and items as bridges across different media , 
TLRec~\cite{Chen2012TLRec} introduces a smoothness constraint and regularization for latent vectors. 
Later, Jiang~\etal introduce an aligned cross-modal user behavior similarity constraint 
via proposing the XPTrans model~\cite{jiang2016little} which exploits a small number of overlapped 
crowds to bridge different media optimally. 
\blue{
	The objective function of XPTrans model is as follows:
}

\begin{equation}
\blue{
	\begin{aligned}
	J &= {\left\Vert \mathbf{W}^1 \odot (\mathbf{R}^1 - \mathbf{U}^1 \mathbf{V}^1) \right\Vert}_{\rm F}^2 \\
	&+ \lambda {\left\Vert \mathbf{W}^2 \odot (\mathbf{R}^2 - \mathbf{U}^2 \mathbf{V}^2) \right\Vert}_{\rm F}^2 \\
	&+ \mu ({\left\Vert \mathbf{W}^{1,2} \mathbf{1}^2 {\mathbf{W}^{1,2}}^T \odot \mathbf{U}^1 {\mathbf{U}^1}^T \odot \mathbf{U}^1 {\mathbf{U}^1}^T \right\Vert} \\
	&+ {\left\Vert {\mathbf{W}^{1,2}}^T \mathbf{1}^1 \mathbf{W}^{1,2} \odot \mathbf{U}^2 {\mathbf{U}^2}^T \odot \mathbf{U}^2 {\mathbf{U}^2}^T \right\Vert} \\
	& -2{\left\Vert \mathbf{U}^1 {\mathbf{U}^1}^T \mathbf{W}^{1,2} \mathbf{U}^2 {\mathbf{U}^2}^T {\mathbf{W}^{1,2}}^T \right\Vert}
	) \\
	& s.t. \mathbf{U}^1>0, \mathbf{V}^1>0, \mathbf{U}^2>0, \mathbf{V}^2>0,
	\label{eq:xptrans}
	\end{aligned}
}
\end{equation}
where the first two lines are traditional loss of matrix factorization on two platforms, and the following three lines are the derived similarity constraint.

\begin{figure}[tb]
	\centering
	\includegraphics[width=0.7\linewidth, height = 10cm]  {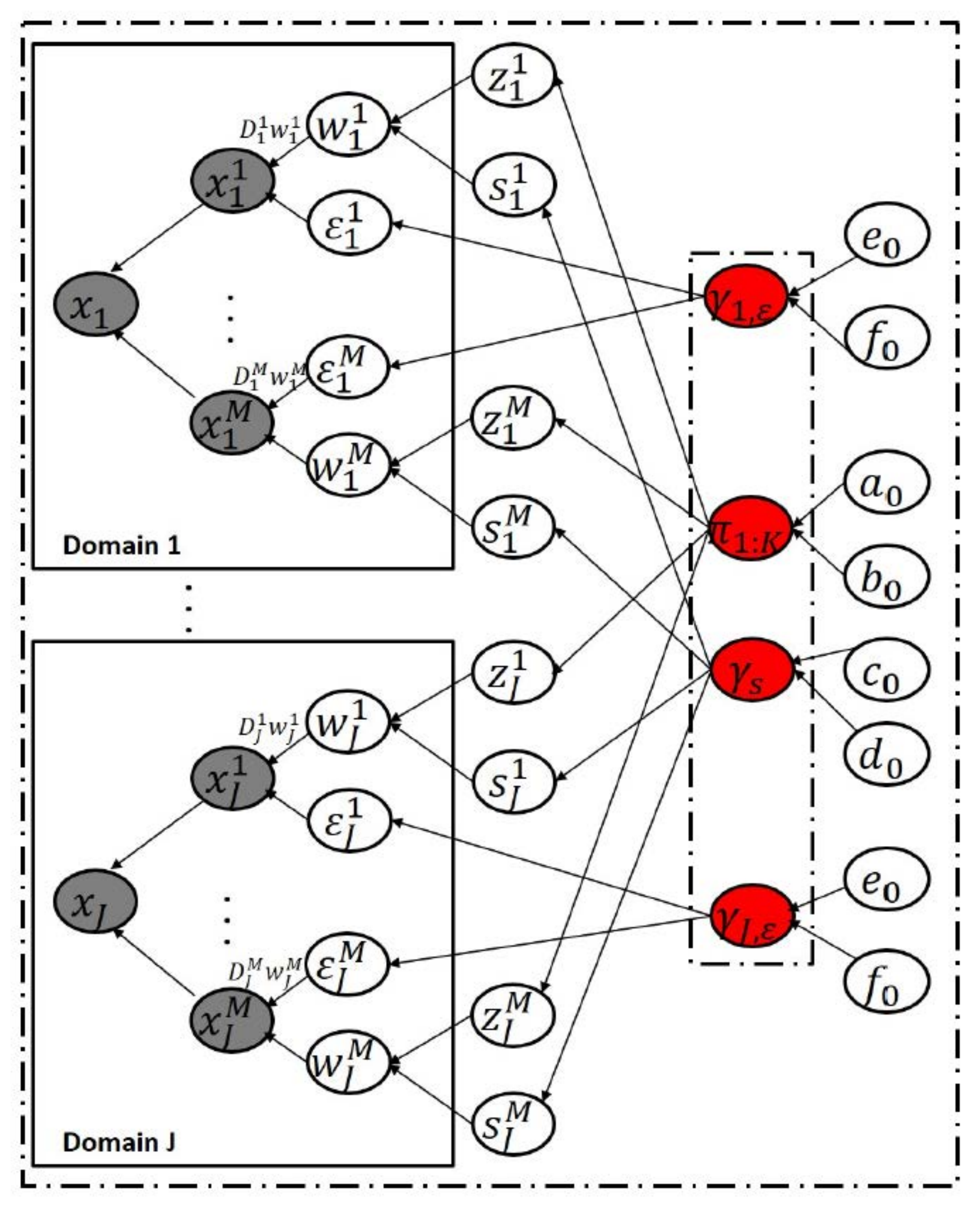}
	\caption{\blue{The graphical representation of the cross-domain collaborative learning (CDCL) algorithm. 
			The red circles represent the shared priors to associate with the relevant information and collaboratively learn the shared feature space in different domains, figure from~\cite{qian2015cross}.}}
	\label{fig:cdcl}
\end{figure}

The other group of methods are devoted to taking advantage of different media characteristics 
towards collaborative applications.
CODEBOOK~\cite{li2009can} investigates behavior prediction across Netflix and MovieLens without 
considering the overlapped users under the assumption that they share the same user-item rating patterns. 
Roy~\etal~\cite{roy2012socialtransfer} exploit 
real-time and socialized characteristics of tweets from Twitter to facilitate video recommendations on YouTube. 
TPCF~\cite{jing2014transfer} integrates three types of data, i.e., aligned users, aligned items and 
user-item ratings , in transfer learning for collaborative filtering. 
Qian~\etal~\cite{qian2015cross} propose a generic cross-domain collaborative learning (CDCL) framework based on
nonparametric Bayesian dictionary learning for cross-modal data analysis \blue{as is shown in Figure~\ref{fig:cdcl}.} 
Min~\etal~\cite{min2015cross} 
develop a multi-modal topic model capable of differentiating topics across modalities.

{\it Grouping by entire structure.}
When looking through existing literature with respect to the entire structure, 
one group of methods are designed to build a unified 
framework~\cite{Chen2012TLRec, jiang2016little, jing2014transfer, qian2015cross, min2015cross}
in which the first two works utilize matrix factorization based techniques and the latter three employ 
probabilistic model based strategies. 
Another group of works adopt a two-step procedure~\cite{yan2014mining, yan2015unified, Man2017Cross} 
by first representing users from different media in their own latent spaces and then jointly associating 
those representations. 

\begin{figure}[tb]
	\centering
	\includegraphics[width=0.98\linewidth]  {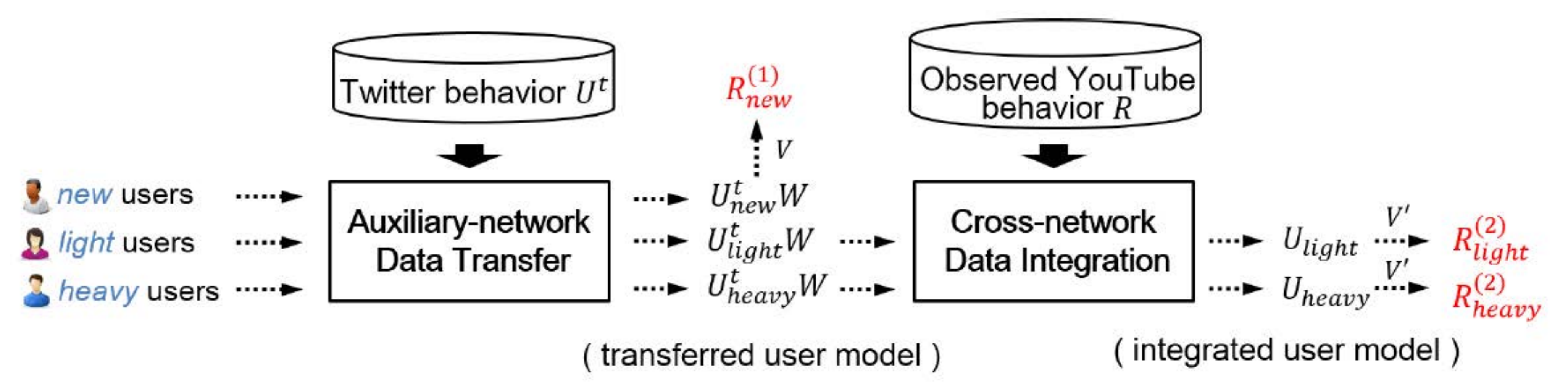}
	\caption{\blue{The proposed cross-network collaboration solution framework for unified YouTube video recommendation, figure from~\cite{yan2015unified}.}}
	\label{fig:yan2015}
\end{figure}

The above mentioned methods hold the same core idea that all cross-modal information is consistent 
and should be aligned. However, a few works~\cite{lu2013selective,yan2015unified} discover 
extra domain knowledge confirming the existence of data inconsistency phenomenon in the procedure of 
associating representations across different media, and attempt to solve this problem through data selection. 
Lu~\etal~\cite{lu2013selective} find that selecting media-consistent 
auxiliary data is important for cross-modal collaborative filtering. 
They propose a novel criterion based on empirical prediction error and variance to assess the consistency, 
and incorporate the criterion into a boosting framework to selectively transfer knowledge.
\blue{As is shown in Figure~\ref{fig:yan2015}}, Yan~\etal~\cite{yan2015unified} divide users into three groups and propose a predefined micro-level 
user-specific metric to adaptively weight data while integrating heterogeneous data across different media. 

\begin{figure}[tb]
	\centering
	\includegraphics[width=0.98\linewidth]  {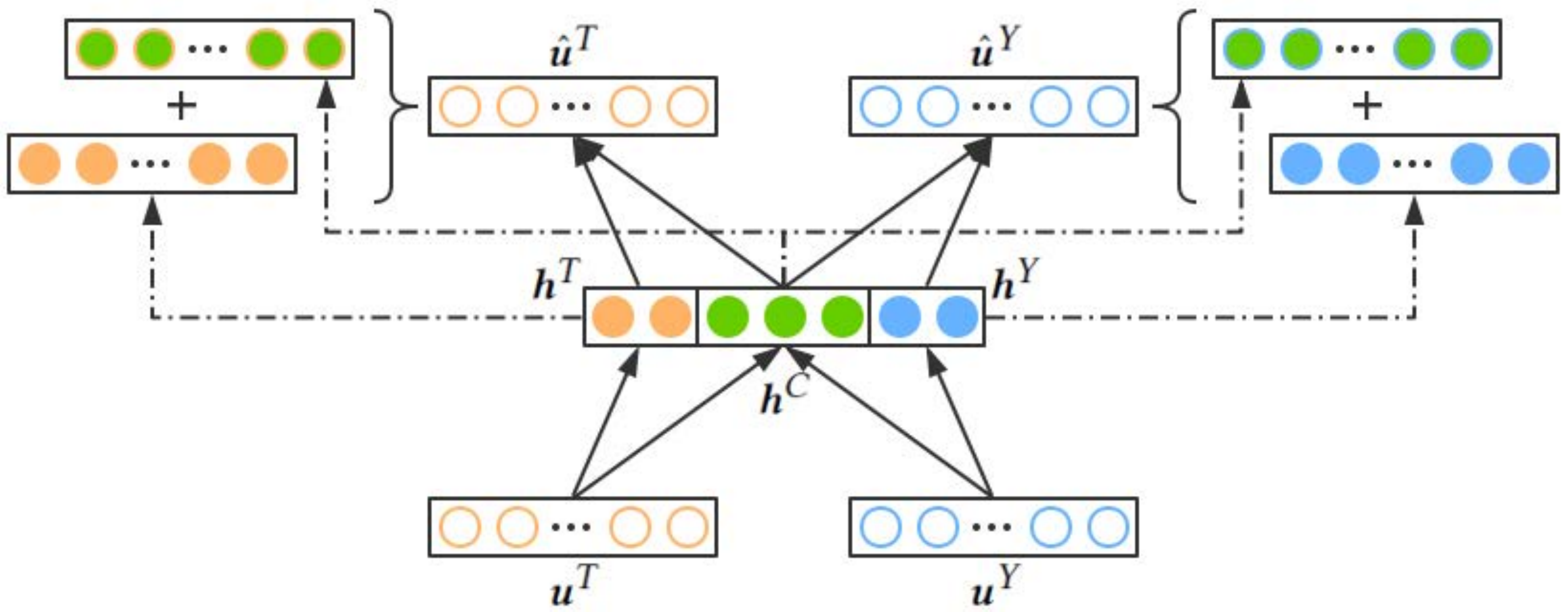}
	\caption{Disparity-preserved Deep Cross-platform Association model. $\bm{u}^T$ and $\bm{u}^Y$ are representations of an overlapped user on Twitter and YouTube. In latent representations, $\bm{h}^T$ and $\bm{h}^Y$ are media-specific parts preserving disparities, while $\bm{h}^C$ is the media-shared part associating representations in different media. The estimated representations $\bm{\hat{u}}^T$ and $\bm{\hat{u}}^Y$ are derived from both media-shared and media-specific parts, figure from~\cite{yu2018disparity}.}
	\label{fig:dca-association}
\end{figure}

\begin{figure*}[tb]
	\centering
	\includegraphics[width=0.8\linewidth]  {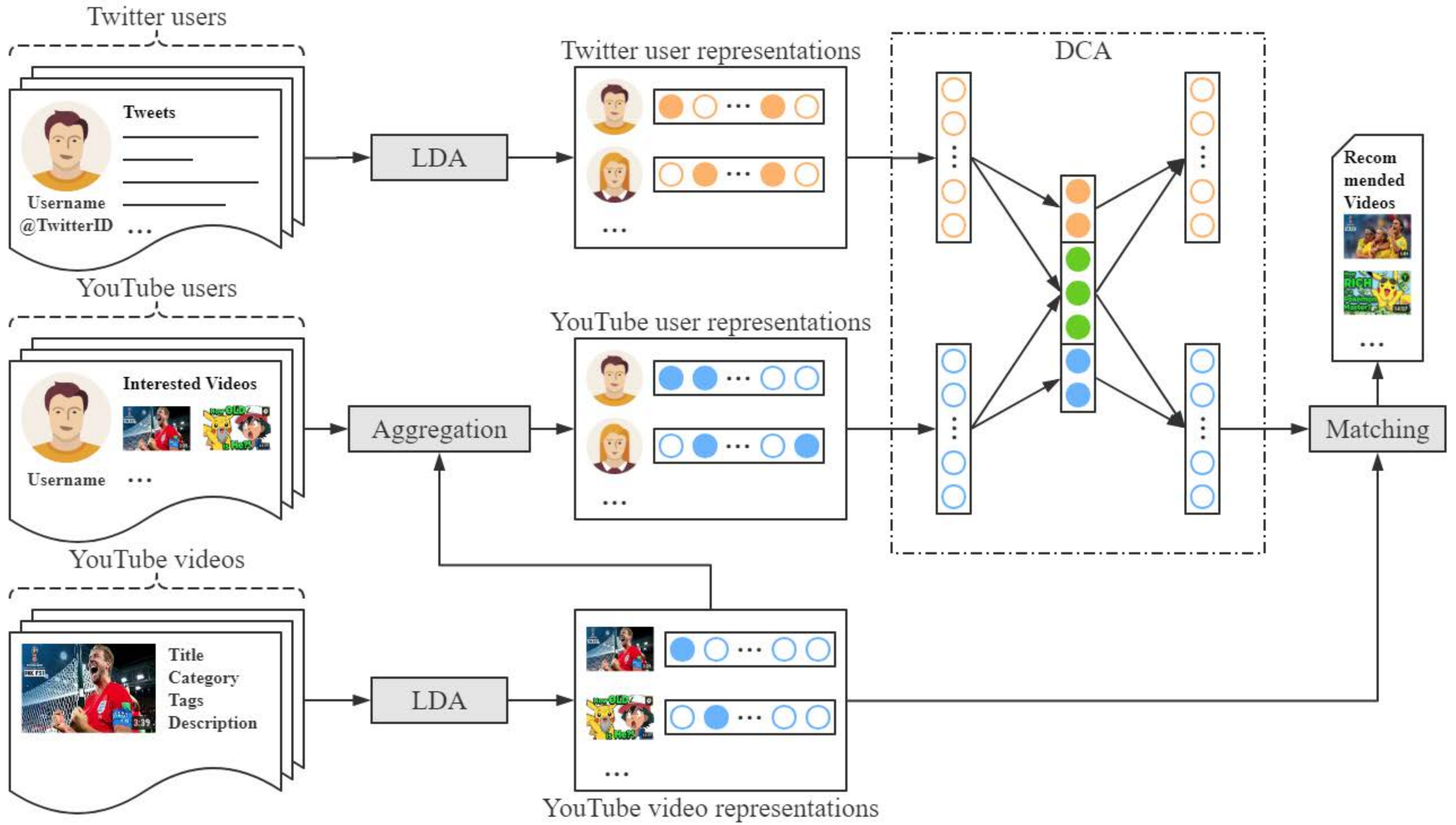}
	\caption{Framwork for disparity-awared deep cross-modal video recommendation, figure from~\cite{yu2018disparity}.}
	\label{fig:dca-whole}
\end{figure*}

In particular, Yu~\etal~\cite{yu2018disparity} analyze the inconsistent behavior patterns of users in Twitter and YouTube by
utilizing the domain knowledge of data inconsistency,  
and discover that the inconsistency is mainly caused by media-specific disparity ---  each individual's inherent 
personal preference consists of a media-shared part and a media-specific part due to 
users' different focuses in different media. To tackle the problem of media-specific disparity and granularity difference, 
they propose a disparity-preserved deep cross-platform association model whose core idea is shown in 
Figure~\ref{fig:dca-association}. Their proposed model contains a partially-connected multi-modal autoencoder 
which explicitly captures and preserves media-specific disparities in latent representations. 
They divide the hidden layer into $\bm{h} = [\bm{h}^T, \bm{h}^C, \bm{h}^Y]$, where $\bm{h}^T$, $\bm{h}^Y$ are Twitter,
YouTube media-specific parts respectively, and $\bm{h}^C$ is the media-shared part. 
Moreover, they also introduce nonlinear mapping functions to associate cross-modal information, 
which is advantageous in handling the granularity difference. 
The detailed structure of multi-modal autoencoder can be written as follows:

\begin{equation}
\begin{split}
\bm{h}&= g \left[ \left( \sum_i \mathbf{W}_1^i \bm{x}^i \right) + \bm{b}_1 \right] \\
\bm{\hat{x}}^i&= g \left( \mathbf{W}_2^i \bm{h} + \bm{b}_2^i \right) ,
\label{eq:cross-model_recommendation_4}
\end{split}
\end{equation}
where $i \in \{T,Y\}$ denotes Twitter or YouTube, and the weights on the unnecessary links are all set to zero. 
Weight matrices $\mathbf{W}$ and bias units $\bm{b}$ are denoted by $\theta$ as parameters of multi-modal autoencoder. $g(\cdot)$ is the Sigmoid activation function.
The total loss consists of reconstruction error, parameter regularizer (regularization penalty) and sparsity constraint,
as is shown in~\eqref{eq:cross-model_recommendation_5}:

\begin{equation}
\begin{aligned}
L(\bm{x}^i;\theta) &= \sum_i {\left\Vert\bm{\hat{x}}^i - \bm{x}^i\right\Vert}_2^2 \\
&+\lambda \sum_{\mathbf{W} \in \theta} {\left\Vert\mathbf{W}\right\Vert}_{\rm F}^2 + \mu {\left\Vert\bm{h}\right\Vert}_1.
\label{eq:cross-model_recommendation_5}
\end{aligned}
\end{equation}
The whole framework of disparity-aware cross-modal video recommendation is presented in Figure~\ref{fig:dca-whole}.

\section{Future Research Directions}
\label{sec:future}

We have presented approaches on multi-modal analysis for multimedia 
and discussed literature on data-driven cross-modal correlational representation 
and knowledge-guided multi-modal fusion. With current approaches,
the fusion of {\it continuous data} and {\it discrete knowledge} has been successfully
handled. However, there are still great challenges in obtaining the ability of reasoning for 
multimedia intelligence. In this section, we share our
insights on future directions for multi-modal research.

\spara{Cross-modal reasoning}
If we take another look at the above two aspects from the perspective that how
close the corresponding approaches/models are to the real intelligence like human beings, the results
would probably be ``both still have a long long way to go'' --- the later one may be closer to 
the real intelligent agent because human can always utilize knowledge from relevant domains to help make
decisions. Moreover, if we think deeper about what makes current algorithms take a further step towards
human intelligence, the answer will be ''reasoning". The ability of reasoning distinguishes
human being from animals. 
One representative embodiment of reasoning lies in the process of communications
among humans --- the ability of reasoning meanings of spoken languages during a conversation or 
major ideas of written articles when reading becomes a vital necessity in understanding each other.
This being the case, cross-modal intelligence reasoning over the evolution of knowledge serves as
a key solution in bridging the gap between current machine learning algorithms and human intelligence.
It will result in a more human-like cognition in cross-modal intelligence. 
Therefore, being capable of performing human-like reasoning over various kinds of knowledge in 
cross-modal analysis may be an great opportunity for the next breakthrough in artificial intelligence.

\spara{Cross-modal cognition}
Let us consider another question: how do human learn and how can infants learn so well?
The ability of cognition through ``real'' understanding the world could be one main answer to the question. 
We continuously learn different skills (tasks)
since we are babies and obtaining new skills (learning new tasks) seldom deteriorates our 
possessions of old skills (learned old tasks). Most existing machine learning algorithms are capable of 
tackling only one single type of task. For instance, an image classification algorithm can hardly solve 
(or achieve a very poor performance on) the trajectory prediction problem, 
although both image classification and trajectory prediction can be handled by human easily.
This indicates that the ability of learning to solve new tasks while maintaining the capacity
to tackle previous tasks (a reflection of cognitive process) plays a crucial role in generating human-like algorithms
for cross-modal analysis.

We remark that as another reflection of cognition, {\it commonsense learning} will be an effective path to the goal of touching
real human intelligence. Just imagine what kind of scene will appear in your mind when seeing the 
following sentence ``Tom picks up his bag and goes out'': Tom is probably a man who is at work, he
stretches out his arm and holds the grip of his bag, stands up and walks to the door, opens it and goes 
out --- Tom does not fly or crawl to the door, nor does he go out by walking through the wall.
It is obvious that none of the existing models are able to obtain the above knowledge which is easily
understood by human given the quoted sentence as input.
We call the process of learning such commonsense knowledge {\it commonsense learning}, which may lead to
another breakthrough in research on cross-modal intelligence.

\spara{Cross-modal collective intelligence}
The concept of collective intelligence (also refers as wisdom of crowd) 
was originally derived from the observations of entomologist William Morton Wheeler.
On the surface, independent individuals can work very closely so that they look like a single organism.
In 1911, Wheeler observed such a collaborative process indeed works on ants. An ant behaves like an animal's cell and processed 
the ability of collective thinking.
He called these collective ants a larger creature, namely the cluster of ant colonies seems to form a ``superorganism''.
In human society, given that decisions made by a single individual tend to be inaccurate compared to decisions made by the majority,
collective intelligence becomes a shared intelligence as well as the process of assembling opinions and turning them into 
decision-making procedure.
Wikipedia, as a type of media that fully demonstrates collective intelligence,
serves as an encyclopedia that can be changed by anyone at almost any time, which connects people on the web to create a 
huge intelligent brain.
All these phenomena or instances confirm one thing, i.e., collective intelligence can produce a more power 
``superorganism'' or brain that possesses more intelligence. 
With abundant cross-modal information, we believe that collective intelligence can be employed for human planning
which is another unique and complex characteristic shared by human being.

In addition, it is also desirable that the advances in cross-media intelligence can indeed make some contributions to human society.
Current approaches have done a good job on modality adaption, but they seldom can achieve good performances 
on {\it cross-modal generation}.
Let's take visual impaired people as an example, people with visual handicap usually wear a special-tailored helmet with a distance
sensor on it. This helmet will produce some noises when there exist some obstacles within a certain distance from the people wearing it.
We believe it will be a significant help towards visual impaired people if the helmet can act as an ``artificial eye'' by describing how far 
and what obstacle(s) are in which direction from him. This could be accomplished by generating logical verbal languages from understanding
sensory data.
In general, there is still a large potential space of improvement for cross-media intelligence in both methodologies and applications.

\section{Conclusion}
\label{sec:conclusion}

In this article, we give a comprehensive and deep investigation on multi-modal
analysis. 
We present two scientific problems on multi-modal analysis for multimedia. 
In order to address these two scientific problems,  
we discuss multi-modal fusion methods in two aspects: 
1) data-driven multi-modal correlational representation and 2) knowledge-guided multi-modal fusion.
We first give a brief summary on multi-task and multi-view learning, and target works on deep representation, 
transfer learning as well as hashing for data-driven correlational representation.
We then present our ideas on potential methods suitable for handling the fusion of multi-modal data and domain knowledge, 
and discuss approaches for four promising applications, 
i.e., visual question answering, video summarization, visual pattern mining and recommendation,
which need diverse domain knowledge for multi-modal fusion of data
with knowledge.
Last but not least, we highlight some insights on future research directions
in the new era of artificial intelligence, and point out a few promising future directions, including: 
cross-modal reasoning , cross-modal cognition and cross-modal collective intelligence,
for further investigation.
We believe these directions have a great potential to lead the next breakthrough in
cross-media intelligence.

\section*{Acknowledgment}
We thank Guohao Li, Shengze Yu and Yitian Yuan for providing relevant materials and valuable
opinions. This work will never be accomplished without their useful suggestions.



\begin{IEEEbiography}[{\includegraphics[width=1in,height=1.25in,clip,keepaspectratio]{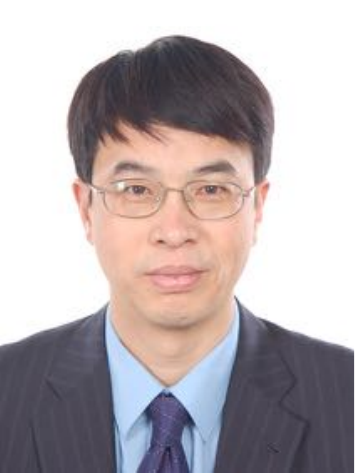}}]{Wenwu Zhu}
is currently a Professor and the Vice Chair of the Department of Computer Science and Technology at Tsinghua University, the Vice Dean of National Research Center for Information Science and Technology, and the Vice Director of Tsinghua Center for Big Data. Prior to his current post, he was a Senior Researcher and Research Manager at Microsoft Research Asia. He was the Chief Scientist and Director at Intel Research China from 2004 to 2008. He worked at Bell Labs New Jersey as Member of Technical Staff during 1996-1999. He received his Ph.D. degree from New York University in 1996 in Electrical and Computer Engineering.

Wenwu Zhu is an AAAS Fellow, IEEE Fellow, SPIE Fellow, and a member of The Academy of Europe (Academia Europaea). He has published over 300 referred papers in the areas of multimedia computing, communications and networking, and big data. He is inventor or co-inventor of over 50 patents. He received seven Best Paper Awards, including ACM Multimedia 2012 and IEEE Transactions on Circuits and Systems for Video Technology in 2001. His current research interests are in the area of Cyber-Physical-Human big data computing, and Cross-media big data and intelligence. 

Wenwu Zhu currently serves as EiC for IEEE Transactions on Multimedia. He served as Guest Editors for the Proceedings of the IEEE, IEEE Journal on Selected Areas in Communications, ACM Transactions on Intelligent Systems and Technology, etc.; and Associate Editors for IEEE Transactions on Mobile Computing, ACM Transactions on Multimedia, IEEE Transactions on Circuits and Systems for Video Technology, and IEEE Transactions on Big Data, etc. He served in the steering committee for IEEE Transactions on Multimedia (2015-2016) and IEEE Transactions on Mobile Computing (2007-2010), respectively. He served as TPC Co-chair for ACM Multimedia 2014 and IEEE ISCAS 2013, respectively. He serves as General Co-Chair for ACM Multimedia 2018 and ACM CIKM 2019, respectively.
\end{IEEEbiography}

\begin{IEEEbiography}[{\includegraphics[width=1in,height=1.25in,clip,keepaspectratio]{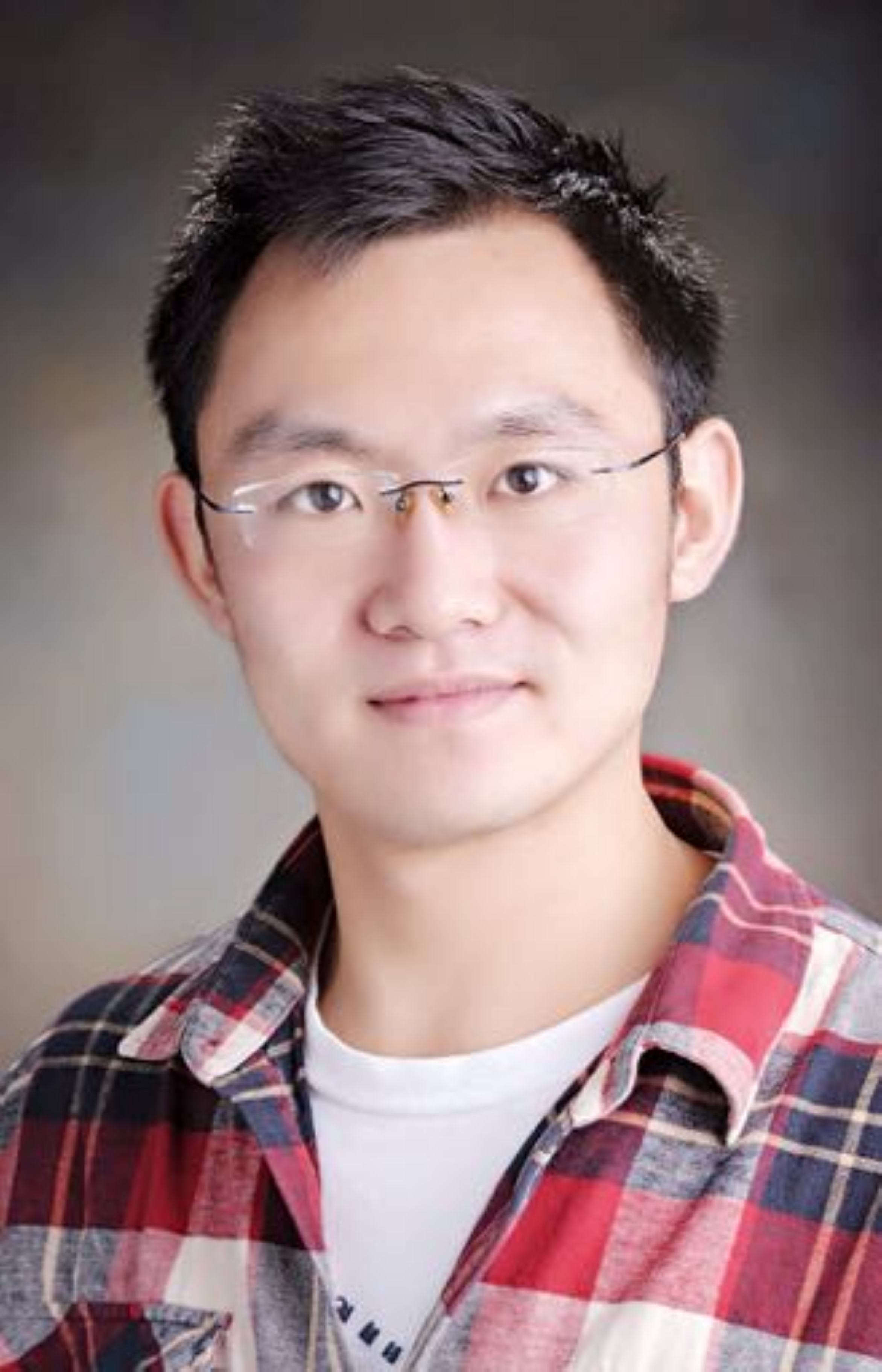}}]{Xin Wang}
is currently an Assistant Researcher at the Department of Computer Science and Technology, Tsinghua University. He got both of his Ph.D. and B.E degrees in Computer Science and Technology 
from Zhejiang University, China. He also holds a Ph.D. degree in Computing Science from Simon Fraser University, Canada. 
His research interests include cross-modal multimedia intelligence and inferable recommendation in social media. 
He has published several high-quality research papers in top conferences including ICML, MM, KDD, WWW, SIGIR etc.
He is the recipient of 2017 China Postdoctoral innovative talents supporting program. 
\end{IEEEbiography}

\begin{IEEEbiography}[{\includegraphics[width=1in,height=1.25in,clip,keepaspectratio]{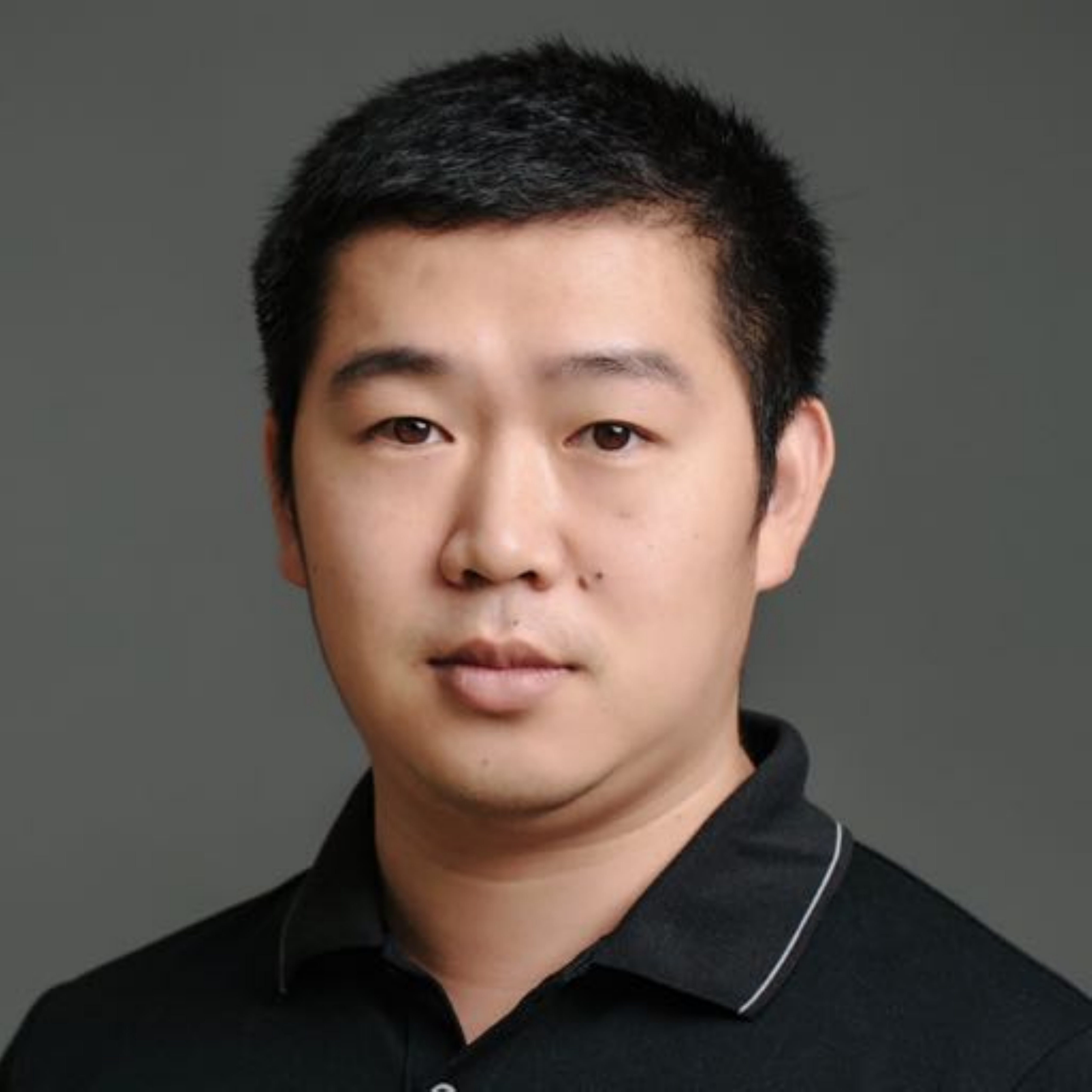}}]{Hongzhi Li}
is principal researcher and research manager
in Microsoft AI \& Research. His research interests are machine learning,
multimodal content analysis and cloud-based computing. His current research is
focused on deep learning for visual intelligence and its applications on cloud
computing platform. Dr. Li received his PhD degree in Computer Science from
Columbia University in 2016. Before that, he received his Bachelor and master’s
degree in computer science from Zhejiang University, China and Columbia
University, US, in 2010 and 2012, respectively. Dr. Li has published in ACM
Multimedia, TMM, ICMR, EMNLP, NAACL, SPIE and other venues. He received best
poster award in ACM ICMR 2018. He is the winner of grand challenge (first
place) in ACM Multimedia 2012. Dr. Li severed in program committee of major
international conferences, including ACM MM, ICME, IJCAI, etc. He also severed
as a reviewer in journals including IEEE TMM, IEEE TCSVT, TPAMI, JVCI, JVIS,
etc.
\end{IEEEbiography}

\end{document}